\documentclass[twocolumn]{aastex631}

\usepackage{afterpage}
\usepackage{amsmath}

\newcommand{\gaia}{\textit{Gaia}}

\shorttitle{Demographics and Structure of the Cep-Her Complex}
\shortauthors{Kerr et al.}

\begin{document}

\title{SPYGLASS. V. Spatially and Temporally Structured Star-Forming Environments in the Cepheus-Hercules Complex}

\correspondingauthor{Ronan Kerr}
\email{rmpkerr@utexas.edu}

\author[0000-0002-6549-9792]{Ronan Kerr}
\affiliation{Department of Astronomy, University of Texas at Austin\\
2515 Speedway, Stop C1400\\
Austin, Texas, USA 78712-1205\\}

\author[0000-0001-9811-568X]{Adam L. Kraus}
\affiliation{Department of Astronomy, University of Texas at Austin\\
2515 Speedway, Stop C1400\\
Austin, Texas, USA 78712-1205\\}

\author[0000-0001-9811-568X]{Daniel Krolikowski}
\affiliation{Steward Observatory, University of Arizona\\
933 N.\ Cherry Avenue\\
Tucson, Arizona, USA, 85721\\}

\author[0000-0002-0514-5538]{Luke G. Bouma}
\altaffiliation{51 Pegasi b Fellow}
\affiliation{Department of Astronomy, California Institute of Technology\\
1216 E. California Boulevard, Pasadena, CA 91125, USA\\}

\author[0000-0002-5851-2602]{Juan P. Farias}
\affiliation{Department of Astronomy, University of Texas at Austin\\
2515 Speedway, Stop C1400\\
Austin, Texas, USA 78712-1205\\}





\begin{abstract}

Young associations record complete histories of star forming events through their demographics and dynamics, and \textit{Gaia} has greatly expanded our knowledge of these associations. We present the first structural and dynamical overview study of the Cep-Her Complex, which has recently emerged as the largest stellar population within 500 pc that still lacks substantial follow-up. We reveal that Cep-Her is not a singular association, but rather an amalgam of four, consisting of the older ($\tau > 100$ Myr) open cluster Roslund 6, in addition to three dynamically coherent and highly substructured young associations that we focus on: Orpheus (25-40 Myr), Cinyras (28-43 Myr), and Cupavo (54-80 Myr). With $9552 \pm 960$ stars in Orpheus, $3872 \pm 455$ stars in Cinyras, and $8794 \pm 1827$ stars in Cupavo, all three are among the largest young associations within 500 pc, rivalling major associations like Sco-Cen. Our novel view of the ages and dynamics of these associations reveals evidence for sequential star formation in Cinyras, in addition to a multi-origin pattern of stellar dispersal in Orpheus that may hint to the presence of multiple clouds at formation. Dynamical simulations indicate that while some embedded open clusters and central components of these associations are converging, they likely lack the mass necessary to capture one another and undergo hierarchical cluster assembly. Our results provide our first view of the complex star-forming environments that gave rise to the associations of Cep-Her, which will serve as important laboratories for future studies of star and planet formation. 

\end{abstract}



\section{Introduction} \label{sec:intro}

After formation, most nearby young stars can be found in largely unbound structures known as associations \citep[][]{deZeeuw99, lada03}. These associations leave a record of the star formation event that produced them, making them essential for revealing the evolution of star formation processes, especially on longer timescales \citep{Krumholz19, Krause20}. Within 50 Myr of formation, many associations are still evolving through a combination of gradual expansion in sparse regions and hierarchical assembly and relaxation in dense regions \citep[e.g.,][]{Lada84, Pecaut16, VazquezSemadeni17}. Studies of young associations across a range of ages therefore trace the evolving morphologies and dynamical processes that lead to longer-lived structures such as open clusters \citep{Dalessandro21}. 

The study of young associations has progressed substantially in the last few years, accelerated by the monumental contributions of the Gaia spacecraft, which is providing precise stellar positions and motions for 2 billion nearby stars in the Milky Way \citep{GaiaDR3}. Many studies to date have taken advantage of \textit{Gaia}'s exquisite photometric and astrometric data, discovering stellar populations over a wide range of ages and scales \citep[e.g.,][]{Kounkel19, Sim19, Prisinzano22, Kerr21,Hunt23}. The surveys conducted through the SPYGLASS (Stars with Photometrically Young Gaia Luminosities Around the Solar System) program have discovered or more deeply characterized dozens of young ($\tau \la 50$ Myr) associations through \textit{Gaia}, detecting new associations across a range of scales and morphologies \citep{Kerr21,Kerr23}. \citet{Kerr21} (hereafter SPYGLASS-I) reported 27 top-level associations within 333 pc, nearly half of which remain largely unstudied to date, and \citet{Kerr23} (hereafter SPYGLASS-IV) expanded the survey to 1 kpc, revealing 116 young associations, 30 of which have no direct literature equivalent. The populations identified by these surveys vary from tenuous, low-density associations that barely emerge from the sky background, to large and massive populations with varied densities within, revealing an intricate network of interconnected star formation. 

Among the major emerging stellar complexes identified in SPYGLASS-IV is the Cep-Her complex, which connects the SPYGLASS-I populations of Cerberus, Lyra, and Cepheus-Cygnus with more distant open clusters like $\delta$ Lyr and RSG 5 \citep{Stephenson59, Roser16}. The presence of a larger population in the area was first noted by \citet{Zari18}, who observed overdensities of young stars in the direction of Cygnus and Lyra. \citet{Kounkel19} grouped these populations into the broadly-defined Stephenson 1 ($\delta$ Lyr Cluster) ``string'', which contains the populations identified by \citet{Zari18}, in addition to Cepheus Far North \citep[e.g.,][]{Kerr22a}. SPYGLASS-IV then provided the first contiguous definition of Cep-Her's extent in space-velocity coordinates, along with a comprehensive list of candidate members present in Gaia DR3. Consisting of 1164 population-defining photometrically young stars in addition to over $1.5\times10^5$ candidate members spanning more than 300 pc, SPYGLASS-IV revealed that Cep-Her may be one of the largest young associations within 500 pc \citep{deZeeuw99, Pecaut16, CantatGaudin19b}. However, unlike other large associations such as Sco-Cen and Vela, Cep-Her has never been the target of a dedicated survey, so its population, structure, dynamics, and star formation history are almost entirely unknown. Furthermore, while we know that the stars in Cep-Her are broadly young and co-moving in transverse velocity, we do not know whether they remain co-moving after the radial velocity component is considered, or whether they have ages consistent with a single, continuous star-forming event. Adding these new dimensions of analysis may therefore reveal entirely new star-forming environments that contribute to the Cep-Her Complex.  

The masses and velocities of dynamically coherent sub-populations in Cep-Her are critical for understanding the history and evolution of the association, which will inform the future fate of other nearby complexes like Sco-Cen, Vela, and Orion. The mass and subgroup sizes inform the virial state, which sets whether stars in a population can diverge freely, or are gravitationally bound within a dense open cluster \citep{Kuhn19}. This distinction is critical during dynamical traceback analyses, as the trajectories of stars in bound clusters often change dramatically due to two-body encounters and the overall gravitational field of the cluster. Unbound populations are then well-suited for traceback, as their motions are largely unchanged since formation. As has been shown in previous \textit{Gaia}-based studies \citep[e.g.,][]{MiretRoig20, MiretRoig22, Galli23} and SPYGLASS publications, the limited dynamical evolution in young associations after gas dispersal allows for their point of closest configuration at formation to be established, providing a straightforward method for computing dynamical ages \citep{Kerr22b, Kerr22a}. The balance between internal motions and gravitational binding can also inform the future evolution on the scale of an entire association, informing whether association subgroups will coalesce and hierarchically assemble into a single clustered population over time, or whether they will drift apart. The understudied nature of Cep-Her also opens the possibility of dynamically and demographically distinct structures that have not emerged in the 5D space-transverse velocity phase space that has been used in past studies. Identifying any such substructure will better focus future studies by identifying manageable sets of closely related stars which can be studied together. By characterizing Cep-Her substructure, we can also take advantage of Cep-Her's substantial overlap with the Kepler field by providing ages and formation environments for numerous young planet hosts, which are critical for studies of planet formation and evolution \citep{Bouma22}. 

In this publication, we complete the first ever comprehensive demographic and structural survey of the Cep-Her Complex.  We progressively divide the association using velocities and age to reveal its constituent associations and subgroups, and revise the association's stellar membership to estimate the total mass and population size of Cep-Her and its substructures. 
In Section \ref{sec:data}, we introduce the data used to complete this survey, which we use to assess memberships in Section \ref{sec:methods}. We then analyze Cep-Her's substructure in Section \ref{sec:clustering}, before analyzing the demographics of those substructures and the population more broadly in Section \ref{sec:demographics}. We then discuss the association's dynamics in Section \ref{sec:dynamics} and address the implications of Cep-Her's structure and dynamics in Section \ref{sec:discussion}, before concluding in Section \ref{sec:conclusion}. 

\section{Data} \label{sec:data}

\subsection{Sample Selection} \label{sec:sampleselect}

\begin{figure}
\centering
\includegraphics[width=8cm]{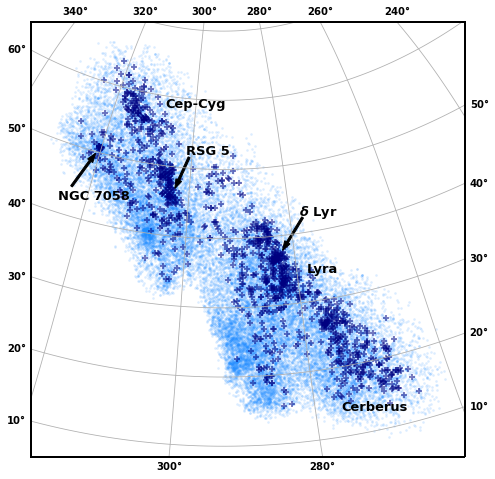}\hfill
\caption{Distribution of Cep-Her photometrically young founding members (dark crosses) and candidate members with ambiguous photometry that have with $P_{\rm spatial}>0.2$ (light dots), shown in RA/Dec sky coordinates. We also label the locations of a few notable open clusters with arrows, and also label the locations of the SPYGLASS-I groups of Cep-Cyg, Lyra, and Cerberus.}
\label{fig:cephermap}
\end{figure}

Our stellar census of Cep-Her is based on the sample that we defined in SPYGLASS-IV. This publication identified over 4$\times$10$^5$ photometrically young stars in Gaia DR3 using a Bayesian identification framework, which were clustered into 116 young populations within 1 kpc using the HDBSCAN clustering algorithm. The Cep-Her Complex (SCYA-96) was one of the largest populations identified in that work, with its base sample of photometrically young founding members consisting of 1164 stars. SPYGLASS-IV used that population of young stars to identify additional stars with positions and velocities consistent with membership, but uncertain photometric youth assessments. Young stars in this broader category are generally earlier type stars which have already merged onto the main sequence at the age of the Cep-Her, or have fallen close enough to the pre-main sequence to be photometrically indistinguishable from the field binary sequence. The fraction of the pre-main sequence that is indistinguishable from main sequence binaries increases as stars age, resulting in increasing rates of inconclusive photometric membership until the pre-main sequence and field binary sequence merge completely for $\tau>50-60$ Myr. To provide a non-photometric method of assessing stellar youth, SPYGLASS-IV includes an assessment of membership probability, $P_{\rm spatial}$ \citep[Section 5;][]{Kerr23}, which compares the relative abundances of likely young and likely old stars in space-velocity coordinates. Stars with $P_{\rm spatial} < 0.05$ are excluded from the extended sample of co-moving candidate members in SPYGLASS-IV, and the complete catalog of potential Cep-Her members contains 152936 stars. The distribution of Cep-Her founding members and candidates on-sky is shown in Figure \ref{fig:cephermap}.

As Cep-Her is a relatively old population \citep[29 Myr on average based on PARSEC isochrones;][]{Kerr23}, the SPYGLASS framework is expected to detect young members at a rate of 10-20\%. This suggests that the total population of the Cep-Her Complex is in the range of 5000 to 10000 members \citep{Kerr21}, although Cep-Her is distant enough that not all photometrically young stars will have high-quality \textit{Gaia} entries, suggesting the total population may be substantially higher. Nonetheless, with more than $1.5 \times 10^5$ stars in the extended population of candidates, we expect substantial field contamination rates, likely caused by a combination of its crowded environment near the galactic plane and spatial and dynamical dispersal of its members into the field since formation. This contamination is reflected in the photometric distribution of these candidate members, where the pre-main sequence does not cleanly separate from a dominant field main sequence. We therefore restrict the sample to only include stars with $P_{\rm spatial} > 0.2$ in our main analysis. This restricts the sample to 34886 candidates, reducing the population of field stars to allow the young association members to be separated. This cut certainly removes genuine members, however this is necessary to ensure manageable contamination rates at the outer edges of our sample. We discuss the population of genuine members that may have been removed by this cut in Section \ref{sec:completeness}.

\subsection{Supplemental Data}

High-resolution optical spectra provide radial velocities in addition to lithium 6708 \AA\ (Li) and H$\alpha$ equivalent widths (EWs), which are essential for membership verification and traceback in addition to dynamical and lithium depletion age measurements. The \textit{Gaia} sample provides RVs for approximately 18\% of the 34886 sources in our initial Cep-Her sample. However, \textit{Gaia} does not provide measurements of spectroscopic youth indicators, and \gaia~RVs have shown inconsistencies at the 1 km s$^{-1}$ level among young stars that prevent accurate dynamical age measurements \citep[e.g.,][]{Kerr22a}. Cep-Her has only recently been discovered in its entirety \citep{Zari18,Kerr23}, making its past spectroscopic coverage less complete compared to other nearby young associations. However, there are still large surveys that have covered Cep-Her candidate members, as well as some targeted studies that have focussed on well-known clusters within the larger sample like $\delta$ Lyrae. The coverage of Cep-Her is nonetheless far from complete, especially in lower-density subgroups, necessitating that these literature measurements be supplemented with new observations. We therefore use a combination of literature spectroscopic measurements and new observations to assemble a robust spectroscopic sample for Cep-Her. 

\subsubsection{Literature RV and Spectroscopic Line Measurements} \label{sec:litRVs}

To assemble a sample of literature spectroscopic measurements, we compiled a list of existing sources that provide RVs, Li EWs, or H$\alpha$ EWs. The search included large surveys like LAMOST, APOGEE, and \textit{Gaia} \citep[e.g., ][]{APOGEEDR17, LAMOSTSCRV, GaiaDR3}, as well as smaller surveys that appear in Vizier and Simbad searches of Cep-Her and other young associations in the SPYGLASS-IV sample. We also collected additional sources from \citet{Luhman22} and \citet{Zerjal23}, who provide their own compilations of recent spectroscopic studies of young stars. We additionally consult the RV source compilation in \citet{Anderson12}. We include sources with an RV quality grade better than a ``C''\footnote{A ``C'' quality source may have uncertainties that do not account for larger systematic errors}, which avoids sources with substantial uncorrected systematic uncertainties \citep{Anderson12}. The compilation from \citet{Gontcharov06} is taken to broadly represent observations which precede it, with the exception of the northern hemisphere survey provided by \citet{Famaey05}, which was listed as an ``A''\footnote{An ``A'' grade indicates reliable uncertainties} quality source separate from \citet{Gontcharov06} in \citet{Anderson12}. Studies that compile measurements from multiple sources are otherwise excluded, so our citations mark the publication where the measurement first appeared.

For targets with multiple observations, precedence was generally given to the source with the lower uncertainty. However, for sources with $\sigma_{RV}<1$ km s$^{-1}$, we de-prioritized sources with known systematic uncertainties at the 1 km s$^{-1}$ level so that high-quality sources are not overwritten in favor of observations with substantial systematic uncertainties \citep{Anderson12, LAMOSTSCRV}. The sources we deprioritized in this manner were \gaia, RAVE DR6 \citep{RAVEKunder17}, and all LAMOST catalogs \citep{LAMOSTDR8VAC} with the exception of the self-consistent MRS RV measurements described in \citet{LAMOSTSCRV}. A complete list of adopted measurements and their sources is provided in Table \ref{tab:spectres}. Most of the 6518 RVs we adopt are from \textit{Gaia} (5456), with the next largest RV sources being LAMOST (506 across LRS and MRS spectra), APOGEE (148), and \citep{Gontcharov06} (54). We also record flags for spectroscopic binaries where available, and include those flags in Table \ref{tab:spectres}.

\begin{deluxetable*}{cccccccccccc}
\tablecolumns{12}
\tablewidth{0pt}
\tabletypesize{\scriptsize}
\tablecaption{Cep-Her data containing both literature values and new observations. The source for each value is included. High-quality literature values supersede our observations in cases of poor RV results or low-signal line measurements. The full table, which contains 6518 entries, is available in the online-only version of this paper. }
\label{tab:spectres}
\tablehead{
\colhead{Gaia DR3 ID} &
\colhead{RA} &
\colhead{Dec} &
\multicolumn{3}{c}{RV (km s$^{-1}$)} &
\multicolumn{3}{c}{EW$_{Li}$ (\AA)} &
\multicolumn{3}{c}{EW$_{H\alpha}$ (\AA)} \\
\colhead{} &
\colhead{(deg)} &
\colhead{(deg)} &
\colhead{val} &
\colhead{err} &
\colhead{src\tablenotemark{a}} &
\colhead{val} &
\colhead{err} &
\colhead{src\tablenotemark{a}} &
\colhead{val} &
\colhead{err} &
\colhead{src\tablenotemark{a}} \\
}
\startdata
2069631150945233280 & 309.7628 & 43.4871 &  -19.92 &   0.04 &   HJST & 0.138 &   0.011 &     HJST &   0.00 &    0.00 &      HJST \\
2069673344703582592 & 309.6291 & 43.8329 &  -35.08 &   3.07 &   GDR3 &    &      &       &     &      &        \\
2069674512934689536 & 309.6933 & 43.8082 &  -14.54 &   4.94 &   GDR3 &    &      &       &     &      &        \\
2069755361399374080 & 308.7965 & 43.7338 &  -22.32 &   3.02 &   GDR3 &    &      &       &     &      &        \\
2069764535449283712 & 308.1051 & 43.3637 &  -21.91 &   1.19 &   GDR3 &    &      &       &     &      &        \\
2069765287066380416 & 308.3421 & 43.2745 &  -24.54 &   4.06 & LAM8LV &    &      &       &     &      &        \\
2069767902703713536 & 308.4111 & 43.4374 &  -20.25 &   0.03 &   HJST & 0.187 &   0.011 &     HJST &   0.00 &    0.00 &      HJST \\
2069836209864147840 & 309.2886 & 43.6723 &  -22.28 &   1.88 & LAM8LV &    &      &       &   2.42 &    0.02 & LAM8LMCAT \\
2069869848049047552 & 309.5119 & 44.1504 &  -21.37 &   3.41 &   GDR3 &    &      &       &     &      &        \\
2069875757923114240 & 309.5231 & 44.1710 &  -35.06 &   6.83 &   GDR3 &    &      &       &     &      &        \\
2069876170239978112 & 309.5263 & 44.2266 &  -20.05 &   0.04 &   HJST & 0.209 &   0.011 &     HJST &   0.00 &    0.02 &      HJST \\
2069890154653506432 & 309.5657 & 44.3327 &  -18.70 &   3.00 &     G6 &    &      &       &     &      &        \\
 \enddata
\tablenotetext{a}{The source for the measurement, either 2.7 Harlan J. Smith Telescope at the McDonald Observatory (HJST), Keck HIRES (HIRES), or an external source. The abbreviations for external sources are: Gaia DR3 (DR3), APOGEE DR17 \citep{APOGEEDR17} (AP17), GALAH DR3 \citep{GALAHDR3} (GAL3), LAMOST DR9 MRS spectra \citep{LAMOSTSCRV} (LAM9M), LAMOST DR8 LRS parameters of A, F, G, and K stars (LAM8L), LAMOST DR8 Catalog of M Stars (LAM8LMCAT), LAMOST DR8 Value Added Catalog \citep{LAMOSTDR8VAC} (LAM8LV), \citet{Famaey05} (F5), \citet{Gontcharov06} (G6), \citet{Grieves18} (G18), \citet{Klutsch20} (K20), \citet{Tautvaisiene22} (T22) } 
\vspace*{0.1in}
\end{deluxetable*}

\subsubsection{New Observations}

To deepen and broaden our spectroscopic coverage of Cep-Her, we performed new spectroscopic observations with two different telescopes and instruments: the Tull Coud\'e Echelle Spectrograph at the McDonald Observatory's 2.7m Harlan J. Smith Telescope \citep[HJST;][]{Tull95}, and the High Resolution Echelle Spectrometer on the Keck I 10m telescope \citep[HIRES;][]{Vogt1994}. Both instruments provide high-resolution optical spectra, with the Tull spectrograph providing non-contiguous coverage from 3400 and 10900 \AA~ at $R\approx60{,}000$, and HIRES coverage spanning 3600 to 7900 \AA~ at $R\approx 30{,}000$. The observational configurations were selected to cover the Li 6708 \AA~and H$\alpha$ lines wherever possible, with resolution sufficient for resolving the rotational broadening of these lines for each star in our sample. The high resolution of these instruments combines with their stable wavelength solutions to enable RV measurements with sub-km s$^{-1}$ precision. 

At the HJST, target selection was optimised to enable dynamical traceback and provide lithium depletion sequences across a diverse set of locations in Cep-Her. We first performed preliminary clustering, similar to what is described in Section \ref{sec:clustering}, identifying stars in regions where subgroup age measurements are likely possible. We then observed a sample of young stellar populations spread evenly throughout the preliminary subgroups. Stars in dense regions such as the $\delta$ Lyrae cluster were de-prioritized due to the higher likelihood of them being gravitationally bound, which prevents the free expansion of members after formation that simplifies dynamical analysis. Our limiting magnitude at the HJST was $m_G<14$, and we focused our observations on later-type stars that have fewer past observations, more informative lithium abundances, and less rotational broadening, which makes precise radial velocities more accessible. Our programs observed a total of 351 stars in Cep-Her.

The HIRES data were acquired as part of a targeted survey focused on two of the Cep-Her subgroups described by \citet{Bouma22} that contain statistically-validated transiting planets: RSG-5 and $\delta$~Lyr.  The main purpose of the program was to acquire spectra for stars spanning spectral types of M1V to G2V in order to improve prospects for lithium-dating the association.  We acquired spectra for 19 stars in $\delta$ Lyr, and 26 in RSG-5.  We selected these stars by beginning with the candidates reported in Table~2 of \citet{Bouma22}. We selected stars with ``weights'' exceeding 0.1, and removed any objects with visual, astrometric, spectroscopic, or photometric indications of binarity based on Gaia DR3.  We then required the stars to have detected rotation periods with either TESS or ZTF, and required them to be in the range of 12.0$<$V$<$16.2, where the faint limit was set by time and S/N constraints.  The final selected objects were a random draw from the set of objects that met these constraints.   The observations themselves were executed between April--July of 2023, without iodine, and otherwise using the standard setup and reduction techniques of the California Planet Survey \citep{Howard2010}.  

The HJST observations were processed and reduced following typical echelle spectroscopy procedures using a publicly available pipeline tailored to the Tull spectrograph\footnote{\url{https://github.com/dkrolikowski/tull_coude_reduction}}. We computed RVs for the HJST targets using spectral line broadening functions from the \texttt{saphires} package \citep{Tofflemire19}. This produced sub-km~s$^{-1}$ radial velocity measurements for 271 stars in our sample. For HIRES, we measured radial velocities using \texttt{SpecMatch-Synth} \citep{Petigura2017}, a synthetic template-matching code built on an interpolation of the model spectra from \citet{Coelho2005}. We also extracted Li and H$\alpha$ equivalent width measurements from both the HIRES and the HJST measurements by fitting those lines with Gaussian profiles, following the method described in SPYGLASS-II. We combine the final results for RV, Li EW, and H$\alpha$ EW from both HIRES and the HJST with the literature results described in Section \ref{sec:litRVs}, making final RV and EW selections according to the rules described there. 

\section{Membership} \label{sec:methods}

By combining the Cep-Her catalog provided in SPYGLASS-IV with new spectroscopic measurements, we can assess the membership probability of candidate members using four different membership markers: the local fractions of young and old stars, photometry, RVs, and Lithium. Together, these markers can be used to produce aggregate membership probabilities that take into account all available data, allowing us to estimate key demographic properties. Furthermore, we can use membership probabilities to remove probable non-members, producing a sufficiently pure sample to analyse the structure of the association. 

\subsection{Spatial-Photometric Membership Probability} \label{sec:photcuts}

SPYGLASS-IV has already produced measurements for probability of youth, $P_{\rm Age<50 Myr}$, which is available for all stars that pass the photometric and astrometric quality flags presented in that paper. SPYGLASS-IV also provides a probability of membership calculation, $P_{\rm spatial}$, which is based on the relative frequencies of high-confidence young and old objects among adjacent stars in space-velocity coordinates. Since $P_{\rm Age<50 Myr}$ is based on a constant star formation rate over the last 11.2 Gyr \citep{Binney00}, it implicitly assumes that $\simeq 0.4$\% of the sample has ages $\la 50$ Myr. However, the stars identified as members of Cep-Her in SPYGLASS-IV exist at points in parameter space where young stars are much more dominant, as indicated by their values of $P_{\rm spatial}$. We can therefore use the ratio between $P_{\rm spatial}$ and the field abundance of young stars to correct the relative abundances of young and old stars, re-calibrating $P_{\rm Age<50 Myr}$ for the environment in which each star is found.  

We start by defining a corrective factor $C_y$ by which the abundance of young stars in the original model must be inflated to recreate the environments in Cep-Her where far more young members are present: 

\begin{equation}
    C_{y} =\frac{P_{\rm spatial}}{P_i}
\end{equation}

\noindent where $P_i$ is the fraction of young stars in the original SPYGLASS model, or $P_i = 50/11200 \simeq 0.004$. This increase to the fraction of young stars in our model in turn reduces the fraction $C_o$ of older model stars by the following factor: 

\begin{equation}
    C_{o} = \frac{1 - P_{\rm spatial}}{1 - P_i}
\end{equation}

\noindent which is equal to the local spatial probability that the star is a non-member divided by the initial model probability that the star is an old non-member. Using these corrections to the fractions of old and young stars in the model, we can compute a renormalized spatial-photometric membership probability $P_{\rm sp}$ which accounts for both the photometric membership probability and local abundance of young members:

\begin{equation}
    P_{\rm sp} = \frac{(C_y \times P_{\rm Age<50 Myr})}{(C_y \times P_{\rm Age<50 Myr}) + (C_o \times (1 - P_{\rm Age<50 Myr}))}
\end{equation}

Before calculating $P_{\rm sp}$, we set the maximum value of $P_{\rm spatial}$ to 0.95 to ensure that stars with values of $P_{\rm spatial}$ that are close to 1 can still be plausibly photometrically removed from the sample. This cap also reflects results from when values of $P_{\rm spatial}$ were initially computed. In SPYGLASS-IV, $P_{\rm spatial}$ was generated by computing the relative abundances of young and old stars across bins defined in clustering proximity, $D$, a proxy for local density. A Gaussian CDF was fit to the curve of $D$ vs  $P_{\rm spatial}$ computed for that range of $D$, effectively smoothing the scatter in those bins. However, even for very high values of $D$, some old contaminants were found to still be present, albeit rare. Setting the maximum $P_{\rm spatial}$ to 0.95 approximately captures the maximum scatter in the computed values of $P_{\rm spatial}$ after the fit results for $P_{\rm spatial}$ exceed 0.99. The resulting values of $P_{\rm sp}$ provide the probability of youth conditioned on the local abundance of young members. We provide these results for all stars that pass the astrometric and photometric cuts that enable the calculation of $P_{\rm Age<50 Myr}$ in SPYGLASS-IV.

\subsection{Spectroscopic Membership} \label{sec:rvselect}

The new and existing spectroscopic measurements we have collected provide two additional ways of assessing membership: lithium depletion and radial velocity. Conclusions of membership or non-membership from velocity or lithium can therefore dramatically raise or lower the probability of membership for a spectroscopically observed star relative to its initial spatial-photometric membership probability. In this section we assess membership using these spectroscopic markers, taking into account the influence of field contamination and false negatives that can affect the results from these methods. We then compute final membership probabilities, which take into account these spectroscopic membership indicators.

\subsubsection{RV Selection}

Stars in our sample were identified as candidate members using the ubiquitous on-sky proper motion measurements from \textit{Gaia}. However, RVs are much less widespread, limiting their use in dynamical studies, especially in the almost unstudied Cep-Her Complex. These RVs complete the 3D stellar velocity vector, establishing whether a star is fully co-moving with the parent association and greatly strengthening membership assessment. 

\begin{figure}
\centering
\includegraphics[width=8.2cm]{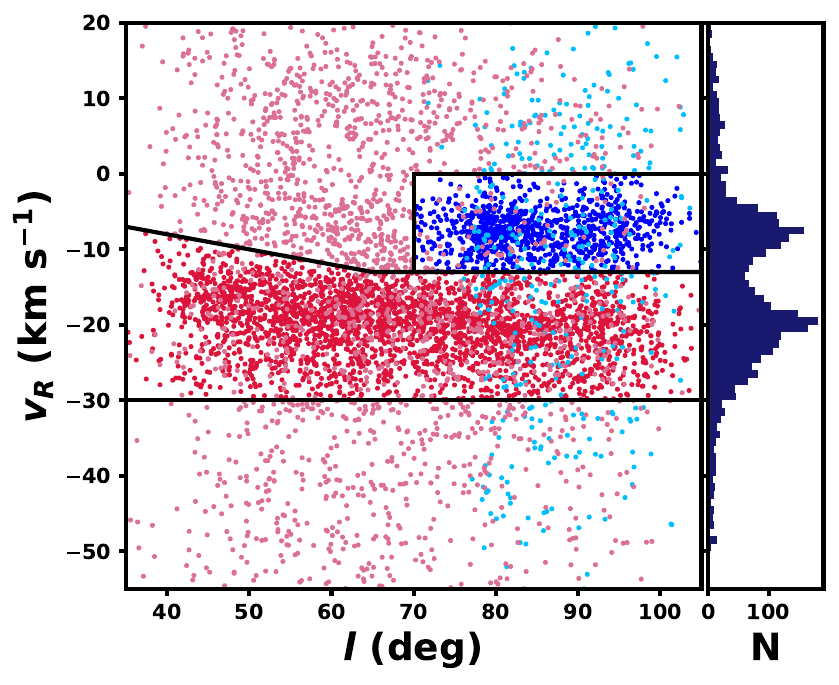}\hfill
\caption{Cep-Her candidate members in galactic $l$ sky coordinates, plotted against RVs from new observations and the literature. The result shows two distinct radial velocity sequences. The black lines represent the manual cuts used to separate Cep-Her Main (red) from Cep-Her North (blue) and the field (lighter shades of red and blue). Probable field stars are assigned to one of the two Cep-Her sub-regions according to their position in 5D space-transverse velocity coordinates for reassessment in the final RV selections (see Fig \ref{fig:rvselect}) and for use in assessments of contamination \ref{sec:contamination}. The histogram at the right shows the distribution of RVs for stars with $l>70$, showing the bimodal distribution of RVs over this range in $l$ where both RV sequences are present.}
\label{fig:rvbimod}
\end{figure}

In Cep-Her, the addition of this new dimension in velocity space reveals a distinct bimodality in Cep-Her's RV distribution. In Figure \ref{fig:rvbimod}, we show the distribution of RVs in Cep-Her as a function of galactic longitude ($l$). This figure reveals two distinct radial velocity sequences separated by more than 10 km/s on average. This RV split is present even among populations that overlap substantially in both space and transverse velocity coordinates. In limited cases, we find that the space-transverse velocity overlap can be so great that HDBSCAN clustering in 5D phase-space, as described in Section \ref{sec:clustering}, can preserve this bimodality within individual subgroups. Therefore, not only is it necessary to perform the RV cuts on the two samples separately, but the two populations must be separated entirely to provide internally consistent clustering results in five dimensions. 

We separate the two sequences with a manual cut consisting of two line segments, as shown in Figure \ref{fig:rvbimod}.  We refer to the larger population which resides below this line as Cep-Her Main, and stars above this line as Cep-Her North. The overdensity associated with Cep-Her North occupies a narrower range in $l$, so we also exclude stars with $l < 70$ from that clump. We do not apply this analysis to stars with poor RV measurements (defined as $\sigma_{RV}>5$ km s$^{-1}$), so that stars with unreliable radial velocity measurements are not grouped into one of those sequences according to those velocities. The stars in these two clumps define regions in 5D space-transverse velocity coordinates that can be used to group stars without reliable RVs into a parent subregion. 

We assign stars with poor RVs or no RVs to whichever population is closest in 5D space-transverse velocity coordinates. We used the clustering proximity $D_{N}$, which is defined as the distance to the Nth nearest member, as the metric to assess closeness. This HDBSCAN-inspired metric has been used in past SPYGLASS publications, where $D_{N}$ serves as a proxy for the inverse cube-root of the density surrounding a given star. We defined this distance in 5D $(X,Y,Z,c*\Delta$v$_{T,l},c*\Delta$v$_{T,b})$ space-transverse velocity anomaly coordinates, where the transverse velocity anomaly $\Delta$v$_{T}$ is defined as the transverse velocity for the star, minus the expected transverse velocity at its location given a median UVW 3D velocity vector. The constant $c$ matches the scales of the space and velocity dimensions, and its value is chosen to produce similar structure sizes in space and velocity coordinates. Here we select $c=12$ pc km$^{-1}$ s, following SPYGLASS-II and -III. The use transverse velocity anomaly instead of transverse velocity removes some projection effects, making populations with different velocities and widely-distributed on-sky locations easier to separate.  The value of $N$ in $D_N$ serves as a smoothing factor that affects sensitivity to small-scale structures, and we find that a value of $N = 20$ works well to separate these large-scale clumps. This value of $N$ provides more smoothing compared to the usage in SPYGLASS-IV, where $N = 10$ was employed to assign outlying members to a range of associations that include smaller-scale features.

We assign each star without a reliable radial velocity (no RV or $\sigma_{RV}>5$ km s$^{-1}$) to the group with which it has the lowest $D_{20}$, measuring $\Delta$v$_{T}$ relative to the median UVW velocity across all of Cep-Her. We apply the same approach to stars with RVs that do not fall in either clump to both reassess their membership with more robust group extents, and to assess field contamination in Section \ref{sec:contamination}. RV-separated groups that are otherwise contiguous in 5D space-velocity coordinates will experience some cross-contamination through this approach, and we discuss potential impacts on population demographics in Section \ref{sec:xcontam}. 

\begin{figure*}
\centering
\includegraphics[width=15cm]{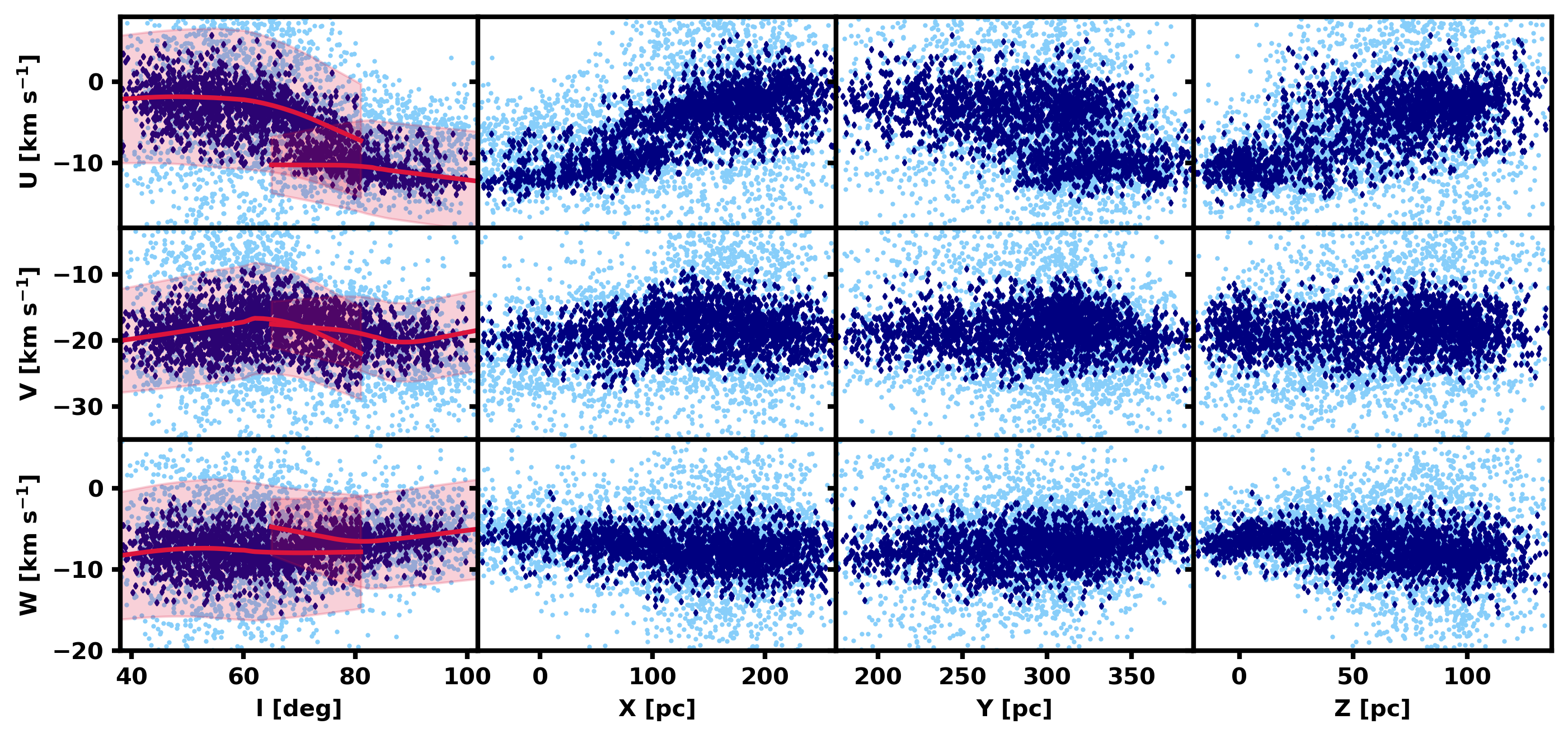}\hfill
\includegraphics[width=15cm]{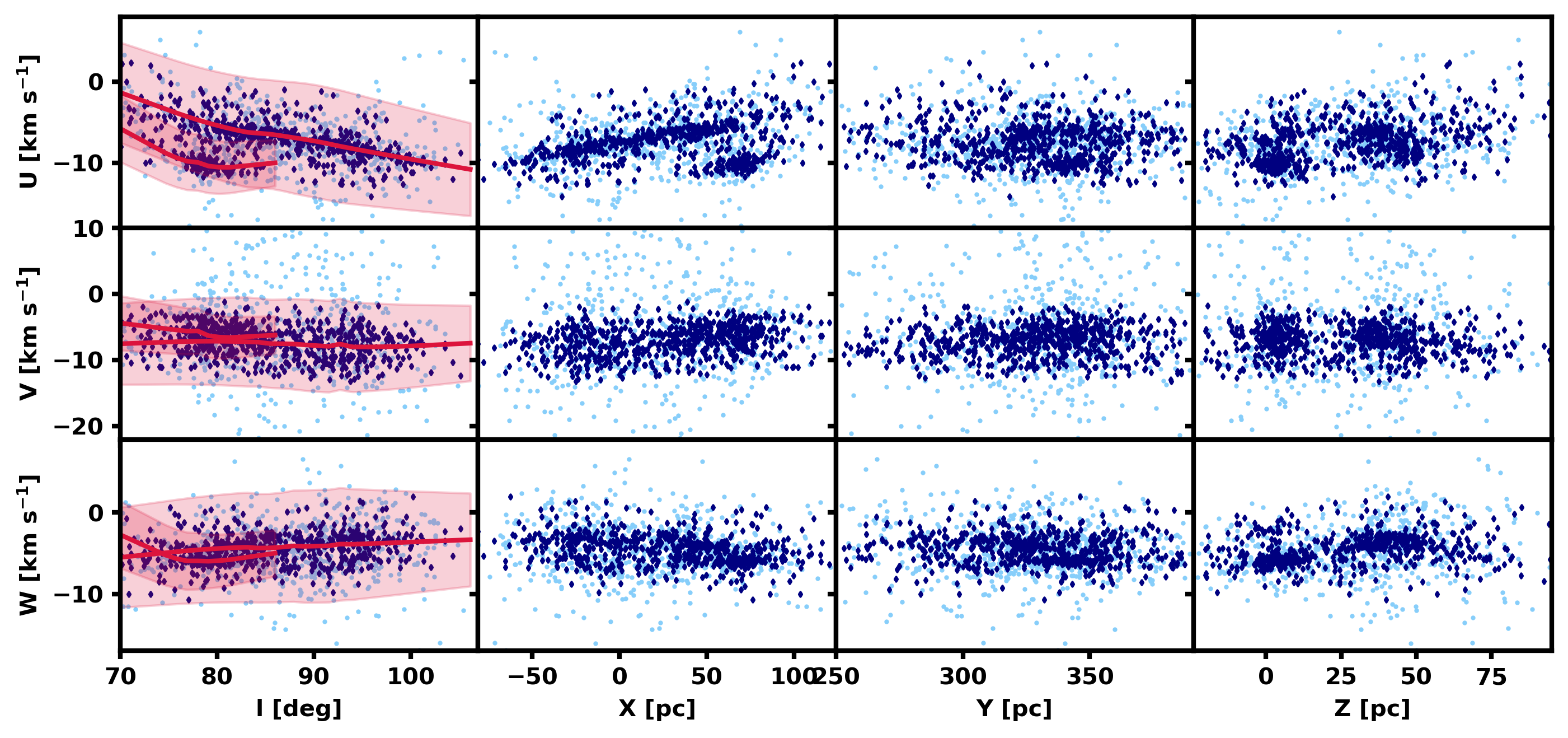}\hfill
\caption{Results of RV selection across a cross-section of space and velocity axes in Cep-Her Main (top) and Cep-Her North (bottom). Dark blue diamonds are accepted as kinematic members, and light blue dots are not. For the plots of velocity against galactic $l$ coordinates, we use a red curve to show the median values of U, V, and W as a function of $l$, which were used to define the velocity loci for the two subtly discontiguous sub-components of Cep-Her. The shaded red region encloses the maximum and minimum velocities which may be identified as members in that axis, as defined by the velocity axis, plus or minus the maximum velocity dispersion. }
\label{fig:rvselect}
\end{figure*}

Once all stars are assigned to a parent population, we make a final 3D kinematic membership assessment for stars with radial velocities. Both RV-defined sub-populations in Cep-Her contain two visibly discontiguous subregions each that emerge in $l$-transverse velocity space (Fig.~\ref{fig:rvselect}), so we draw a manual cut between them and consider their velocity distributions independently. We split each of the four components across Cep-Her Main and Cep-Her North into between 10 and 25 bins each in $l$, and in each bin we compute median values of transverse velocity and 3D UVW velocity. We also compute the maximum variance $\sigma_{v_{max}}$ in transverse velocity relative to the median, which informs the velocity spread in the plane of the sky. In regions of parameter space where Cep-Her Main and Cep-Her North overlap, we use the 95th percentile in transverse velocity rather than the maximum variance to reduce cross-contamination between the two populations. The number of bins is chosen by hand to both capture apparent structure and minimize scatter in the velocity results caused by small sample sizes. We smooth the results with a Savitsky-Golay filter in $l$ to further reduce the variations from bin to bin. Taking the transverse velocity extent to be a 2D projection of a spherical 3D distribution in UVW space, we assume that the maximum variance in 2D space is also the maximum variance in 3D space. We therefore identify stars within $\sigma_{v_{max}}$ of the UVW median as members with a velocity membership flag $V=1$, and identify stars outside this cut as velocity non-members ($V=-1$). We use an inconclusive membership flag $V=0$ for stars with $\sigma_{RV} > \sigma_{v_{max}}/2$, as their large uncertainties provide no strong membership constraints. We show the results of this selection in Figure \ref{fig:rvselect}. 

\subsubsection{RV False Positives from Field Contamination} \label{sec:contamination}

Much of Cep-Her and its adjacent associations lie close to the galactic midplane, making them subject to substantial field contamination. As a result, many field stars have velocities matching those of Cep-Her, resulting in their incorrect classification as RV members. Pre-main sequence stars and O and B stars can be reliably identified as members based on photometry alone, however stars between the pre-main sequence turn-on (PMSTO) and the tip of the main sequence have no photometric means of membership assessment. Furthermore, many of the stars in the same range lack conclusive measurements of the Lithium line, which is a common alternative means of assessing membership. RV measurements are necessary to assess the membership of these stars, and it is therefore important to quantify the probability of field contamination. We account for this possibility by estimating a false positive rate for our RV membership assessment, and factoring that into our membership probabilities. 

\begin{figure}
\centering
\includegraphics[width=3.7cm]{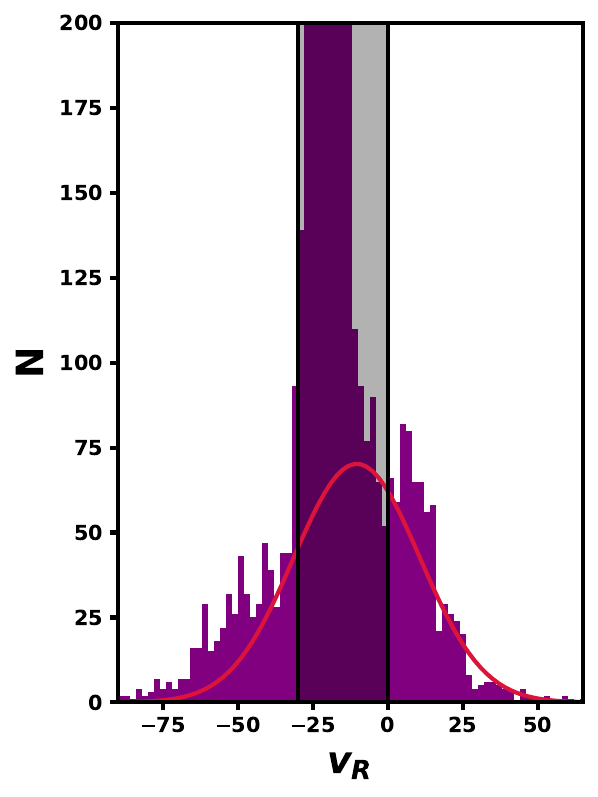}
\includegraphics[width=3.7cm]{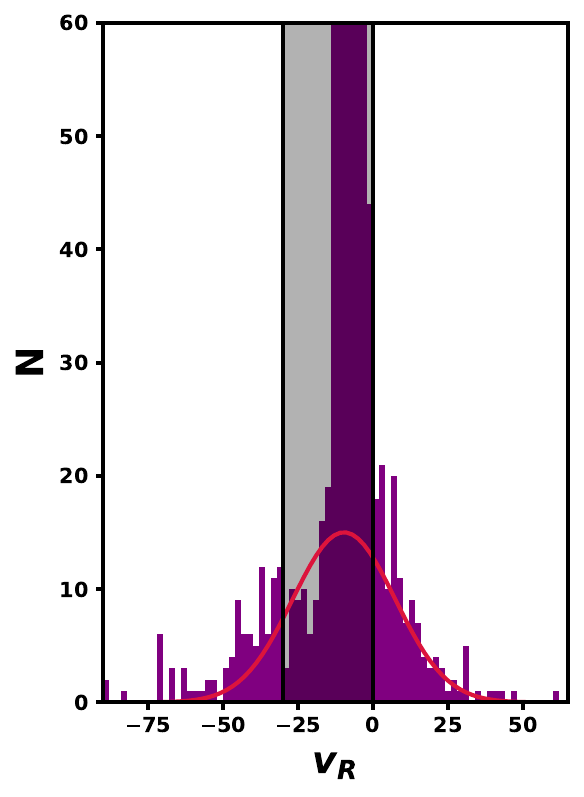}\hfill
\caption{Fits to the field contamination in Cep-Her Main and Cep-Her North. The purple histogram shows the distribution of RVs, while the red curve fits the sections of the distribution with ($v_R < -30$ and $v_R > 0$) where the field dominates. We apply dark shading to the Cep-Her-dominated range of RVs that we do not include in these fits. }
\label{fig:fieldcontam}
\end{figure}

To assess this contamination, we must compute a false positive probability $P_{\rm fp}$, which we define as the conditional probability of RV membership, given that the object is actually a non-member field star ($P_{\rm fp} = P(V=1|\textrm{nm})$, where ``nm'' limits the calculation to non-members only). In Section \ref{sec:amp}, we compute final probabilities of membership conditioned on the $V$ membership flags, $P_{\rm fin}(\textrm{mem}|V=x)$, so $P_{\rm fp}$ becomes an important input into Bayes Theorem. To produce these probabilities, we separately fit the distributions of the field star RVs in the samples for Cep-Her Main and Cep-Her North with a gaussian, including only stars with $\sigma_{RV}<5$ km s$^{-1}$ to remove poor RV measurements, and excluding stars with $-30 < v_R < 0$ km s$^{-1}$ to limit the fit to the RV range where the field dominates. These fits are shown in Figure \ref{fig:fieldcontam}. We can then compute the false positive rate for a field star as the area under the fit across the range of RVs where stars are marked as RV members, divided by the total area under the fit. 

The range of RVs where stars are marked as members is proportional to the radius  $\sigma_{v_{max}}$ used to flag velocity membership for stars at a given position in $l$. However, since we identify stars as RV members using a spherical cut in velocity space, stars with outlying transverse velocities have stricter cuts on RV than stars closer to the transverse velocity average. As a result, not all stars within the radius $\sigma_{v_{max}}$ of the mean RV are marked as members. We must therefore compute an effective range in RV for computing $P_{\rm fp}$. The field velocity distribution should be nearly isotropic over the range where stars are identified as velocity members, so any shape in 3D velocity space with the same volume should contain approximately the same number of stars. The initial velocity selection of potential members in SPYGLASS-IV was performed in transverse velocity space with an effective 2D search radius of $\sigma_{v_{max}}$. Restricting that population in the RV dimension turns this 2D bounded circular area into a 3D cylindrical bounded volume, so we can define the effective RV range $\Delta_{RV}$ as the length of a cylinder of radius $\sigma_{v_{max}}$ and length $\Delta_{RV}$ that has the same volume as a sphere of radius $\sigma_{v_{max}}$. This corresponds to $\Delta_{RV} = \frac{4}{3}\sigma_{v_{max}}$, and we use this value as the width of the window where field stars would be identified as members, centered on the mean RV. In ranges of galactic longitude $l$ where two velocity sequences are present (see Fig. \ref{fig:rvselect}), the velocity range is taken to contain the maximum extent across both sequences. 

The false positive rate for a star with a given value of $l$, $\sigma_{v_{max}}$, and mean RV is then the integral over the fit to the field RVs across a window of width $\Delta_{RV}$ centered on the mean RV, divided by the integral across the entire fit to the field RVs. The resulting values for $P_{\rm fp}$ range from 12\% to 22\% in Cep-Her Main and 17\% to 21\% in Cep-Her North. Stars on the main sequence, where we largely rely on velocities for membership, can have values of $P_{\rm sp} \la 0.2$, consistent with a field population that is dominant over Cep-Her members at a rate of 4:1 or more. With up to 21\% of field stars being assigned as velocity members in certain locations, this implies that even among stars with $V=1$, outlying sections of Cep-Her may still host members and field stars at similar rates. It is therefore important that we do not treat velocity membership as conclusive, and consider the field contamination when assessing membership. 

\subsubsection{RV False Negatives} \label{sec:falseneg}

Due to a combination of poor RV measurements and apparent motions from the influence of a binary companion, our RV membership cut may occasionally mark real members as non-members. We account for this possibility by estimating a false negative rate $P_{\rm fn}$ for our RV membership assessment, which we define as the rate at which true members are marked as velocity non-members ($P_{\rm fn} = P(V=-1|\textrm{mem})$). Like for our definition of the false positive rate in Section \ref{sec:contamination}, this form is chosen so that it can be directly inputted into our final membership probability calculations in Section \ref{sec:amp}, $P_{\rm fin}(\textrm{mem}|V=x)$, which are conditioned on the $V$ membership flags. 

Since this probability is conditional on membership, we must estimate it over an essentially pure sample of members. We first select a sample of stars with $P_{\rm spatial} = 0.95$, which is the maximum value we allow $P_{\rm spatial}$ to assume. In this set, 6.2\% of stars with conclusive velocity membership $V \neq 0$ are classified as velocity non-members. Some of these are likely to be genuine non-members, as a very weak field main sequence is visible even among this set. However, of the 5 stars in the section of the CMD where the pre-main sequence and field main sequence separate, 4 are closer to the pre-main sequence. This suggests that most of the stars in this set are true false negatives. We also investigate a set of stars with a spatial-photometric $P_{\rm sp}>0.99$, in which 18\% of stars with $V \neq 0$ are classified as velocity non-members. This is a much higher value, likely caused by the fact that it is based almost entirely on a set of photometrically young O and B stars, which have the highest binarity rates and rapid rotation \citep{Sullivan21}, both of which significantly weaken the accuracy of our velocity membership assessment. Due to the biased selection in the sample chosen by $P_{\rm sp}$, we take the sample with $P_{\rm spatial} = 0.95$ to be a closer representation of the overall RV false membership statistics, and adopt a false negative rate of $P_{\rm fn} = 0.05$. This roughly reflects the 6.2\% value computed from that sample, with a 1\% correction to account for potential field contamination in that otherwise pure sample. This result is also broadly consistent with the false negative rates reported in previous studies of young associations, such as those provided for $\beta$ Pic in \citet{Shkolnik17}. 

\subsubsection{Lithium Selection} \label{sec:limem}

\begin{figure}
\centering
\includegraphics[width=7cm]{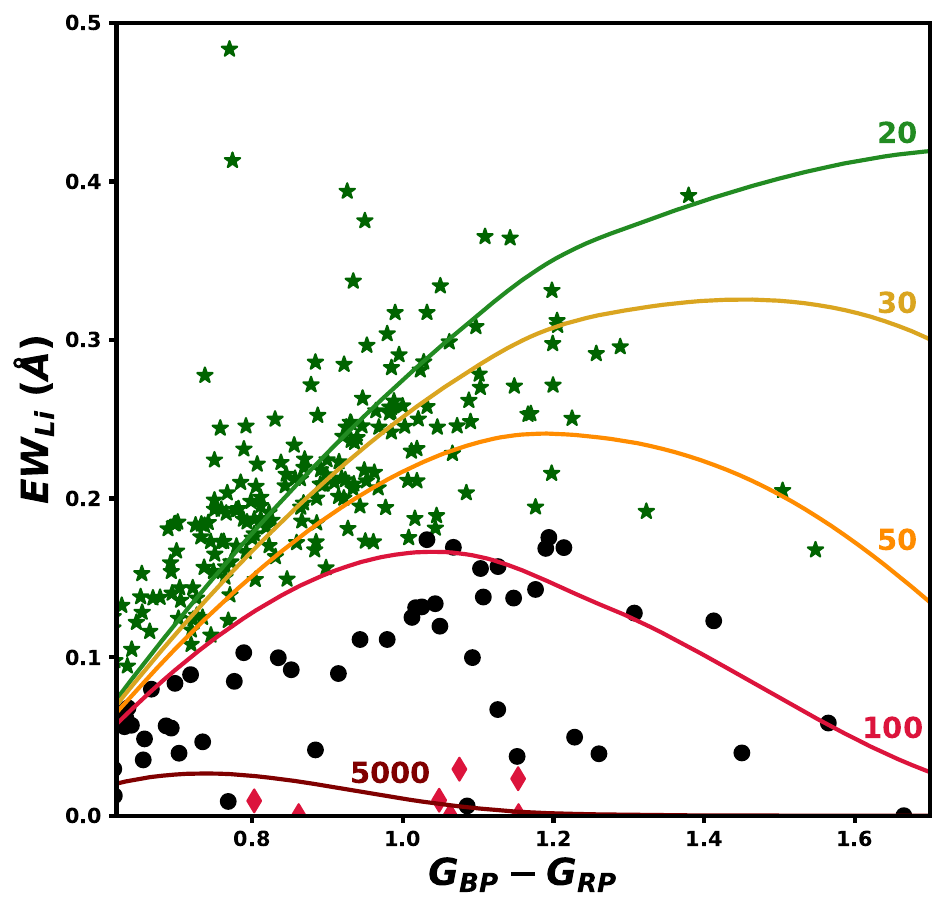}\hfill
\caption{Li selection in both Cep-Her sub-components, showing stars that are identified as Li members as green stars, Li non-members as red diamonds, and stars with inconclusive Li membership as black circles. We also include 20 Myr, 30 Myr, 50 Myr, 100 Myr, and 5 Gyr Lithium sequences for reference. }
\label{fig:limems}
\end{figure}

Lithium abundances are a strong indicator of stellar youth, as Lithium depletes rapidly after star formation for stars less massive than the Sun \citep{Bodenheimer65}. 
We therefore use the EAGLES package \citep{Jeffries23, EAGLES23} to assess membership for stars with a Lithium equivalent width (EW) measurement. For a given Li EW and $T_{eff}$, EAGLES computes age posteriors, which can be converted into a probability of youth for each star. \citet{Jeffries23} also includes a formula for converting from $G_{BP}-G_{RP}$ to $T_{eff}$, making results attainable directly from Gaia photometry. As SPYGLASS-IV uses the PARSEC 80 Myr isochrone as the maximum age for inclusion in the SPYGLASS catalog of young associations \citep{PARSECChen15}, we identify non-members by requiring that $P_{age<80 Myr}<0.005$ according to the EAGLES age posteriors. This strict cut for excluding non-members reduces the probability of false negatives due to to photometric contamination from binary companions, which can reduce Li EWs. To further limit false negatives, we also require that stars that receive a negative membership assessment also have a value of \textit{Gaia}'s Renormalized Unit Weight Error RUWE$>1.2$. This is a common condition used to exclude unresolved binaries \citep{Bryson20}, which we discuss further in Section \ref{sec:idbin}. We then classify all stars with a lithium age $\tau_{mean}<80$ Myr as Li-verified members. Unlike for identifying non-members, we do not apply an RUWE restriction to our identification of members, as any binary contaminant to a Li-rich star would almost certainly lower the measured Li EW, not increase it.  All other stars receive an ambiguous Lithium membership assessment. 

\subsubsection{Aggregate Membership Probabilities} \label{sec:amp}

To produce a final membership probability $P_{\rm fin}$, we must combine our initial spatial-photometric membership probability estimates $P_{\rm sp}$ with our spectroscopic membership flags from lithium and velocity. For stars that lack conclusive spectroscopic membership flags from velocity or Lithium, we set $P_{\rm fin}$ equal to $P_{\rm sp}$, which takes into account all of our non-spectroscopic means of membership assessment. Due to the well-established reliability of Lithium as an indicator of youth \citep[e.g.,][]{Binks14}, we take Li membership to be conclusive, and set the $P_{\rm fin}$ for Li members to 1, and $P_{\rm sp}$ for Li non-members to 0. However, as discussed in Sections \ref{sec:contamination} and \ref{sec:falseneg}, velocity false negatives and false positives are both possible, and it is therefore not appropriate to simply set the probability to 1 or 0 based on a velocity membership flag.  

\begin{figure*}[ht]
\centering
\includegraphics[width=8cm]{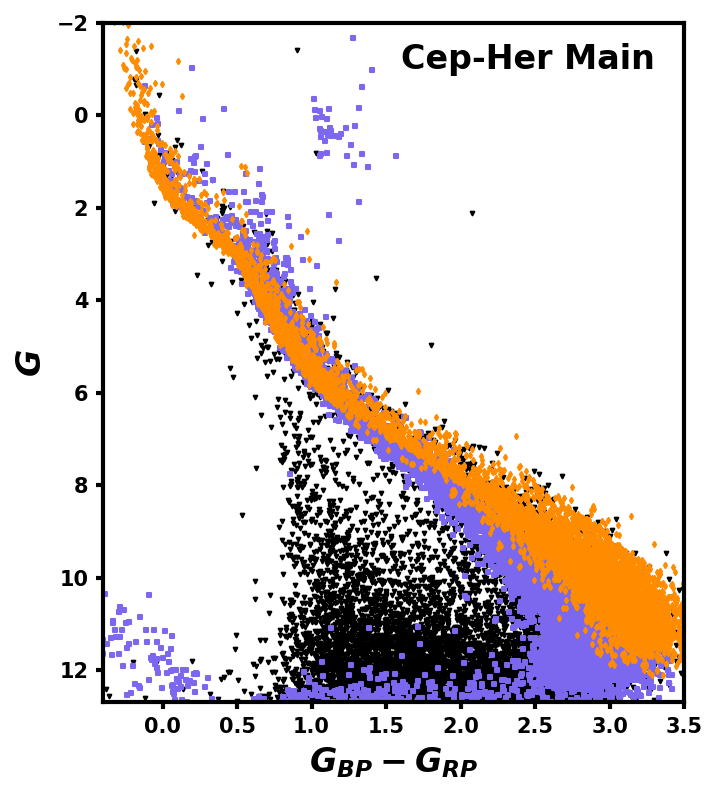}
\includegraphics[width=8cm]{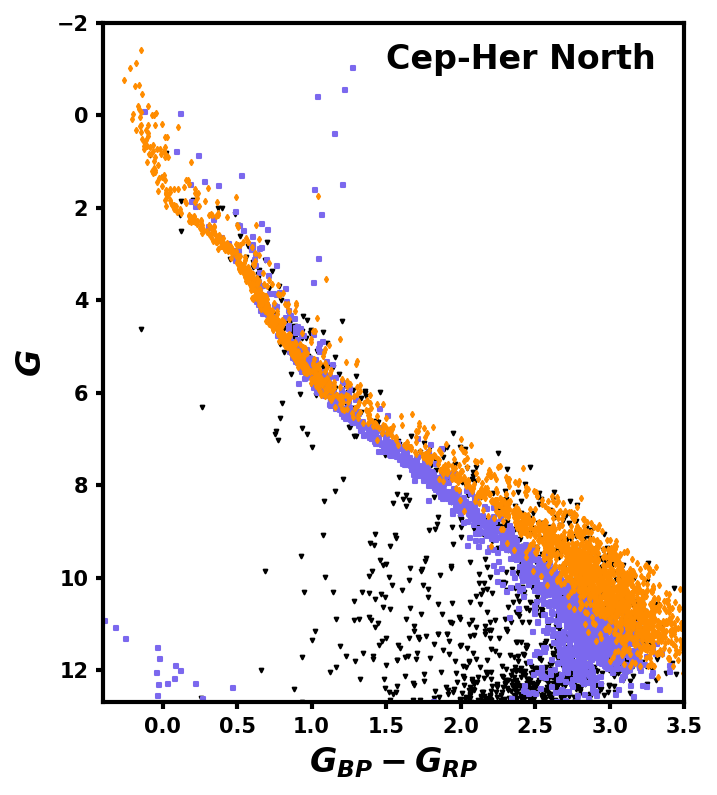}\hfill
\caption{Color magnitude diagram of Cep-Her Main (left), and Cep-Her North (right), colored by membership results. Orange diamonds indicate objects with $P_{\rm fin} > 0.5$ that are included in our probable member sample, while blue squares indicate objects that are not in that sample, but pass the photometric and astrometric quality cuts. Black triangles fail at least one of the astrometric and photometric quality cuts.} 
\label{fig:CMD}
\end{figure*}

We can compute the membership probability $P_{\rm fin}$ for a star with a good RV ($V \neq 0$) as the posterior probability $P(\textrm{mem}|V=x)$, or the probability of membership given a certain velocity membership flag. We adopt the spatial-photometric membership probability $P_{\rm sp}$ as the prior membership probability. Then, for $V=1$, the likelihood is provided by $P(V=1|\textrm{mem}) = 1 - P_{\rm fn}$, which follows from that fact that $P_{\rm fn} = P(V=-1|\textrm{mem})$, as described in Section \ref{sec:contamination}. Bayes theorem for $P(\textrm{mem}|V=1)$ can then be written as:

\[P_{\rm fin}(\textrm{mem}|V=1) = \frac{(1-P_{\rm fn})P_{\rm sp}}{P(V=1)}\]

\noindent and where the marginal probability $P(V=1)$ can be written as:

\[P(V=1) = (1-P_{\rm fn})P_{\rm sp}+P_{\rm fp}(1-P_{\rm sp})\]

\noindent In the second term, we use the fact that $P_{\rm fp} = P(V=1|\textrm{nm})$, as defined in Section \ref{sec:contamination}. 

For stars with a $V=-1$, we can apply an analogous calculation for assessing membership probability given velocity non-membership, which produces the following:

\[P_{\rm fin}(\textrm{mem}|V=-1) = \frac{P_{\rm fn}P_{\rm sp}}{P_{\rm fn}P_{\rm sp}+(1-P_{\rm fp})(1-P_{\rm sp})}\]

\noindent This formula also follows Bayes theorem, this time using $P_{\rm fn} = P(V=-1|\textrm{mem})$ and $1 - P_{\rm fp} = P(V=-1|\textrm{nm})$. 

For each star with a conclusive velocity membership flag ($V=1$ or $V=-1$), we set the aggregate probability $P_{\rm fin}$ equal to the output of the corresponding $P_{\rm fin}$ formula above for its value of $V$. For $V=0$, we adopt the original value of $P_{\rm sp}$. In cases with a definitive Lithium flag, we prioritize the lithium result, which returns a $P_{\rm fin}$ value of either 1 or 0. These membership adjustments for the velocity and lithium flags complete our results for $P_{\rm fin}$. 

In both Cep-Her North and Cep-Her Main, the sum over $P_{\rm fin}$ is within 5\% of the sum over $P_{\rm sp}$, demonstrating that these results do not deviate significantly from our probability prior of $P_{\rm sp}$ at a population level. The main impact of including false positive and false negative rates is to refine membership in cases where contamination and false negatives are especially common. F and G stars rely almost entirely on velocities to assess membership, so stars in this regime that have a low $P_{\rm sp}$ prior may still have relatively ambiguous values of $P_{\rm fin}$ even after being assigned $V=1$, as described in Section \ref{sec:contamination}. On the O and B sequence, where $P_{\rm sp}>0.9$ is common, the non-zero value of $P_{\rm fn}$ can become significant, as $P(mem|V=-1) = 0.54$ for $P_{\rm sp} = 0.95$, $P_{\rm fp} = 0.2$ and $P_{\rm fn} = 0.05$. These corrections therefore condition our membership probabilities for velocity members on the relative dominance of members and non-members in its section of parameter space, which is a function of the spatial-photometric membership probability, velocity false negative rate, and false positive rate from field contamination. 

We use the final values of $P_{\rm fin}$ to select a purer probable member sample that can be used for identifying substructure. We define this sample by removing all stars that fail the astrometric and photometric quality cuts (and therefore lack $P_{\rm fin}$), and require $P_{\rm fin}>0.5$. This $P_{\rm fin}$ cut balances completeness and purity to produce robust substructure-defining samples for both Cep-Her Main and Cep-Her North. In Cep-Her Main, 7228 stars pass these conditions and the sum of $P_{\rm fin}$ across those stars is 6268, so we expect that 87\% of objects in that sample are members and 13\% are field stars. The sum over all $P_{\rm fin}$ in Cep-Her Main is 7351, so these restrictions accept a 15\% loss rate of members. The same calculations as above predict similar contamination and loss rates in Cep-Her North, where the probable member sample is expected to contain 89\% members and 11\% field stars with an 11\% loss rate. Our probable member samples therefore contain the vast majority of association members with good photometry and astrometry, while having very limited field contamination. 

We summarize the membership selection in Figure \ref{fig:CMD}, showing the distribution of probable members, probable non-members, and stars that fail our astrometric and photometric quality cuts and therefore lack a result for $P_{\rm fin}$. We also compile all membership flags and probabilities for all candidate Cep-Her members in Table \ref{tab:members}, alongside some basic properties for each star.

\begin{deluxetable*}{ccccccccccccccccc}
\tablecolumns{18}
\tablewidth{0pt}
\tabletypesize{\scriptsize}
\tablecaption{List of Cep-Her members, including basic IDs and properties, substructure assignments, membership probabilities and other flags. The complete version of this table is available in the online-only version, which contains 34886 members}
\label{tab:members}
\tablehead{
\colhead{Gaia DR3 ID} &
\colhead{AS\tablenotemark{a}} &
\colhead{SG\tablenotemark{a}} &
\colhead{AA\tablenotemark{a}} &
\colhead{AG\tablenotemark{a}} &
\colhead{RA} &
\colhead{Dec} &
\colhead{d} &     
\colhead{g} &  
\colhead{M} &  
\colhead{$V$\tablenotemark{b}} &  
\colhead{Li\tablenotemark{c}} &  
\colhead{F\tablenotemark{d}} &  
\colhead{$P_{\rm Age<50Myr}$} &  
\colhead{$P_{\rm spatial}$} &
\colhead{$P_{\rm sp}$} &
\colhead{$P_{\rm fin}$} \\
\colhead{} &
\colhead{} &
\colhead{} &
\colhead{} &
\colhead{} &
\colhead{(deg)} &
\colhead{(deg)} &
\colhead{(pc)} &
\colhead{} &
\colhead{M$_{\odot}$} &
\colhead{} &
\colhead{} &
\colhead{} &
\colhead{} &
\colhead{} &
\colhead{} &
\colhead{}  \\
}
\startdata
2125758539488457344 & CUPV &      -1 &   CINR &         1 & 292.3272 & 42.6479 & 314 & 15.91 &  0.59 &       0 &       0 &     0 &  0.0095 &  0.35 &  0.54 &  0.54 \\
2125766751463294976 & CUPV &      -1 &     &        & 292.2994 & 42.8453 & 352 & 20.26 &  0.18 &       0 &       0 &     0 &      &  0.23 &    &    \\
2125769294084167936 & CUPV &      -1 &     &        & 292.3366 & 42.9828 & 334 & 19.82 &  0.18 &       0 &       0 &     0 &  0.0000 &  0.24 &  0.00 &  0.00 \\
2125781831091630208 & CUPV &       7 &   CINR &         1 & 291.9657 & 42.6111 & 286 & 19.42 &  0.25 &       0 &       0 &     0 &  0.0000 &  0.23 &  0.00 &  0.00 \\
2125788428159994368 & ORPH &      -1 &     &        & 291.5863 & 42.5099 & 309 & 11.76 &  1.09 &       1 &       0 &     8 &  0.0033 &  0.27 &  0.21 &  0.61 \\
2125805779829542784 & ORPH &      -1 &     &        & 292.0603 & 42.6919 & 297 & 17.83 &  0.33 &       0 &       0 &     0 &  0.0014 &  0.22 &  0.08 &  0.08 \\
2125812381199146112 & ORPH &      -1 &     &        & 291.9605 & 42.9027 & 370 & 18.98 &  0.09 &       0 &       0 &     1 &      &  0.24 &    &    \\
2125842033652058368 & ORPH &      -1 &     &        & 292.8799 & 42.7379 & 315 & 18.12 &  0.30 &       0 &       0 &     0 &  0.0024 &  0.45 &  0.30 &  0.30 \\
2125856361663343232 & CUPV &       9 &   CINR &         1 & 292.7672 & 43.0986 & 345 & 15.76 &  0.60 &       0 &       0 &     0 &  0.0012 &  0.22 &  0.07 &  0.07 \\
2125861717484155008 & ORPH &      -1 &     &        & 292.5439 & 43.0932 & 301 & 20.52 &  0.16 &       0 &       0 &     8 &      &  0.26 &    &    \\
2125864161317485824 & ORPH &      -1 &     &        & 292.4818 & 43.0602 & 326 & 20.66 &  0.16 &       0 &       0 &     0 &      &  0.27 &    &    \\
2125887663380213504 & ORPH &      -1 &     &        & 293.1205 & 43.4306 & 279 & 18.94 &  0.25 &       0 &       0 &     8 &      &  0.21 &    &    \\
\enddata
\tablenotetext{a}{AS, SG, AA, and AG indicate membership in parent structures: AS for association-level assignment, SG for the subgroup, AG for alternate subgroup assignment in the case of membership in the opposite top-level group, and AA for the association that contains the assigned alternate subgroup. -1 indicates that it does not have an assigned subgroup.} 
\tablenotetext{b}{Velocity membership assessment. 1 is a kinematic member, 0 is inconclusive, and -1 is a non-member} 
\tablenotetext{c}{Lithium membership assessment. 1 is a kinematic member, 0 is inconclusive, and -1 is a non-member} 
\tablenotetext{d}{General flag: 1 indicates that the star has a resolved companion within 10,000 au in the plane of the sky, 2 indicates a bad broadening function solution, 4 indicates a bimodal line profile likely indicative of spectroscopic binarity, 8 indicates an RUWE$>$1.2, indicating likely unresolved binarity, and 16 indicates that the RV recorded was ambiguously attributed to two components of a binary pair. The flags are added in cases where multiple are true; for example, flag 6 indicates both flags 2 and 4.} 
\end{deluxetable*}

\section{Structure} \label{sec:clustering}

\subsection{Clustering}

Cep-Her is a large and complicated association that contains diverse substructure, ranging from dense star clusters to diffuse associations that blend with both the field and adjacent populations. These populations provide basic units of co-natal star formation within the broader population that can trace the structure of the parent cloud, while providing suitable populations for measuring accurate local ages. 

We use our probable member sample described in Section \ref{sec:amp} to identify subgroups in 5-dimensional $(X,Y,Z,c*\Delta$v$_{T,l},c*\Delta$v$_{T,b})$ space-transverse velocity anomaly coordinates, with the constant $c$ set to $c=6$ pc km$^{-1}$ s, following the choices for initial group identification in SPYGLASS-II. For both Cep-Her Main and Cep-Her North, we cluster this 5D parameter space using the HDBSCAN clustering algorithm with the clustering parameters {\tt min$\_$samples} and {\tt min$\_$cluster$\_$size} set to 8. This provides clustering that is slightly less sensitive compared to the selection in SPYGLASS-II, which sets these parameters to 7. This choice reflects the greater risk of false groups introduced by our less sensitive astrometry at the distance of Cep-Her, which is approximately twice as distant as Cepheus Far North \citep[CFN;][]{Kerr22a}. We perform this clustering in leaf mode, which identifies the smallest scale populations in HDBSCAN's clustering tree. EOM clustering, which identifies the populations that persist over the largest range of clustering scales, was used in SPYGLASS-I to hint at possible connections between subgroups, although SPYGLASS-II revealed that true primordial star-forming sites do not necessarily reflect EOM groups. We therefore consider connections between leaf subgroups manually.

HDBSCAN is optimised for identifying subtle overdensities within a complex background, and therefore most objects in a sample are assigned to the background rather than any group. However, since we are clustering on samples of candidate members which exclude most contaminants, HDBSCAN's conservative group assignments mark many outlying subgroup members as part of the background, and stars with $P_{\rm fin}<0.5$ are excluded entirely. We can therefore produce more complete samples for each subgroup by assigning outlying members and stars that are not in our probable member sample to the association that agrees most closely in space-velocity coordinates. To assign these outlying members to a parent subgroup, we compute the clustering proximity $D_N$ with $N = 8$ for each star in the outlying sample relative to the space-velocity distributions of each HDBSCAN cluster in 5D $(X,Y,Z,c*\Delta$v$_{T,l},c*\Delta$v$_{T,b})$ space. Here we use $c=12$ pc km$^{-1}$ s, following SPYGLASS-II's selection for assigning outlying members to a parent subgroup. Stars are assigned to a given group if the clustering proximity of that group is 15\% lower than that of any other group. This assigns most stars to a subgroup, while leaving stars that are not closely connected to any one group as unclustered Cep-Her members. The 15\% clustering proximity cut allows most stars to be assigned to a subgroup, while preventing cases where substantial portions of Cep-Her's extended halo are assigned to outlying subgroups, which can give those groups unphysical shapes. The final membership assignments are listed in Table \ref{tab:members}. 

We compile information on the groups identified by HDBSCAN in Table \ref{tab:groupstats}. We identify 50 subgroups in total, including 38 in Cep-Her Main and 12 in Cep-Her North. Half of the 34886 Cep-Her Candidate members are assigned to a subgroup, and stars in a subgroup account for 66\% of the sum over $P_{\rm fin}$, indicating that probable members are much more likely to be assigned a subgroup compared to probable field stars. Before reintroducing outlying stars and those outside of the probable membership sample ($P_{\rm fin}<0.5$), only 2135 stars were included in these subgroups, demonstrating the importance of broadening these groups' membership beyond the initial HDBSCAN selection.

\begin{deluxetable*}{cccccccccccccccccc}
\tablecolumns{18}
\tablewidth{0pt}
\tabletypesize{\scriptsize}
\tablecaption{Demographics and mean properties of the subgroups in Cep-Her, identified by their parent association and ID.}
\label{tab:groupstats}
\tablehead{
\colhead{ASSOC} &
\colhead{ID} &
\colhead{N} &
\colhead{M} &
\colhead{$\sigma_{dem}$\tablenotemark{a}}&
\colhead{RA} &
\colhead{Dec} &
\colhead{l} &
\colhead{b} &     
\colhead{d} &  
\colhead{$\overline{v_{T,l}}$} &  
\colhead{$\overline{v_{T,b}}$} &  
\colhead{$\overline{v_r}$} &  
\colhead{$R_{hm}$} &  
\colhead{$\sigma_{1D}$} &  
\colhead{$\sigma_{vir}$} &
\colhead{Vir. Rat} &
\colhead{Age} \\
\colhead{} &
\colhead{} &
\colhead{} &
\colhead{(M$_{\odot}$)} &
\colhead{} &
\multicolumn{2}{c}{(deg)} &
\multicolumn{2}{c}{(deg)} &
\colhead{(pc)} &
\multicolumn{3}{c}{(km s$^{-1}$)} &
\colhead{(pc)} &
\multicolumn{2}{c}{(km s$^{-1}$)} &
\colhead{} &
\colhead{(Myr)}
}
\startdata
 CUPV &    1 &  129 &  50.4 &           0.379 & 292.6 & 31.0 & 64.7 &  6.1 & 347.0 &  -6.8 & -5.4 & -18.2 &      14.1 &           0.96 &            0.06 &       12.2 $\pm$         6.1 &   68.2 $\pm$    9.2 \\
 CUPV &    2 &  187 &  68.6 &           0.364 & 291.4 & 30.5 & 63.8 &  6.8 & 309.0 &  -5.6 & -2.5 & -19.4 &      17.5 &           0.52 &            0.06 &        6.3 $\pm$         5.2 &   62.6 $\pm$   11.8 \\
 ORPH &    1 &  185 &  76.3 &           0.100 & 286.5 & 19.0 & 51.5 &  5.5 & 359.0 &  -8.9 & -5.2 & -14.3 &      11.8 &           0.94 &            0.07 &        8.9 $\pm$         2.6 &   27.9 $\pm$    2.4 \\
 ORPH &    2 &   88 &  35.2 &           0.100 & 284.8 & 18.7 & 50.5 &  6.8 & 361.0 & -10.3 & -8.9 & -14.2 &      11.9 &           0.59 &            0.05 &        8.3 $\pm$         5.4 &   26.5 $\pm$    2.8 \\
 ORPH &    3 &  272 & 102.9 &           0.140 & 295.5 & 48.8 & 81.8 & 12.6 & 294.0 &   3.5 & -4.7 & -21.9 &      22.4 &           1.74 &            0.07 &       18.9 $\pm$         5.3 &   39.0 $\pm$    1.8 \\
 CUPV &    3 & 1435 & 561.3 &           0.224 & 320.4 & 51.2 & 93.1 &  0.9 & 362.0 &  12.1 & -5.6 & -19.3 &       3.0 &           0.31 &            0.42 &        0.5 $\pm$         0.2 &   55.3 $\pm$    2.3 \\
 CUPV &    4 &  401 & 137.0 &           0.219 & 287.0 & 24.9 & 57.1 &  7.7 & 332.0 &  -5.8 & -3.5 & -14.6 &      15.2 &           1.23 &            0.09 &        9.9 $\pm$         3.0 &   68.4 $\pm$    8.8 \\
 CUPV &    5 &  140 &  51.3 &           0.257 & 315.9 & 46.4 & 87.6 & -0.2 & 295.0 &  11.6 & -6.3 & -18.4 &      12.8 &           0.48 &            0.06 &        5.8 $\pm$         2.2 &   71.0 $\pm$    6.5 \\
 CUPV &    6 &  371 & 158.1 &           0.291 & 315.2 & 54.1 & 93.2 &  5.2 & 307.0 &  13.1 & -4.0 & -19.5 &      17.3 &           0.75 &            0.09 &        5.8 $\pm$         2.1 &   68.1 $\pm$    6.7 \\
 CUPV &    7 &  391 & 143.7 &           0.325 & 302.9 & 44.3 & 80.5 &  5.7 & 295.0 &   6.1 & -2.6 & -20.4 &      16.5 &           0.55 &            0.09 &        4.6 $\pm$         2.2 &   $>$ 80.0  \\
 ORPH &    4 &  180 &  73.7 &           0.100 & 276.9 & 15.4 & 44.2 & 12.1 & 363.0 & -13.2 & -2.8 & -14.8 &      13.0 &           0.32 &            0.07 &        3.3 $\pm$         1.5 &   34.5 $\pm$    2.5 \\
 ORPH &    5 &  155 &  64.7 &           0.100 & 281.1 & 22.6 & 52.6 & 11.6 & 375.0 & -10.5 & -2.8 & -15.3 &      11.8 &           0.52 &            0.07 &        5.3 $\pm$         1.9 &   30.0 $\pm$    1.9 \\
 CUPV &    8 &  887 & 397.2 &           0.258 & 309.3 & 43.3 & 82.3 &  1.4 & 341.0 &   8.1 & -6.6 & -20.1 &      17.1 &           0.69 &            0.15 &        3.4 $\pm$         1.0 &   55.5 $\pm$    4.1 \\
 CUPV &    9 &  892 & 360.9 &           0.243 & 300.9 & 39.3 & 75.5 &  4.3 & 354.0 &   3.3 & -4.1 & -20.7 &      16.5 &           0.78 &            0.14 &        3.9 $\pm$         1.2 &   53.8 $\pm$    4.7 \\
 CUPV &   10 &  559 & 250.1 &           0.242 & 303.4 & 34.4 & 72.4 &  0.0 & 325.0 &   4.3 & -6.7 & -18.5 &       7.1 &           0.21 &            0.18 &        0.8 $\pm$         0.3 &   58.4 $\pm$    3.0 \\
 ORPH &    6 &  109 &  65.2 &           0.100 & 272.1 & 18.0 & 44.7 & 17.4 & 344.0 & -12.1 & -0.7 & -15.8 &      12.1 &           0.53 &            0.07 &        5.5 $\pm$         2.9 &   33.7 $\pm$    3.8 \\
 ORPH &    7 &  164 &  65.5 &           0.100 & 276.5 & 30.1 & 58.1 & 18.3 & 307.0 & -10.2 & -1.7 & -20.2 &      14.2 &           0.49 &            0.06 &        5.5 $\pm$         2.8 &   39.8 $\pm$    3.3 \\
 ORPH &    8 &  197 &  67.2 &           0.100 & 288.6 & 32.4 & 64.5 &  9.8 & 324.0 &  -3.5 & -6.4 & -16.8 &      17.1 &           0.54 &            0.06 &        6.5 $\pm$         2.8 &   27.1 $\pm$    1.3 \\
 ORPH &    9 &  229 & 107.0 &           0.100 & 275.9 & 27.2 & 55.0 & 17.7 & 364.0 & -11.2 & -0.8 & -18.1 &      10.2 &           0.80 &            0.09 &        5.9 $\pm$         1.6 &   32.8 $\pm$    2.8 \\
 ORPH &   10 &  199 &  70.7 &           0.100 & 274.8 & 23.6 & 51.1 & 17.3 & 386.0 &  -9.6 & -4.3 & -18.4 &      15.4 &           0.57 &            0.06 &        6.4 $\pm$         2.1 &   24.9 $\pm$    1.9 \\
 ORPH &   11 &  268 & 110.6 &           0.100 & 290.8 & 35.5 & 68.1 &  9.5 & 372.0 &  -2.5 & -4.7 & -19.0 &      13.7 &           0.39 &            0.08 &        3.3 $\pm$         1.5 &   32.4 $\pm$    2.0 \\
 ORPH &   12 &  151 &  73.9 &           0.100 & 276.4 & 20.9 & 49.1 & 14.9 & 304.0 & -10.5 & -2.9 & -17.3 &       8.7 &           0.22 &            0.09 &        1.8 $\pm$         1.0 &   27.7 $\pm$    1.6 \\
 ORPH &   13 &  143 &  55.4 &           0.100 & 275.4 & 20.0 & 47.9 & 15.4 & 345.0 & -11.7 & -5.4 & -16.8 &       9.8 &           0.38 &            0.07 &        3.8 $\pm$         1.3 &   26.5 $\pm$    3.8 \\
 ORPH &   14 &  108 &  34.5 &           0.100 & 275.7 & 26.6 & 54.3 & 17.7 & 323.0 &  -7.6 & -3.4 & -18.6 &       5.0 &           0.14 &            0.08 &        1.3 $\pm$         1.3 &   26.0 $\pm$    2.3 \\
 ORPH &   15 &  172 &  76.4 &           0.100 & 277.1 & 26.2 & 54.5 & 16.3 & 347.0 &  -7.7 & -4.7 & -18.5 &       2.2 &           0.19 &            0.17 &        0.8 $\pm$         0.4 &   26.5 $\pm$    2.1 \\
 ORPH &   16 &  461 & 185.5 &           0.100 & 282.7 & 33.0 & 62.9 & 14.5 & 282.0 &  -3.6 & -4.4 & -17.8 &       7.1 &           0.18 &            0.15 &        0.8 $\pm$         0.2 &   34.5 $\pm$    1.6 \\
 ORPH &   17 &  162 &  58.2 &           0.100 & 273.0 & 26.7 & 53.5 & 19.9 & 282.0 &  -7.2 & -0.6 & -18.2 &      10.6 &           0.58 &            0.07 &        6.0 $\pm$         2.2 &   27.9 $\pm$    2.8 \\
 ORPH &   18 &  119 &  40.8 &           0.100 & 271.8 & 19.5 & 46.0 & 18.2 & 329.0 & -11.4 & -3.6 & -16.1 &       9.1 &           0.34 &            0.06 &        3.9 $\pm$         1.4 &   30.0 $\pm$    2.5 \\
 ORPH &   19 &  202 &  68.9 &           0.100 & 266.9 & 16.0 & 40.7 & 21.1 & 307.0 & -13.5 & -3.0 & -16.0 &       7.6 &           0.25 &            0.09 &        2.0 $\pm$         0.8 &   27.3 $\pm$    1.3 \\
 ORPH &   20 &   79 &  24.5 &           0.100 & 271.0 & 18.9 & 45.1 & 18.7 & 307.0 & -11.4 & -3.0 & -16.3 &       4.4 &           0.33 &            0.07 &        3.4 $\pm$         1.2 &   32.6 $\pm$    2.0 \\
 ORPH &   21 &  141 &  49.4 &           0.100 & 269.4 & 20.9 & 46.4 & 20.8 & 286.0 & -10.5 & -2.3 & -17.5 &       8.1 &           0.29 &            0.07 &        2.8 $\pm$         1.3 &   31.4 $\pm$    1.9 \\
 ORPH &   22 &  189 &  71.3 &           0.100 & 279.9 & 31.1 & 60.2 & 16.0 & 361.0 &  -6.1 & -2.8 & -16.1 &      11.0 &           0.26 &            0.07 &        2.5 $\pm$         1.0 &   31.7 $\pm$    2.7 \\
 ORPH &   23 &  342 & 134.3 &           0.100 & 285.0 & 31.6 & 62.4 & 12.1 & 363.0 &  -4.0 & -6.1 & -15.9 &      10.0 &           0.21 &            0.11 &        1.4 $\pm$         0.4 &   30.1 $\pm$    1.6 \\
 ORPH &   24 &  277 & 116.8 &           0.100 & 281.3 & 34.2 & 63.6 & 16.0 & 391.0 &  -5.0 & -3.0 & -17.2 &      10.6 &           0.29 &            0.10 &        2.1 $\pm$         0.7 &   27.4 $\pm$    1.9 \\
 ORPH &   25 &  300 & 115.7 &           0.100 & 286.3 & 41.7 & 72.4 & 15.3 & 328.0 &  -0.7 & -3.1 & -19.4 &       9.4 &           0.33 &            0.10 &        2.3 $\pm$         0.5 &   34.9 $\pm$    2.6 \\
 ORPH &   26 &  237 & 101.5 &           0.100 & 285.2 & 39.7 & 70.1 & 15.2 & 353.0 &  -1.7 & -3.3 & -17.9 &       9.3 &           0.47 &            0.10 &        3.4 $\pm$         0.9 &   28.0 $\pm$    2.7 \\
 ORPH &   27 &  117 &  45.3 &           0.100 & 282.8 & 34.6 & 64.5 & 15.0 & 361.0 &  -4.2 & -3.4 & -17.8 &       6.3 &           0.28 &            0.08 &        2.5 $\pm$         1.1 &   28.7 $\pm$    1.8 \\
 ORPH &   28 &  615 & 247.4 &           0.100 & 283.6 & 36.9 & 66.9 & 15.3 & 352.0 &  -4.0 & -3.8 & -18.4 &       1.3 &           0.27 &            0.41 &        0.5 $\pm$         0.1 &   32.1 $\pm$    1.2 \\
 ROS6 &    1 & 1030 & 467.8 &           0.230 & 306.9 & 40.0 & 78.5 &  0.9 & 346.0 &   8.8 & -5.8 &  -7.9 &       6.6 &           0.27 &            0.25 &        0.8 $\pm$         0.2 &   $>$ 80.0  \\
 CINR &    1 &  139 &  63.4 &           0.346 & 295.7 & 51.4 & 84.2 & 13.6 & 276.0 &   5.0 & -2.6 & -10.1 &      16.3 &           0.41 &            0.05 &        5.6 $\pm$         3.6 &   27.9 $\pm$    1.9 \\
 CINR &    2 &  151 &  79.0 &           0.212 & 322.2 & 50.2 & 93.3 & -0.6 & 355.0 &   8.6 & -6.4 &  -6.9 &      17.1 &           0.98 &            0.06 &       11.5 $\pm$         3.5 &   37.1 $\pm$    4.5 \\
 ROS6 &    2 &  246 & 100.4 &           0.385 & 318.1 & 49.5 & 90.9 &  0.8 & 301.0 &   7.9 & -6.3 &  -7.6 &      12.3 &           0.63 &            0.08 &        5.6 $\pm$         3.4 &   71.2 $\pm$    6.4 \\
 CINR &    3 &   96 &  42.3 &           0.281 & 314.1 & 47.8 & 87.9 &  1.6 & 373.0 &   6.2 & -5.5 &  -6.1 &       8.1 &           0.35 &            0.07 &        3.6 $\pm$         1.9 &   29.3 $\pm$    2.6 \\
 CINR &    4 &  135 &  55.6 &           0.311 & 310.3 & 43.3 & 82.7 &  0.8 & 354.0 &   4.2 & -5.5 &  -6.7 &      12.3 &           0.31 &            0.07 &        3.2 $\pm$         1.8 &   38.5 $\pm$    4.8 \\
 CINR &    5 &  475 & 177.9 &           0.109 & 312.7 & 55.5 & 93.3 &  7.2 & 330.0 &   8.5 & -2.4 &  -7.1 &       9.0 &           0.27 &            0.13 &        1.5 $\pm$         0.5 &   31.4 $\pm$    1.4 \\
 CINR &    6 &  375 & 142.0 &           0.111 & 318.7 & 61.0 & 99.5 &  8.5 & 314.0 &  10.6 & -2.7 &  -6.1 &      11.3 &           0.29 &            0.10 &        2.0 $\pm$         0.4 &   29.9 $\pm$    1.6 \\
 CINR &    7 &  606 & 225.6 &           0.126 & 305.8 & 49.3 & 85.8 &  6.9 & 354.0 &   5.6 & -2.0 &  -7.2 &       8.1 &           0.26 &            0.16 &        1.2 $\pm$         0.3 &   32.9 $\pm$    1.2 \\
 CINR &    8 &  149 &  52.7 &           0.268 & 306.0 & 49.3 & 85.8 &  6.7 & 314.0 &   6.4 & -3.1 &  -8.5 &       9.0 &           0.25 &            0.07 &        2.4 $\pm$         1.1 &   39.3 $\pm$    4.5 \\
 CINR &    9 &   93 &  35.3 &           0.537 & 302.7 & 44.3 & 80.4 &  5.9 & 312.0 &   4.3 & -3.3 &  -7.0 &      10.8 &           0.28 &            0.06 &        3.2 $\pm$         2.0 &   43.1 $\pm$    3.5 \\
 CINR &   10 &  486 & 213.4 &           0.112 & 303.1 & 45.0 & 81.1 &  6.0 & 339.0 &   5.0 & -3.6 &  -7.0 &       3.6 &           0.19 &            0.23 &        0.6 $\pm$         0.2 &   35.7 $\pm$    1.5 \\
\enddata
\tablenotetext{a}{$\sigma_{dem}$ is the fractional uncertainty in both the number of members $N$ and the mass $M$.}
\vspace*{0.1in}
\end{deluxetable*}

\subsection{Ages} \label{sec:isoages}

\begin{figure}
\centering
\includegraphics[width=8cm]{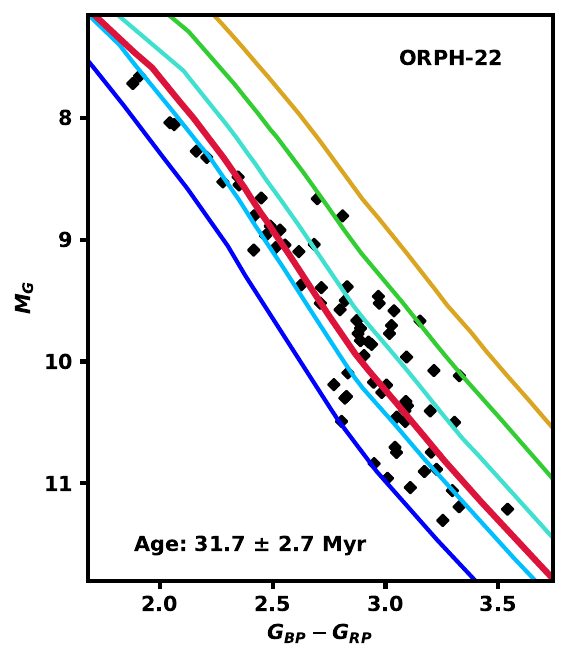}\hfill
\caption{Age solution for Orpheus subgroup 22, best fit indicated by the thick red curve. We mark stars used in the fit as black diamonds, and include isochrones of 10 Myr, 20 Myr, 40 Myr, 80 Myr, and 1 Gyr, from top to bottom, for reference. The age fits for all other Cep-Her subgroups are provided in the online-only version of this figure.}
\label{fig:isoages}
\end{figure}

Broad age categories are a useful diagnostic for verifying common origins. While complete dynamical and lithium depletion boundary ages are beyond the scope of this paper and will be published in an upcoming publication \citep{Kerr24c}, isochronal ages are readily available using \textit{Gaia} photometry and astrometry. These ages provide a reliable relative age scale, which supports the identification of broader formation trends. These ages can therefore identify stars that formed in the same generation, helping us to identify co-natal structures within the larger Cep-Her Complex. 

\begin{figure*}
\centering
\includegraphics[width=17cm]{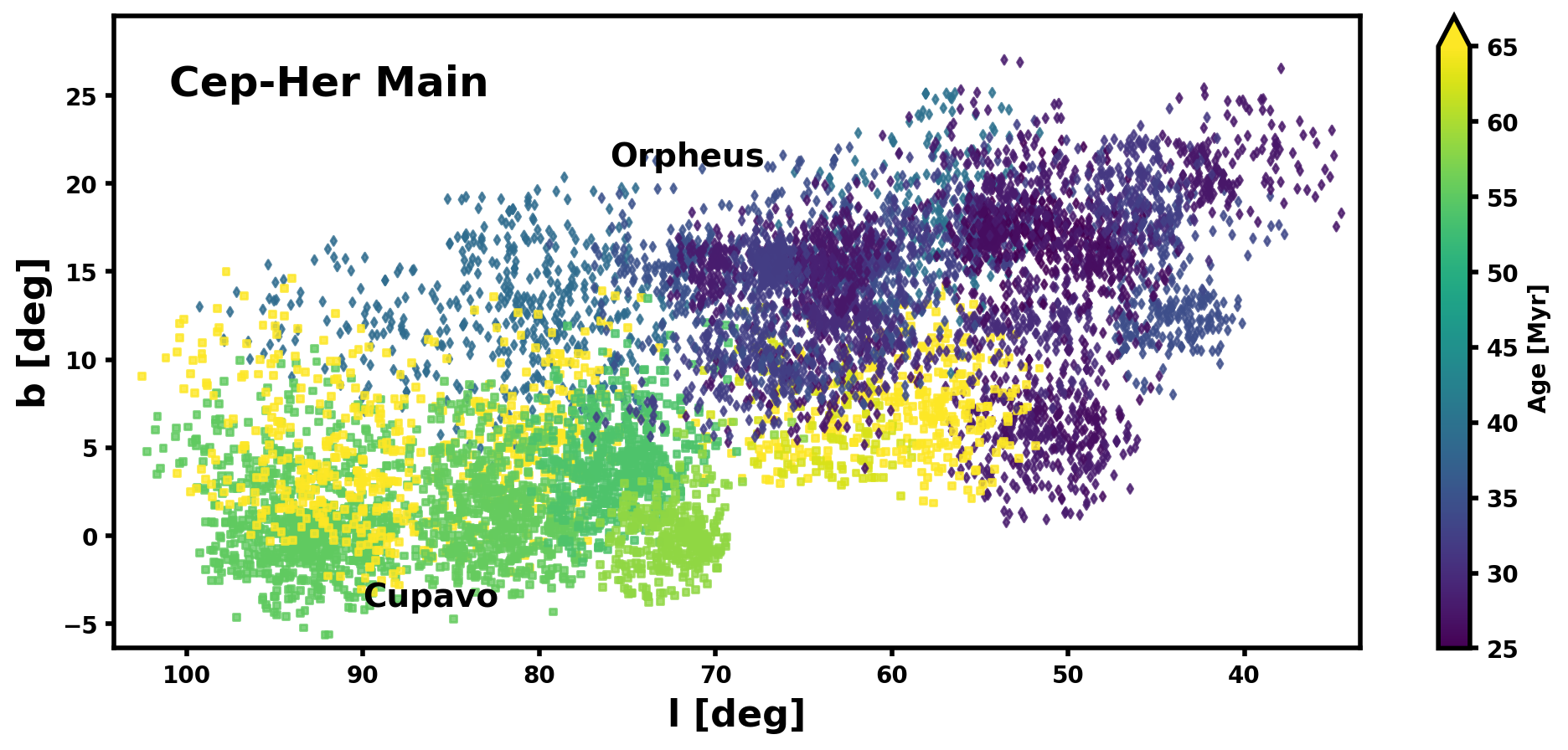}\hfill
\caption{Distribution of ages in Cep-Her Main by subgroup, revealing a bimodality in age that we use to split Cep-Her Main into the older Cupavo association and younger Orpheus association. Stars associated with Orpheus are marked with diamonds, and stars associated with Cupavo are marked with squares.} 
\label{fig:CHM_Ages}
\end{figure*}

\begin{figure*}
\centering
\includegraphics[width=16cm]{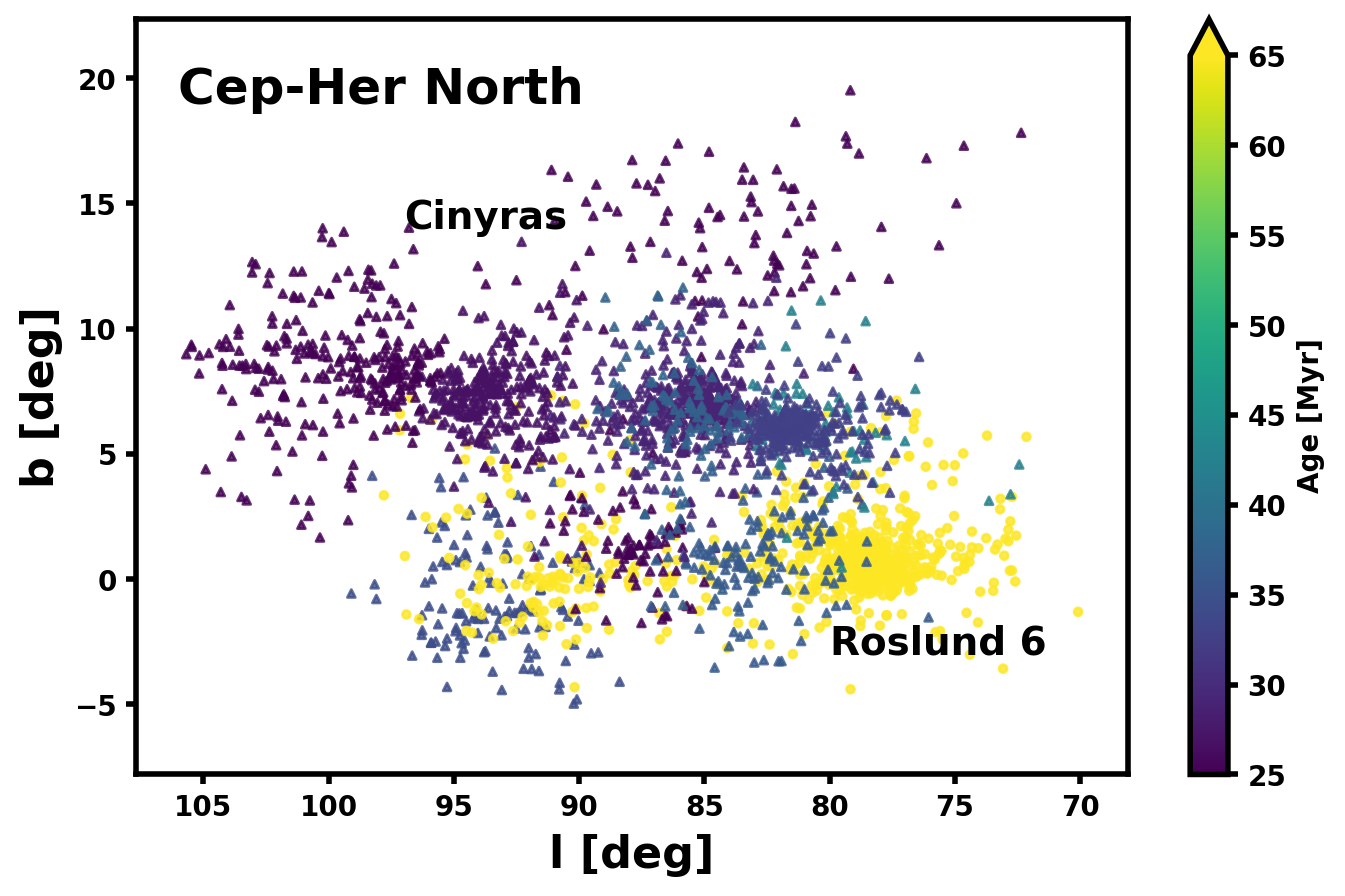}\hfill
\caption{Distribution of ages in Cep-Her North by subgroup. Like in Cep-Her Main, Cep-Her North also shows an age bimodality between the older Roslund 6 and younger Cinyras association. Stars associated with Roslund 6 are marked with dots, and stars associated with Cinyras are marked with triangles.}
\label{fig:CHNF_Ages}
\end{figure*}

We compute age by fitting the absolute magnitudes of stars in each subgroup against PARSEC v1.2S isochrone grids \citep{PARSECChen15}. We use \textit{Gaia} photometry in $M_G$ /$G_{BP}-G_{RP}$ space for these fits, focusing on the pre-main sequence with $1.8<G_{BP}-G_{RP}<4$, and correcting for distance and reddening using \textit{Gaia} distances \citep{BailerJones18} and \citet{Lallement19} reddening maps. We limit the stellar selection to stars that satisfy the astrometric and photometric quality cuts in SPYGLASS-IV. We also exclude stars with $RUWE > 1.2$ to removing likely unresolved binaries, as the additional flux from the companion makes these objects appear brighter and therefore photometrically younger than single stars. We further restrict the sample to RUWE $< 1.1$ in subgroups where at least 15 stars would remain after applying the cut. This is an even more restrictive binarity cut that removes nearly all stars with astrometrically detectable influence from binaries \citep{Bryson20}, provided that the population has enough stars to support a strong fit after the imposition of this harsh restriction. Finally, we require that stars included in the fit lie above the 80 Myr PARSEC isochrone, which is well below the sequence for Cep-Her groups in most cases. We then employ a bootstrapping routine to both limit the impact of outlying field and binary stars and to improve our handling of the systematic uncertainties in our age solutions. For each population and model grid, we select half of the stars in the sample at random with a minimum selection size of 8, with selection probability weighted by their $P_{\rm spatial}$. We do this 10000 times for each population, defining the 2-$\sigma$ clipped mean as the age solution and the corresponding standard deviation as the uncertainty. 

In most cases, the solutions fit the densest part of the pre-main sequence, with little impact from apparent spectroscopic binaries that lie above the pre-main sequence, or probable field contaminants below. However, there is a limited set of populations with age solutions exceeding 40 Myr where the cut on the 80 Myr isochrone may remove a substantial number of photometrically older good members, artificially lowering the age solution. For groups with age $\tau>40$ Myr, we therefore remove this cut and re-calculate ages using a different set of restrictions designed to avoid introducing age biases. We reduce contamination by requiring that all stars used in the fit lie above a 1 Gyr PARSEC isochrone (which corresponds to a typical field sequence), and also require that $P_{\rm spatial}>0.5$, provided that the result has at least 20 stars. This latter restriction is meant to counteract the increased risk of field contamination with the looser isochrone cut. We find this choice reduces contamination well where possible, while still producing healthy sample sizes for bootstrapping.
Ages $\tau>80$ Myr lie in a range where the pre-main sequence becomes an increasingly poor age indicator, so we cap solutions at this value and record them as having ages older than 80 Myr. We present the age solutions for all Cep-Her subgroups in Figure \ref{fig:isoages}. The results closely follow the visually identifiable young sequences. While these ages are effective for comparisons between Cep-Her subgroups, we caution against the use of these solutions in absolute terms, as isochronal ages are highly model-sensitive, especially in this age range. 

We show the distribution of ages in Cep-Her Main and Cep-Her North in Figures \ref{fig:CHM_Ages} and \ref{fig:CHNF_Ages}, respectively. There we show a regional bimodality in ages in both components of Cep-Her. These regions with common age correspond roughly to the subtly separated sequences shown in our velocity selection fits (Figure \ref{fig:rvselect}), indicating that these different ages also correspond to physical structures. We discuss the structures indicated by this division in Section \ref{sec:midlevelstructures}.

\subsection{Association-Level Structures} \label{sec:midlevelstructures}

Through the wide gap between Cep-Her Main and Cep-Her North in RV space and the age and kinematic discrepancies between different regions of those populations, our analysis suggests the presence of four distinct regions of star formation within the broader Cep-Her Complex. In Cep-Her North, we see a clear age bimodality, with the larger, young component having ages between 28 and 43 Myr, and the older component having ages near our 80 Myr limit for analysis. The larger of the two subgroups in the older component represents the known cluster Roslund 6 \citep{Roslund60}, which is thought to have a much older age \citep[470 Myr; ][]{Kharchenko13}. Roslund 6 is also the sole representative of the smaller Cep-Her North velocity sequence that we fit in Figure \ref{fig:rvselect}, so its distinction from the rest of the association appears to be supported through dynamics as well as age. The other old subgroup in Cep-Her North does not show the same distinct dynamics, and its wide range of stellar magnitudes suggests it may contain some stars from a younger component.  However, due to the apparent dominance of the older stars, we group it and Roslund 6 into a population distinct from the rest of Cep-Her North. The additional old structure may be related to the tidal tails that have already been proposed around Roslund 6 \citep{Battacharya22}, although it is far enough from the Roslund 6 core that it could be just a component of a larger dissolving association.

At the age of Roslund 6, SPYGLASS-IV was only able to detect populations with a strong binary sequence. These binary-dominated populations were flagged as old, and Roslund 6 would likely have been included in this category if SPYGLASS-IV identified it separately from the rest of Cep-Her. Older ages reduce the accuracy of both pre-main sequence isochronal ages and traceback, substantially reducing our ability to reconstruct its star formation history. We therefore exclude Roslund 6 and its companion from the discussion of association dynamics in Section \ref{sec:dynamics}. 

In Cep-Her Main, the sub-populations divide into two broad components. The younger component is centered on $\delta$ Lyr, and has relatively consistent ages between 25 and 40 Myr with a median of $\tau = 30$ Myr. The older component has a broader age range from 54 Myr to our 80 Myr mass limit. This division is most evident in Figure \ref{fig:CHM_Ages}, where two regions of differing ages emerge in spatial coordinates. Most of this age division also reflects the split in the space-velocity sequences shown in Figure \ref{fig:rvselect}. We therefore use this wide separation in age as the basis to divide Cep-Her Main into a younger group ($\tau < 45$ Myr) and an older group ($\tau > 45$ Myr). This simple age cut alone divides the association into two largely contiguous populations in space and velocity coordinates.

The subgroups of the older component that are furthest to the galactic west ($l \sim 60\degr$, see Fig. \ref{fig:CHM_Ages}) are potential space-velocity outliers within that component, and like the companion group to Roslund 6, these populations appear to contain some young stars in a region overall dominated by older objects. Further spectroscopic follow-up may therefore reassign some objects to the young component. However, for the purpose of this analysis, we group them into the old component. 

The gaps in age and velocity between the four regions identified here suggest that these populations, not the full Cep-Her complex, are closest to the standard definition of an association as a largely unbound population of common origin. We therefore refer to the four populations that emerge from this analysis as ``associations'', while referring to Cep-Her as a ``complex'' of intertwined populations. The definition of Cep-Her nonetheless remains useful due to the difficulty in distinguishing between members of each association contained within. To reflect their distinct nature, we introduce new nomenclature to refer to the four association-level populations. The population connected to Roslund 6 is dominated by that cluster, so we refer to it using the name of that cluster. The other three associations have extents unlike any presented in the literature, so we assign new names to each. We refer to the young component of Cep-Her Main that contains $\delta$ Lyr as the Orpheus Association\footnote{This name refers to Greek hero that lulls the multi-headed dog Cerberus to sleep with a lyre, referencing the two regions covered by this association - Lyra, the constellation of the lyre, and the obsolete constellation Cerberus, which is found in SE Hercules.}, the old component of Cep-Her Main as the Cupavo Association, and the remaining component of Cep-Her North as the Cinyras Association\footnote{Cinyras and Cupavo are the children of Cycnus, the king most often associated with the story of Cygnus in Greek mythology, and due to the youth of these associations the title of ``children of Cygnus'' is fitting. The use the story of King Cycnus also makes reference to the portion of these associations in Cepheus, the constellation of the king.}. 

\begin{figure*}[!ht]
\centering
\includegraphics[width=16cm]{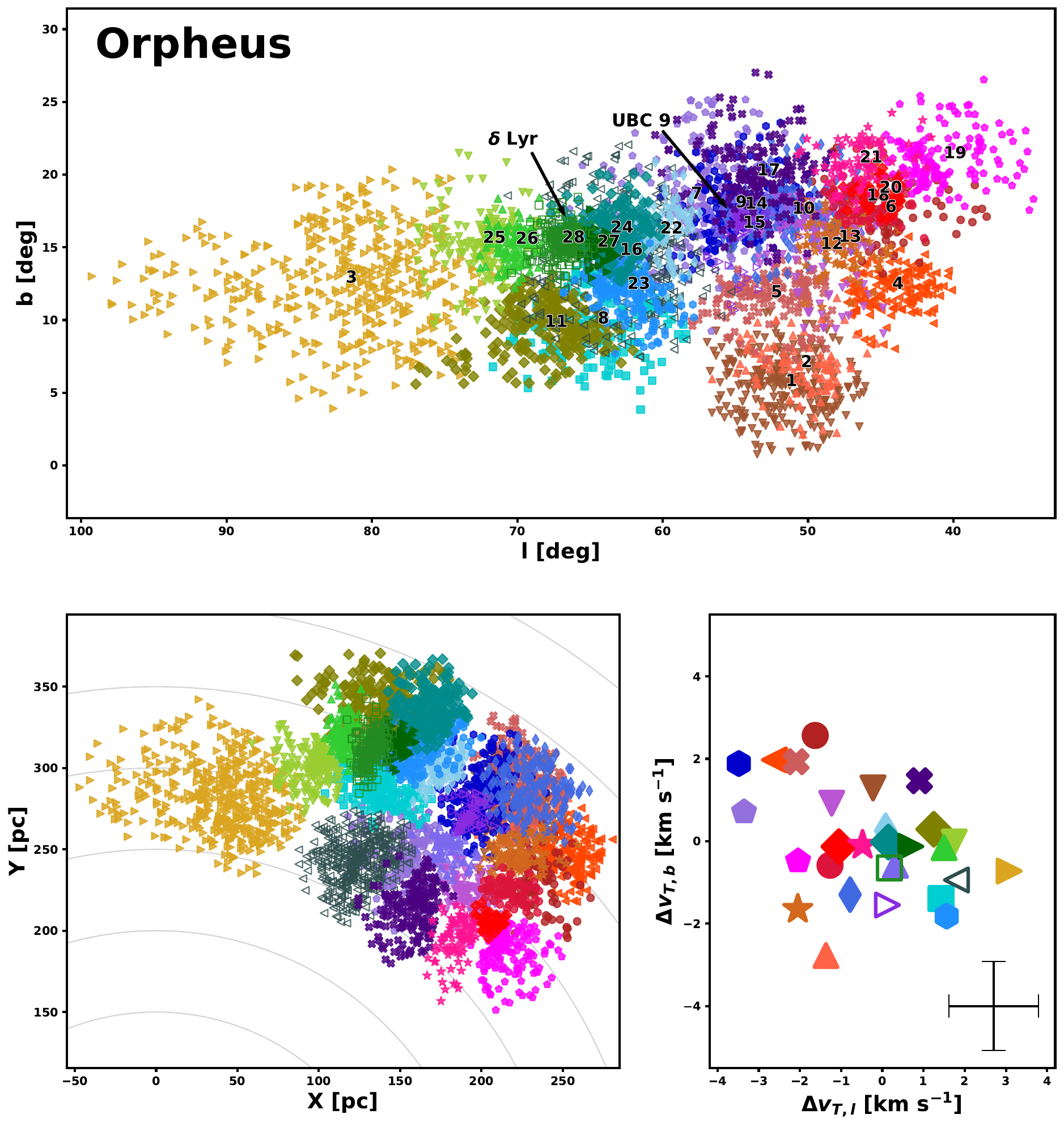}\hfill
\caption{Overview plot for the Orpheus Association, showing the distribution of groups in $l$/$b$ sky coordinates (top), X/Y galactic coordinates (bottom left), and $\Delta v_{T,l}$/$\Delta v_{T,b}$ transverse velocity anomaly (bottom right), measured relative to the mean UVW velocity vector. The transverse velocity anomaly plot shows average values to improve legibility, as the scatter of the groups overlap heavily. The mean velocity standard deviation for a subgroup is marked with the black error bars in the lower right of that panel.  We annotate the subgroup IDs in the top panel. Members of bound open clusters or groups with a bound core as assessed in Section \ref{sec:boundedness} are marked with open icons. We assign similar colors to qualitatively similar populations in spatial coordinates and age to aid visual comprehension. The grey curves are lines of equal distance drawn at 50 pc intervals, which indicate both the location of the sun and the distance spread of the association. We also annotate the location of known bound clusters. Online-only, we include an interactive figure in galactic XYZ cartesian space that covers Orpheus as well as the other associations covered in Figures \ref{fig:cinyras}, \ref{fig:cupavo}, and \ref{fig:roslund6}. In that figure, associations can be shown and hidden using the legend and the display can be rotated or zoomed as desired. }
\label{fig:orpheus}
\end{figure*}

\begin{figure*}[!ht]
\centering
\includegraphics[width=16cm]{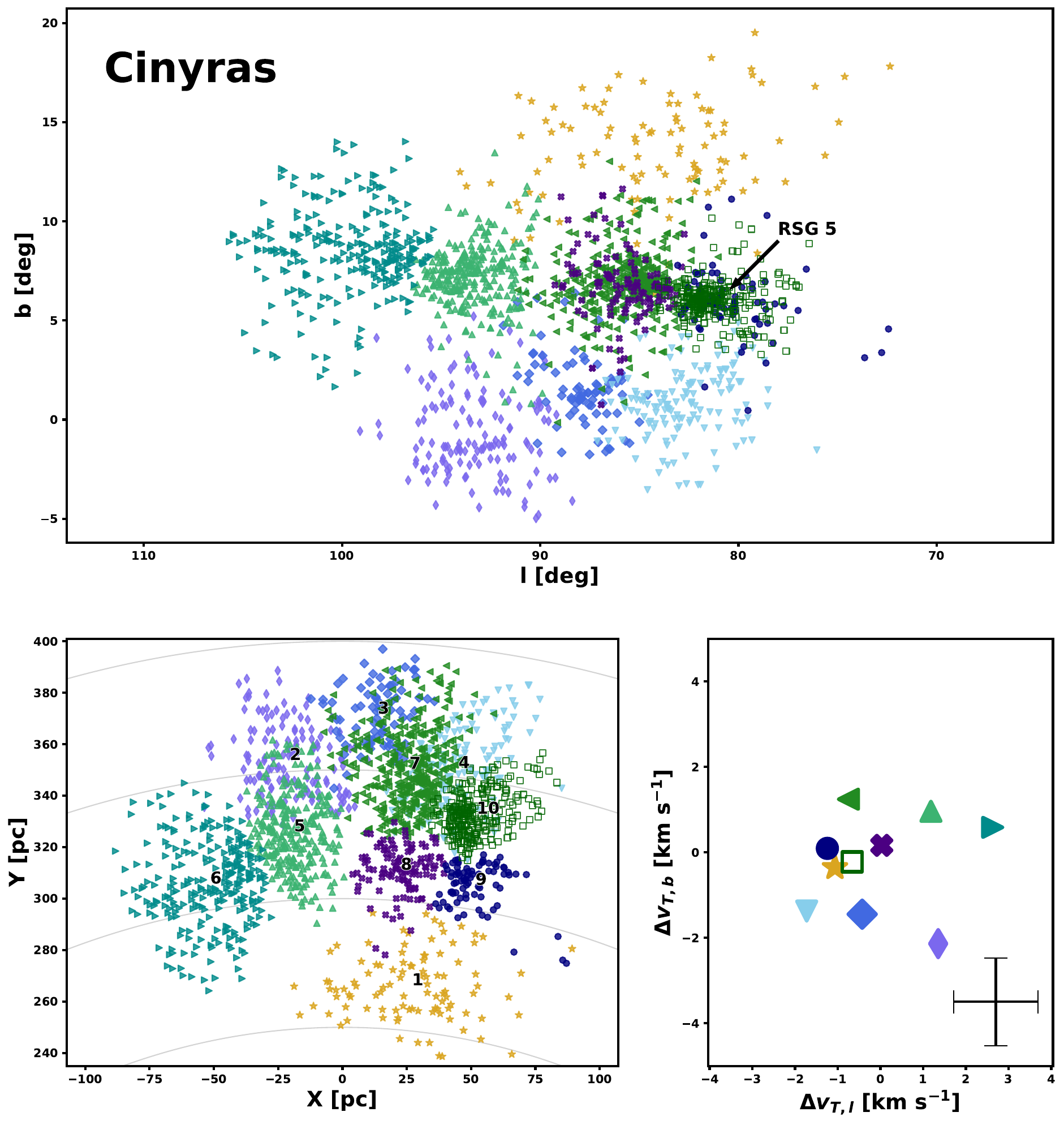}\hfill
\caption{Same as Figure \ref{fig:orpheus}, but for Cinyras. The subgroup IDs are annotated in the XY galactic coordinates panel, as the groups in Cinyras are better separated in that parameter space.}
\label{fig:cinyras}
\end{figure*}

\begin{figure*}[!ht]
\centering
\includegraphics[width=16cm]{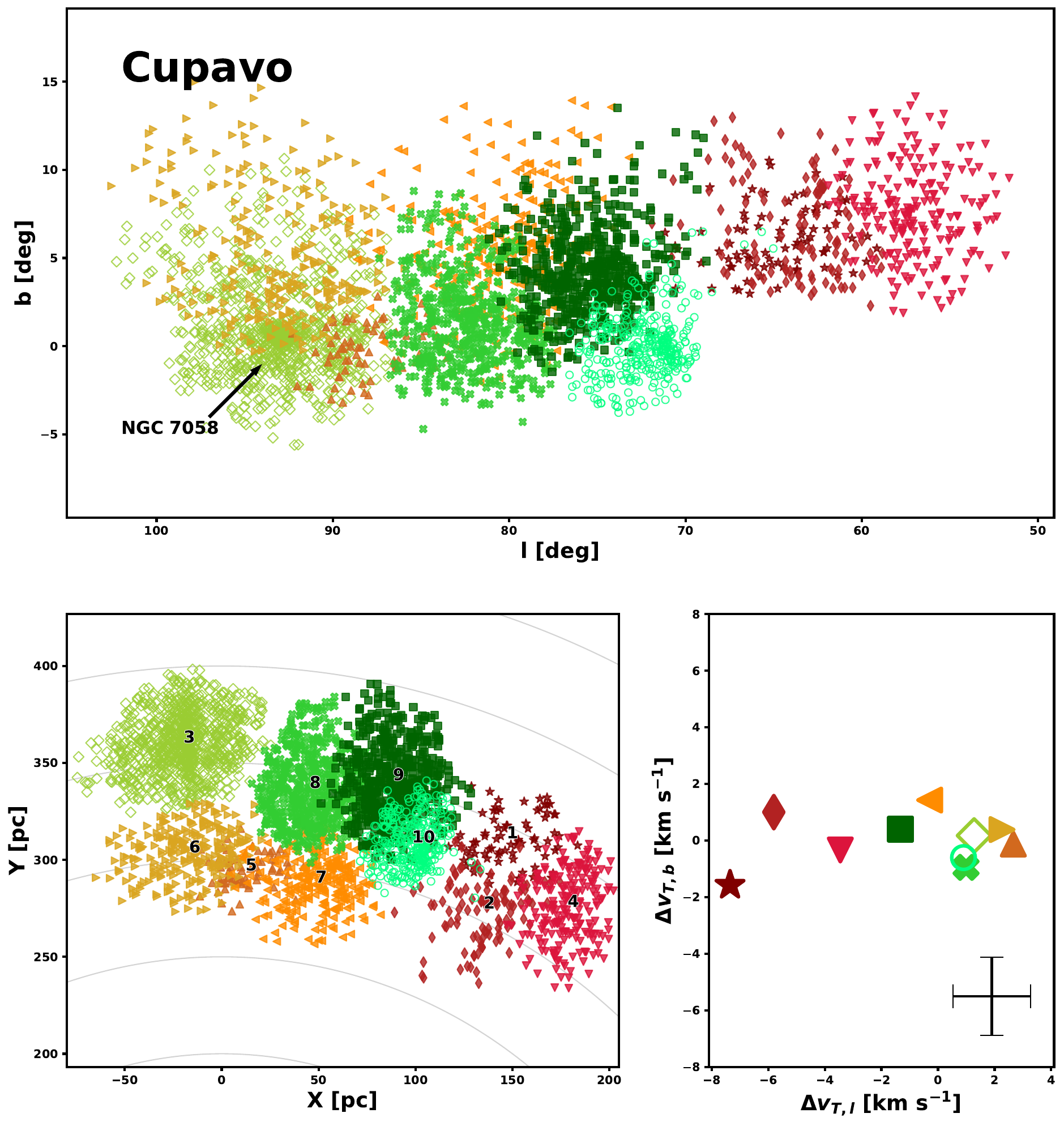}\hfill
\caption{Same as Figure \ref{fig:orpheus}, but for Cupavo. Group IDs are annotated in the XY galactic coordinates panel. }
\label{fig:cupavo}
\end{figure*}

\begin{figure}[!ht]
\centering
\includegraphics[width=8cm]{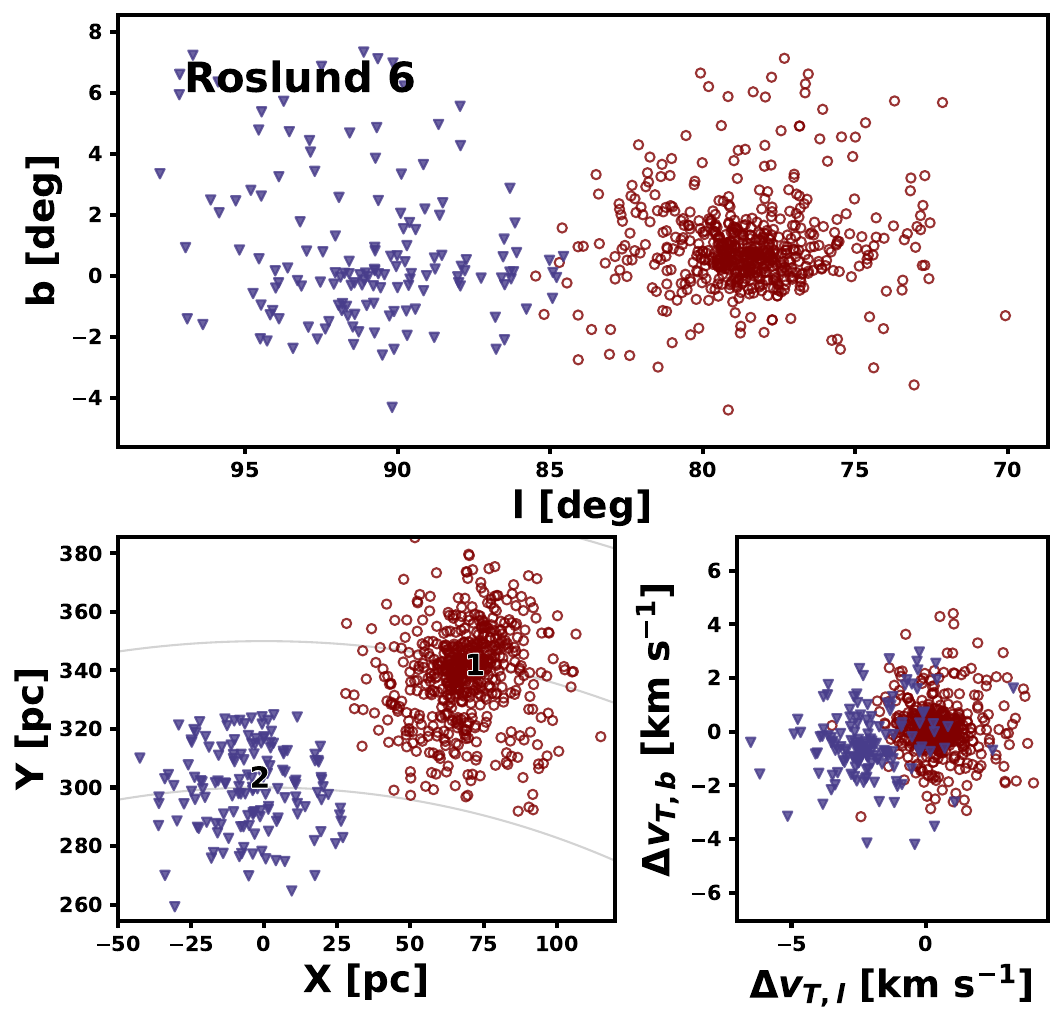}\hfill
\caption{Overview plot for Roslund 6. The axes and layout are the same as Figure \ref{fig:orpheus}, however since there are only two subgroups we show the entire velocity distribution.}
\label{fig:roslund6}
\end{figure}

To enable more complete demographic studies for these newly-defined associations, we assign all unclustered stars to whichever association has the highest local density as defined using the same clustering proximity-based metric that was used to assign outlying members in Section \ref{sec:clustering}. We provide summary plots for each association in Figures \ref{fig:orpheus}, \ref{fig:cinyras}, \ref{fig:cupavo}, and \ref{fig:roslund6}. These show the extents of association subgroups in on-sky space-velocity coordinates, galactic XY coordinates, and transverse velocity anomaly relative to the mean UVW vector of each association. Online-only, we also provide an interactive view of the Cep-Her Complex as part of Figure \ref{fig:orpheus}, which allows for the exploration of the region in sky-plane $l$/$b$ coordinates, 3D galactic XYZ coordinates, and transverse velocity anomaly, with the ability to independently show or hide each association. These figures therefore show the overall structure of these associations, while also allowing the comparison of trends in velocity with the spatial positions of the association, allowing dynamical trends to be unveiled, which we explore further in Section \ref{sec:dynamics}.

\section{Demographics} \label{sec:demographics}

Our substructure identification and association membership assignment enables the first estimates of Cep-Her's demographics for both the constituent associations and the subgroups within. Of particular interest is the mass, which helps to inform to virial state of Cep-Her substructures. Masses of individual stars can be derived from stellar models, however corrections must be made to account for missing stars, such as unresolved binary companions and stars beyond the sensitivity limit of \textit{Gaia}. In this Section, we compute individual stellar masses, and then correct for factors that influence these calculations to produce population-wide results for mass and stellar population size. We then use the masses and velocity dispersions of these groups to perform virial analyses, assessing whether any of our newly-identified substructures can be classified as open clusters. 

\subsection{Stellar Masses}

We compute stellar masses using solar-metallicity PARSEC v1.2S isochrones \citep{PARSECChen15}. Cep-Her's large overall mass makes the presence of high-mass stars likely, so we assemble a new isomass grid from the PARSEC v1.2S isochrones which are spaced every 0.005 M$_{\odot}$ between 0.09 and 1 M$_{\odot}$, every 0.01 M$_{\odot}$ between 1 and 2 M$_{\odot}$, every 0.02 M$_{\odot}$ between 2 and 4 M$_{\odot}$, and every 0.05 M$_{\odot}$ for between 4 and 20 M$_{\odot}$. We assign stars the mass associated with the nearest isomass track in \textit{Gaia} de-reddened $M_G$ /$G_{BP}-G_{RP}$ color-absolute magnitude space. 

The highest mass we report for any star in Cep-Her is 9.55 $M_{\odot}$ for 102 Herculis in Orpheus, which, based on those same PARSEC models, places an upper age limit for the youngest stars in the population at $\sim 24$ Myr \citep{PARSECChen15}. This matches closely with our ages computed using the pre-main sequence in Orpheus (Section \ref{sec:isoages}). As expected, the older populations of Cupavo and Roslund 6 both have lower maximum masses at $7.15 M_{\odot}$ and $7.6 M_{\odot}$, respectively, while Cinyras, which has a younger average age similar to Orpheus, has a larger maximum mass of $8.15 M_{\odot}$. These maximum masses correspond to an upper age limit for the youngest generation of 45 Myr in Roslund 6, 51 Myr in Cupavo, and 40 Myr for Cinyras. Of these, all but the result for Roslund 6 are consistent with their corresponding pre-main sequence isochronal age solutions. The unexpectedly massive star in Roslund 6 is highly outlying within that subgroup and a plausible member of Cinyras, which would better explain its mass. The next most massive star in Roslund 6 is only 4.65 M$_{\odot}$, which would imply a more plausible upper age limit at 140 Myr. 

The stellar sample in Cep-Her contains 204 stars that have masses consistent with O and B stars \citep[$M>2.7$ M$_{\odot}$;][]{Pecaut13}. This is substantially more than the 150 O and B stars that were used to define Sco-Cen \citep{deZeeuw99}, suggesting that the total population of Cep-Her at least exceeds the $10^4$ members that are estimated to reside that population \citep{Pecaut16}. Of these high-mass stars in Cep-Her, 101 are in Orpheus, 56 are in Cupavo, 35 are in Cinyras, and 12 are in Roslund 6.

\subsection{Older Subgroups in Cep-Her}

Our age calculations in Section \ref{sec:isoages} revealed 12 subgroups across Cupavo and Roslund 6 with ages exceeding 50 Myr. The youth probability calculations in SPYGLASS are based on the probability of an age younger than 50 Myr ($P_{\rm Age<50 Myr}$), and as a result, stars older than this limit can fall completely outside of the range of photometry where SPYGLASS identifies stars as young. $P_{\rm spatial}$ is also affected by this issue, as that calculation is based on a ratio of near-certain old stars to near-certain young stars. In these older populations, members occasionally fall in that near-certain old category, which lowers our results for $P_{\rm spatial}$. The result is that very low photometric and spatial-photometric membership probabilities are possible among genuine members of these older populations. The straightforward use of $P_{\rm fin}$ for assessing membership probability will underestimate the mass in these older populations, necessitating an alternate means of assessing membership. We therefore compute subgroup-wide values of membership probability $P_{bulk}$ for populations with $\tau > 50$ Myr that account for the minimal separation between the field main sequence and pre-main sequence in this age regime. 

To do this, we select stars in each subgroup with $0.25M_{\odot}<M<0.5M_{\odot}$, which is roughly the range where the pre-main sequence of these older populations has some separation from the field sequence, while avoiding the dimmest stars, which have higher photometric uncertainties that can result in higher rates of misclassification. We exclude unresolved binaries from these samples by applying the maximally strict cut $RUWE < 1.1$, which prevents binaries above the main sequence from breaking the symmetry of the field stars around the 1 Gyr isochrone. We then mark all stars located below a 1 Gyr PARSEC isochrone in this mass range as field stars. Taking the 1 Gyr isochrone to be roughly representative of the field main sequence, we assume that single field stars are distributed symmetrically on either size of the field sequence. We therefore take the number of field stars ($N_{field}$) in the sample to be double the number of stars found below the 1 Gyr isochrone. The mean membership probability in the population $P_{bulk}$ is then 1 - ($N_{field}$/$N_{tot}$), where $N_{tot}$ is the total number of potential association members in the chosen mass range.  

This result for $P_{bulk}$ assumes that stars of a given mass are represented equally in the association and the field, however this is not true for stars more massive than $\sim 1 $M$_{\odot}$, where field stars may be old enough to evolve into compact objects and thus leave our sample. We must therefore down-weight the expected field contribution in this mass range. Following \citet{Binney00}, we assume a constant star formation in the field over the last 11.2 Gyr. We take a sequence of PARSEC solar-metallicity isochrones from 1 Myr to 11.2 Gyr, and record the maximum stellar mass in each isochrone. This provides a lifespan for a star of a given mass. If that lifespan is less than 11.2 Gyr, a star of that mass may have evolved into a compact object. We therefore define a corrective factor $f_{alive}$ as the lifespan of a star at a given mass, divided by 11.2 Gyr. This value approximates the fraction of field stars of a given mass that have yet to become compact objects, assuming a constant star formation rate over the age of the solar neighborhood \citep{Binney00}. The inclusion of this corrective factor changes our bulk membership probability to $P_{bulk} = 1 - \frac{f_{alive}N_{field}}{N_{tot}}$. Uncertainties in these membership probabilities are dominated by the binomial uncertainty in the number of field stars $N_{field}$, and we propagate these uncertainties through to our mass and population results.

We calculate $P_{bulk}$ for all subgroups in the older associations of Cupavo and Roslund 6. We also calculate $P_{bulk}$ for the outlying unclustered regions separately in each of these associations. These calculations are used to produce approximate overall membership rates in these older populations, which we use to assess association mass and population size in Section \ref{sec:finmass}. These values therefore replace $P_{\rm fin}$ when calculating cluster demographics, however we do not change the values of $P_{\rm fin}$ in Table \ref{tab:members}. 

Populations older than 50 Myr are only marginally detectable through SPYGLASS, and are typically identified through their binary sequence alone. These populations therefore tend to be quite dominant in their local parameter space if detectable through SPYGLASS. As such, the bulk membership probabilities in most old populations are high, with stars in all subgroups being assigned $P_{bulk}>0.7$. Outlying components of Cupavo and Roslund 6 receive $P_{bulk}=0.72$ and $P_{bulk}=0.78$ for low-mass stars, respectively, both of which are quite high for an unclustered halo. This may suggest that there are substantial lower-density populations in Cupavo beyond the SPYGLASS sensitivity limit for groups in this age range. 

\subsection{Binaries} \label{sec:binaries}

Binaries are a major source of uncertainty in the mass estimates and overall stellar demographics of a young association. Unresolved companions require dedicated spectroscopic measurements for their full characterization, resulting in their mass being difficult to ascertain and often overlooked. We therefore identify probable binaries in Cep-Her, and use these to correct our demographic calculations.

\subsubsection{Identifying Binaries} \label{sec:idbin}

To identify resolved binaries, we search an on-sky radius of $10^4$ AU around each candidate member in \textit{Gaia}, marking all stars with $\Delta v_T < 3$ km s$^{-1}$ and $\frac{\Delta\pi}{\pi} < 0.2$ as a probable binary companion. This is similar to the selection in SPYGLASS-II, only changing the velocity cut to adjust for the distance of Cep-Her. 3 km s$^{-1}$ corresponds roughly to the circular velocity of a companion at 1 arcsecond separation from a G or F star at the near edge of Cep-Her, with the addition of a 1 km s$^{-1}$ buffer to account for the typical \textit{Gaia} measurement uncertainties for dim objects at the distance of Cep-Her. The 1 arcsecond value corresponds to the smallest separations resolvable through \textit{Gaia} that have the highest orbital speeds, so this 3 km s$^{-1}$ velocity limit represents the maximum velocity difference expected for a resolved stellar companion \citep{rizzuto_zodiacal_2018}. Changing the selection limit to 2 km s$^{-1}$ or 4 km s$^{-1}$ only changes the number of binaries by $\sim$20\%, verifying that most of the binaries are found within  2 km s$^{-1}$, and that 3 km s$^{-1}$ provides a reasonable buffer for high-uncertainty objects without introducing substantial contamination. 

We avoid duplicate systems by removing searches if a brighter Cep-Her member is found within our search radius, only recording the result from the search around the brighter object, which we mark as the primary. Due to subtly different distances to different stars in a system, it is possible for the projected on-sky distance between two stars to disagree on whether a star is within the 10000 AU limit depending on which distance is adopted. In these cases, the distance to the brighter star is adopted, and cases where the primary is already included in another system are removed. In cases where one companion is assigned to multiple systems but the primary of one system is not assigned as a companion of the other, the primaries of both systems are marked as primaries, and common companions are listed under both possible parent systems. We record probable resolved binary or multiple systems in Appendix \ref{app:bin}. 

For completeness, we also mark unresolved binaries, which are identifiable using other available metrics. As mentioned in Section \ref{sec:limem}, the Renormalized Unit Weight Error (RUWE), which is a metric based on the goodness of fit from \textit{Gaia}'s photometric solution, is a strong indicator of an unresolved companion. We therefore mark stars with RUWE$>1.2$ as probable unresolved binaries, following \citet{Bryson20}. We also mark stars as unresolved binaries if they have spectroscopic line profiles consistent with a double-lined spectroscopic binaries (SB2), either through the visual identification of a bimodal line profile in our observations, or through SB2 flags provided in literature RV catalogs. Stars with evidence of spectroscopic binarity are marked with a flag in Table \ref{tab:members}. 

\subsubsection{Mass in Binaries}


To compute the mass in stellar companions, we follow the methods employed in SPYGLASS-II, first stripping each probable binary system identified in Section \ref{sec:idbin} to only its primary. This ensures that no systems are double-counted in our statistics, as the population-wide contribution of binaries, resolved and unresolved, can be inferred using the combination of known companion demographics and primary mass. The presence of an unresolved companion moves stars in a direction near-perpendicular to the pre-main sequence in the HR diagram, which has a relatively limited effect on the photometric mass estimate of the primary. These mass estimates therefore tend to accurately reflect the mass of the primary regardless of the presence of a companion, allowing us to use these masses to set the binary demographics regardless of unresolved binarity. 

For each primary star, we use existing measurements of binarity rate and mean mass ratio to compute an expected number of companions and total companion mass, following the methods of SPYGLASS-II. We use the binarity rates as a function of mass from \citet{Sullivan21}, interpolating a binarity rate for each star off of that curve. We then use the power law indices for mass ratio to compute an expected mass ratio for a hypothetical companion to each star in the sample, again interpolating the values of the expected mass ratio between mass bins \citep{Sullivan21}. For stars with $M>2$M$_{\odot}$, we adopt the mean mass ratio of the highest mass bin from \citet{Sullivan21}. We also include the \citet{Duchene13} companion fraction for stars with $1.5M_{\odot}<M<5M_{\odot}$ and adopt that result for a mean mass of 3.25 $M_{\odot}$. We interpolate binarity rate and expected mass ratio for a star of a given mass using the resulting curves. The expected number of binaries is then provided by the binarity rate, while the expected mass in binaries is the binarity rate multiplied by the expected mass ratio and the mass of the primary. On average, the inclusion of stellar companions increases the number of stars by a factor of 1.42, and increases the mass by a factor of 1.24.

\subsection{Correction for Low-Mass Stars} \label{sec:lowmasscorr}

\begin{figure}
\centering
\includegraphics[width=8cm]{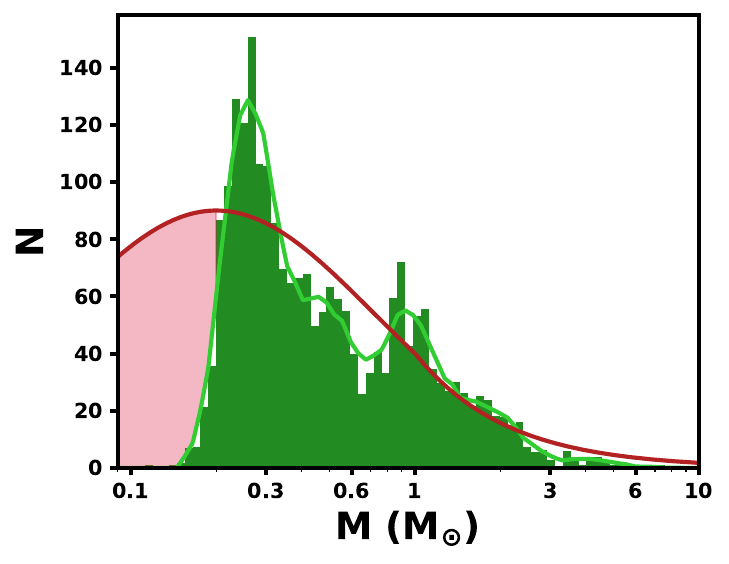}\hfill
\caption{Our correction for stars below \textit{Gaia}'s sensitivity limit. The histogram shows the distribution of stars in Cep-Her, which we show alongside a smoothed version of that histogram and the \citet{Chabrier05} individual star IMF. The shaded area between the curves for low-mass objects indicates the missing mass which we correct for. The deficit of stars at $M \sim 0.5$ and overabundance at $M \sim 0.25$ is a known issue, likely caused by model inaccuracies in stellar evolution times that displace stars to lower masses \citep{Kraus14}.}
\label{fig:IMF}
\end{figure}

Due to the distance of Cep-Her, the lowest-mass stars lack quality Gaia entries. Many of these stars have marginal detections, likely forming a substantial fraction of the widely-scattered \textit{Gaia} measurements in Figure \ref{fig:CMD} that fail our astrometric and photometric quality cuts. Other dim stars in Cep-Her lack $G_{BP}$ and/or $G_{RP}$ photometry entirely, especially M dwarfs where $G_{BP}$ falls below the Gaia flux limit. In these cases, it is impossible to assess their membership through \textit{Gaia} photometry. In Figure \ref{fig:IMF}, we show a mass histogram for all stars in Cep-Her that pass our astrometric and photometric quality cuts. This histogram provides the number of stars per mass bin, weighted by the membership probability $P_{\rm fin}$. We compare that result to the individual object Initial Mass Function (IMF) from \citet{Chabrier05}, which we scale to best fit to the mass histogram using least-squares optimization. 

The form of our mass distribution for Cep-Her is broadly similar to the canonical IMF, however also has a notable deficit of stars at $M \sim 0.5 M_{\odot}$ and an overabundance around $M \sim 0.25 M_{\odot}$. This pattern has been seen in other young populations, and is likely caused by slight model inaccuracies in the evolution of stars in this section of the pre-main sequence \citep{Kraus14}. We also see a deficit of stars more massive than $\sim 2 M_{\odot}$, although we caution against the overinterpretation of this result, as both the saturation of stars with apparent magnitude $m_G < 6$ in \textit{Gaia} and the overabundance of binaries in this mass range can contribute to a lower rate of quality \textit{Gaia} astrometry \citep{GaiaDR2ASLindegren}. Cep-Her may therefore benefit from a follow-up survey targeting bright stars that fail the astrometric and photometric quality cuts to confirm whether this deficit of massive stars is real. However, the most significant deficit of stars that we see is for masses $M \la 0.2 M_{\odot}$, where the mass histogram drops well below the canonical IMF, and no stars have $M < 0.11 M_{\odot}$.

This deficit of low-mass stars relates directly to the sensitivity limits of \textit{Gaia}, and we must therefore apply a correction to account for the mass and population of Cep-Her members in this mass range. To do this, we split our stellar sample into 4 distance bins, and for stars in each of those bins we create a separate weighted mass histogram. All of these have the same form as the histogram shown in Figure \ref{fig:IMF}, and we smooth each result to  suppress unphysical variations using a Savitsky-Golay filter with a window size of 11 and a polynomial of order 2. We assign an IMF to each result, which is scaled to produce the best fit with the histogram for $M>0.25 $M$_{\odot}$ via least-squares optimization. We then set the lowest-mass crossover point between the scaled IMF and the smoothed mass histogram as the point at which mass starts being lost. The region of parameter space where stars are lost to the \textit{Gaia} sensitivity limit is then provided by the region between the IMF and the smoothed mass histogram for stars less massive than this crossover point, as illustrated in Figure \ref{fig:IMF}. The corrective factor for the number of stars is then the sum across the entire stellar histogram, plus the sum over the region where stars are missing (red shaded area in Fig. \ref{fig:IMF}), all divided by sum of the stellar histogram. We limit this calculation to $M>0.09 M_{\odot}$, which is the mass limit in the PARSEC isochrones. We compute the missing mass in a similar way, simply multiplying the histograms and corresponding IMFs by the mass at each bin element and following the same calculation as for the number of stars. We do not limit this calculation in mass, as mass in objects below our $M>0.09 M_{\odot}$ cut can still contribute to the boundedness of clusters. 

Upon computing the missing number and missing mass fractions for each distance bin, we see a gradual linear trend decreasing completeness and increasing corrective factors with distance, as expected for a deficit caused by \textit{Gaia} sensitivity limits. For each star in our sample, we assign a correction by interpolating on the number of missing stars and the missing mass between the distance bin centers. This produces corrective factors that raise the number of members by a factor of between 1.35 and 1.41, and raise the total mass by a factor of between 1.12 and 1.13. The sum of spatial-only membership probabilities $P_{\rm spatial}$ for objects that fail our astrometric and photometric quality cuts implies that 4725 of them are members, compared to the 6798 stars added by this correction. This suggests that more than two-thirds of the stars added through this correction may have \textit{Gaia} entries, but lack the quality photometry necessary to assess their membership.

\subsection{Association-Level Cross-Contamination} \label{sec:xcontam}

A consequence of the substantial overlap between the sky-plane velocities and positions of Cep-Her Main and Cep-Her North is that the membership in one over the other cannot always be confidently established without RVs. In Section \ref{sec:rvselect} we separated the two populations in RV space, assigning stars without RVs to the closest population in 5D space-velocity coordinates. While this procedure assigns all stars to their most plausible group, there remains a risk of cross-contamination, or the assignment of a star to the wrong parent group due to the lack of an RV measurement. 

To assess the impact of this cross-contamination on our measurements of population size and mass, we investigate the rate at which we assign stars to one population versus another with two-dimensional velocities and compare it to the same rate among RV-confirmed members, producing a corrective factor that we can apply to our mass and population size estimates. We first pass each star in both Cep-Her Main and Cep-Her North through the procedure used to assign outlying population members to a parent group in Section \ref{sec:clustering}, this time finding an alternate subgroup assignment, or the most likely parent subgroup in the RV-defined region that each star was not initially assigned to. We assign stars to an alternate subgroup if they pass the clustering proximity dominance metric used in Section \ref{sec:clustering}, provided that the $D_{8}$ for that star within its best-fit subgroup is less than the maximum value of $D_{8}$ for the assigned subgroup members. For stars that receive RV follow-up, these alternate group assignments provide a straightforward way to update membership, as their assigned parent population can be easily updated from their current assigned group to their alternate group if membership in the other RV-defined region is confirmed. We therefore record these alternate group assignments in Table \ref{tab:members}. 

The combination of the original assigned members and alternate members provide a complete list of stars that can be plausibly connected to a given subgroup. The ratio between the number of stars assigned to the subgroup and the combined number of assigned and alternate stars therefore provides the fraction of stars with consistent space-velocity positions that are included in that subgroup's stellar sample. We can compute the same fraction using only stars with reliable RVs ($\sigma_{RV}<5$), which provides the fraction of stars with plausible positions in 5D space-velocity coordinates that remain consistent with the assigned subgroup when an RV is available. We take this latter ratio to represent the true membership rate in the subgroup, and therefore compute the cross-contamination corrective factor as the fraction of members among stars with good RVs, divided by the fraction of all plausible members that are assigned to the subgroup rather than its alternate. 

We assign the cross-contamination factor of the parent subgroup to each stellar primary. For stars without a parent subgroup, we re-compute the value of cross-contamination across the entire association, and assign the unclustered stars this averaged value. The use of these fractions introduces a binomial uncertainty into these corrections, which we propagate through to our mass and population size measurements in Section \ref{sec:finmass}. 

In this correction, values less than one imply that more stars are assigned to the subgroup than is merited by the membership rates among stars with good RVs. Our cross contamination corrective factors are usually at or near 1, however they can be as low as 0.72 and as high as 1.23. The corrective factor of 1.23 is assigned to CINR-1, where there are far more RVs consistent with the population than its relatively small assigned membership would predict. The corrective factor of 0.72 is assigned to CINR-9, which overlaps with several RV-dominant populations in central Cupavo. Most of the corrective factors that are not near 1 are found in Cinyras and Cupavo, as these associations overlap the most in 2D velocity space. 

\subsection{Expected Masses and Population Sizes} \label{sec:finmass}

Expected masses and population sizes are computed by combining our stellar populations and masses with the expected binary contribution, losses due to stars without adequate \textit{Gaia} coverage, and corrections for cross contamination. The expected number of stars present for a stellar single or primary of a given mass is provided by the following formula:

\begin{equation}
N_x = P_{\rm fin} * (1 + f_{bin}) * c_{N,lm} * c_{cc}
\end{equation}

where $f_{bin}$ is the binarity rate for a star given its mass, $c_{N,lm}$ is the correction for the number of stars below the mass detection limit, and $c_{cc}$ is the cross contamination correction for the star's parent region. The expected mass is then computed as:

\begin{equation}
M_x = M_* * P_{\rm fin} * (1 + (f_{bin}*\overline{r_M})) * c_{M,lm} * c_{cc}
\end{equation}

where $\overline{r_M}$ is the expected mass ratio of a star given its mass, and $c_{M,lm}$ is the mass corrective factor for objects below our detection limit. The total number of stars $N_{tot}$ and mass $M_{tot}$ in an association or subregion can then be computed by summing $N_x$ and $M_x$, respectively, across all singles or system primaries in the population. We report the expected masses and population sizes for each subgroup in Table \ref{tab:groupstats}, and use the population masses to assess the virial states of subgroups in Section \ref{sec:boundedness}.

We also provide fractional uncertainties in Table \ref{tab:groupstats}, which combine the binomial uncertainties in the cross-contamination correction and membership probability calculation for older subgroups with systematic uncertainties. The mass models, IMF correction for low-mass stars, and the binarity correction all contribute to the systematic uncertainties, and we adopt a value of 10\% in most cases, which combines the approximate maximum discrepancies between the IMF and our distribution of stellar masses, and typical literature variation in binary statistics \citep{Duchene13, Sullivan21}. We raise the systematic uncertainty to 20\% for the old populations where we use $P_{bulk}$ in place of $P_{\rm fin}$, as the use of a limited stellar mass range in the $P_{bulk}$ calculation and potential for biases from binary contaminants introduce additional systematic uncertainties. The values of fractional uncertainty we report average the upper and lower uncertainty intervals in cases where the uncertainty is asymmetric. In cases of young subgroups without plausible cross-contaminants, only the 10\% systematic uncertainties affect the reported values. 

The binomial uncertainties in the cross contamination decrease substantially at the association level, due to the large number of stars on that scale. We therefore use the binomial uncertainty in the association-wide cross contamination calculation as the statistical uncertainty in the cross contamination rate for an association, calculated using the same methods employed in Section \ref{sec:xcontam}. 
We also compute association-level uncertainties in $P_{bulk}$ for the older associations, computing binomial uncertainties based on the number of field stars across all subgroups in both Roslund 6 and Cupavo. These two sources of binomial uncertainties are added in quadrature with our systematic uncertainties to produce statistics at the association level. 

After all corrections, we find that Orpheus is the largest association in the Cep-Her complex with $9552\pm960$ members totalling $3881\pm390$ M$_{\odot}$. This gives it a membership rivalling the $10^4$-star population of Sco-Cen \citep{Pecaut16}, which is substantially larger than the $\sim 2000$ M$_{\odot}$ Orion Nebula Cluster \citep{Hillenbrand97}.  Cupavo is next largest, with $8794\pm1827$ members containing $3620\pm752$ M$_{\odot}$, followed by Cinyras with $3872\pm455$ members and $1604\pm188$ M$_{\odot}$, and finally Roslund 6, with $1507\pm347$ members and $700\pm161$ M$_{\odot}$. With nearly $10^4$ M$_{\odot}$ and over $2.3 \times 10^4$ stars across all four component associations, Cep-Her as a whole may be larger than the entire Orion Complex, which is thought to contain less than $2 \times 10^4$ stars \citep{Bally08-ori}. 

\subsection{Boundedness} \label{sec:boundedness}


We assess the boundedness of our subgroups following the methods used in SPYGLASS-II, with some optimizations to better automate the process and to minimize the effects of contamination. Bound clusters satisfy $\sigma_{1D} < \sqrt{2}\sigma_{virial}$, where $\sigma_{1D}$ is the square root of the average variance in a multi-dimensional velocity-space, and the virial velocity, $\sigma_{virial}$, is defined as follows \citep{PortegiesZwart10, Kuhn19}:

\[\sigma_{virial} = \left(\frac{GM}{r_{hm}\eta}\right)^{-\frac{1}{2}}\]

\noindent where $M$ is the subgroup mass, $r_{hm}$ is the half-mass radius, and $\eta$ is a value that represents the cluster's mass profile. For a Plummer profile, $\eta = 10$, however young clusters often have lower values of $\eta$ due to having broader density profiles \citep{Kuhn19}. We take $\eta = 5$ for our calculations, typical of these broader profiles. We define $r_{hm}$ in 2D $l$-$b$ coordinates to remove the substantial uncertainties in the distance dimension, and use two different methods to compute $r_{hm}$. First, we identify $r_{hm}$ as the radius out from the subgroup center that contains half of the mass, and use the full mass of the subgroup as $M$. Second, we compute the density profile of the subgroup using a KDE, fit a bivariate gaussian to the resulting profile, and compute $r_{hm}$ using the mass expected by the fit, rather than the measured mass. This secondary method accounts for dense populations with wide halos, where a core is bound but the outer reaches are not. At the age of Cep-Her, clusters are already dissolving, so the presence of a plausibly bound region indicates that the subgroup as a whole was likely fully bound at formation, with all of the mass associated with a cluster's widely-scattered halo in the central core. Our quoted $\sigma_{virial}$ uses the smaller of the two $r_{hm}$ values. The uncertainty in $\sigma_{virial}$ then combines the fractional uncertainty in mass with a 10\% systematic uncertainty for $r_{hm}$. 

For stars within the selected value of $r_{hm}$, we compute $\sigma_{1D}$ in $\Delta$v$_{T}$ coordinates, excluding probable binaries, and list the results in Table \ref{tab:groupstats}. The restriction to single stars within $r_{hm}$ ensures a maximally pure sample of association members, while minimizing any internal sources of velocity uncertainty. In cases where the binarity restrictions would leave fewer than 20 stars, we remove the binarity cuts, which prevents unphysical results for velocity dispersion that can emerge through small number statistics. For the chosen subset of stars on which to compute $\sigma_{1D}$, we then bootstrap, selecting half of the stars at random to a minimum of 10 and a maximum of 60. For each bootstrapped sample, we compute a standard deviation, and subtract the median velocity uncertainty from the result in quadrature to remove the contribution from uncertainty to the velocity dispersion. To further limit the effect of outliers, we sigma-clip each result at $2\sigma$. The value for $\sigma_{1D}$ is then the median of the sigma-clipped standard deviation across 1000 bootstrapped samples, and its uncertainty is the corresponding standard deviation.

\begin{figure}
\centering
\includegraphics[width=8cm]{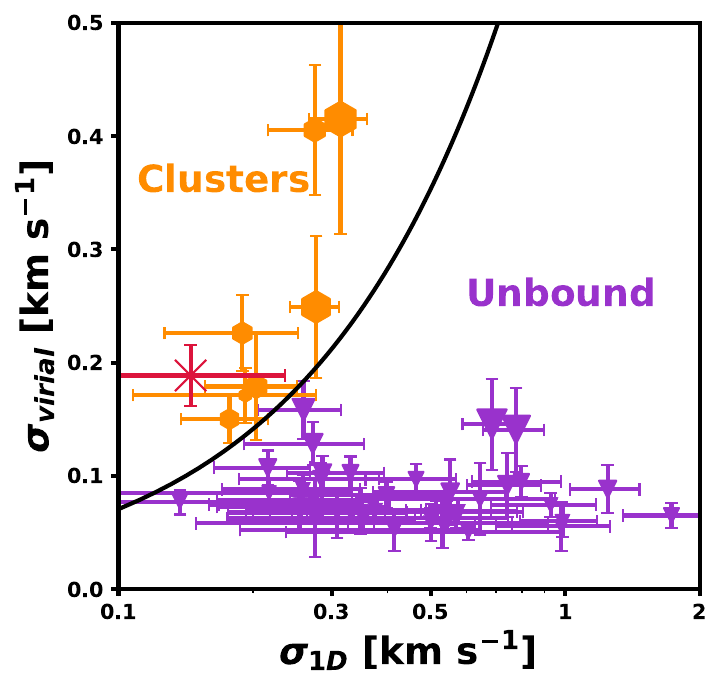}\hfill
\caption{The virial velocities and 1D velocity dispersions for Cep-Her subgroups. The curve shows $\sigma_{1D} = \sqrt{2}\sigma_{virial}$, which is the boundary between bound and unbound populations. Subgroups above this curve are open clusters, marked with orange hexagons, and stars below the line are unbound substructures, marked with purple triangles. Marker sizes are scaled with the subgroup mass. The bound substructure found in ORPH-23 is marked with a red X.}
\label{fig:virialrat}
\end{figure}

From the framework in \citet{Kuhn19}, populations with $\sigma_{1D} < \sqrt{2}\sigma_{virial}$ are bound, so we can define a virial ratio as $\sigma_{1D}/\sqrt{2}\sigma_{virial}$, with bound populations having a virial ratio less than 1. For groups with an initial virial ratio less than 2, we inspect the population by hand to check whether a small high-density core is present that may be smaller than the kernel size for the fit to $r_{hm}$. In seven cases where the kernel appears to be too large for the core that we find, we reduce the kernel width and re-calculate. This is mainly necessary in cases like CUPV-3, which has a dense core that is only about 3 pc across (0.5$^\circ$ on-sky), however some populations with slightly looser cores such as ROS6-1 are also included. In cases with a low virial ratio but a kernel width that appears suitable for the cluster, we experiment further with changes to the kernel size, and find that our choice of kernel does not substantially change our virial results. 

The final virial ratios are presented in Table \ref{tab:groupstats}. We present the distribution of $\sigma_{virial}$ vs $\sigma_{1D}$ in Figure \ref{fig:virialrat}, coloring each population according to their most probable virial state.  There we demonstrate the presence of two distinct sequences - one containing virialized clusters roughly consistent with with a linear trend elevated above the boundedness limit, and a sequence of unbound subgroups, which have similar values of $\sigma_{virial}$ but varied values of $\sigma_{1D}$. The virial ratio of 1 that we include as a black curve in Figure \ref{fig:virialrat} roughly separates the two sequences, enforcing this limit as a transition between two different dynamical states.

Three populations, CUPV-3, ORPH-28, and CINR-10, are high-confidence open clusters, meaning that they have a low virial ratio that is not within uncertainties of 1. Four more, ROS6-1, CUPV-10, ORPH-15, and ORPH-16, are probable clusters, as they have virial ratios less than 1, but have uncertainties high enough that they are consistent with an unbound state. Finally, four populations, CINR-7, ORPH-12, ORPH-14, and ORPH-23 appear unbound in our analysis, however their uncertainties do not rule out the possibility of them being bound. ORPH-23 is a unique case, as it appears to contain an embedded high-density substructure. Since this overdensity is near the edge of OPRH-23, it does not appear to represent a central core, and we therefore do not take the virial state of this region to represent the entire subgroup. However, its size and shape is similar to a series of emerging low-mass and extremely compact clusters that have been discovered recently, such as EE Dra in CFN and $\eta$ Cha in Sco-Cen \citep{Mamajek99, Kerr22a}. By limiting our $\sigma_{virial}$ and $\sigma_{1D}$ calculations to only this core, we found a virial ratio of $0.54 \pm 0.28$, here, indicating that this substructure is bound. ORPH-23 therefore contains a embedded open cluster, which may disrupt stellar motions. We include this bound substructure separately from ORPH-23 in Figure \ref{fig:virialrat}.

Six of the probable bound structures we identify in Cep-Her have already been recognized in published open cluster catalogs \citep[e.g.,]{CantatGaudin20}. The largest of these is the dense open cluster NGC 7058, which is found in our catalog as CUPV-3. The other two high-confidence open clusters, ORPH-28 and CINR-10, are also known structures, corresponding to the $\delta$ Lyrae cluster and RSG-5, respectively \citep{Stephenson59, Roser16}. Of the probable bound structures, ROS6-1 represents the core of the open cluster Roslund 6 \citep{Roslund60}, and ORPH-15 corresponds to UBC 9 \citep{CastroGinard18}. The other two probable bound structures, CUPV-10 and ORPH-16, have not yet been recognized as open clusters. Finally, the dense core we identify in ORPH-23 appears to correspond to the small open cluster ASCC 100 \citep{Piskunov06}.  

\subsection{Completeness} \label{sec:completeness}

While our expected mass and population estimates account for most sources of missing mass within the boundaries of Cep-Her, there are certain to be additional members located beyond the limits to $P_{\rm spatial}$ which we have imposed. Stars in these outer reaches are likely to have been ejected soon after or even during formation, limiting their effect on the long-term virial state of Cep-Her substructures. However, the number of members in these outer reaches may constitute a substantial fraction of Cep-Her's total population \citep[e.g.,][]{Fujii16}. 

The sum of $P_{\rm spatial}$ across all stars beyond our $P_{\rm spatial}>0.2$ limit is over $1.5\times10^4$, suggesting that as much as half of the mass of Cep-Her may be outside of our $P_{\rm spatial}$ limit. However, we also notice that 63\% of the stars in these outer reaches have poor photometric quality or high RUWEs, compared to just 45\% in the core. This suggests that some of these stars may be poor-quality Cep-Her members with sufficiently high uncertainties that they are not readily associated with Cep-Her's core populations. These are already partially accounted for by our correction for stars without quality \textit{Gaia} entries, although only $\sim2000$ stars predicted by this correction are unaccounted for based on $P_{\rm spatial}$ for stars in our core Cep-Her sample (see Sec. \ref{sec:lowmasscorr}). This would leave approximately $1.3\times10^4$ stars in this outer halo as plausible additional members of Cep-Her, which would raise the total number of stars in the complex by over 50\%. 

The outlying stars likely disproportionately originate in Cupavo and Roslund 6, as their older ages give stars more time to reach these widely-distributed locations. Stars in these older populations are also more likely to be missed in the initial clustering in SPYGLASS-IV, so the stars in this outlying region could form entire coherent sub-populations that SPYGLASS is unable to detect. Better constraints on the sizes of these outlying populations will require traceback to reveal the origins of these outlying objects, as well as the modelling of any hidden coherent associations found outside the extent of Cep-Her we explore here. Updates to the \textit{Gaia} astrometric solution are also likely to help, as they will improve the overabundant low-quality \textit{Gaia} solutions that dominate this extended sample. However, true in-depth studies of these populations are likely to require entirely new methods that are more suitable for isolating these older populations. One possible approach may be to select stars based on the \textit{Gaia} excess variability metric, which has recently been shown to predict age within 10-20\% for stars younger than 2.5 Gyr \citep{Barber23}. 

\section{Population Dynamics} \label{sec:dynamics}

\subsection{Expansion Velocities} \label{sec:dyncohere}

Young associations gradually expand after their gas cloud disperses, as they typically have densities far too low to overcome the initial internal velocity dispersions of stars, except in relatively localized overdensities like embedded open clusters \citep[e.g.,][]{deZeeuw99}. Dynamically coherent young associations are therefore expected to have stars with velocity vectors pointed radially outwards from the association center, with higher velocities for more outlying members. Deviations from a divergent trend can indicate the presence of expanding substructures within the broader population, or contaminant populations with velocities that do not place them close to the rest of the association at formation. 

\begin{figure*}
\centering
\includegraphics[width=15cm]{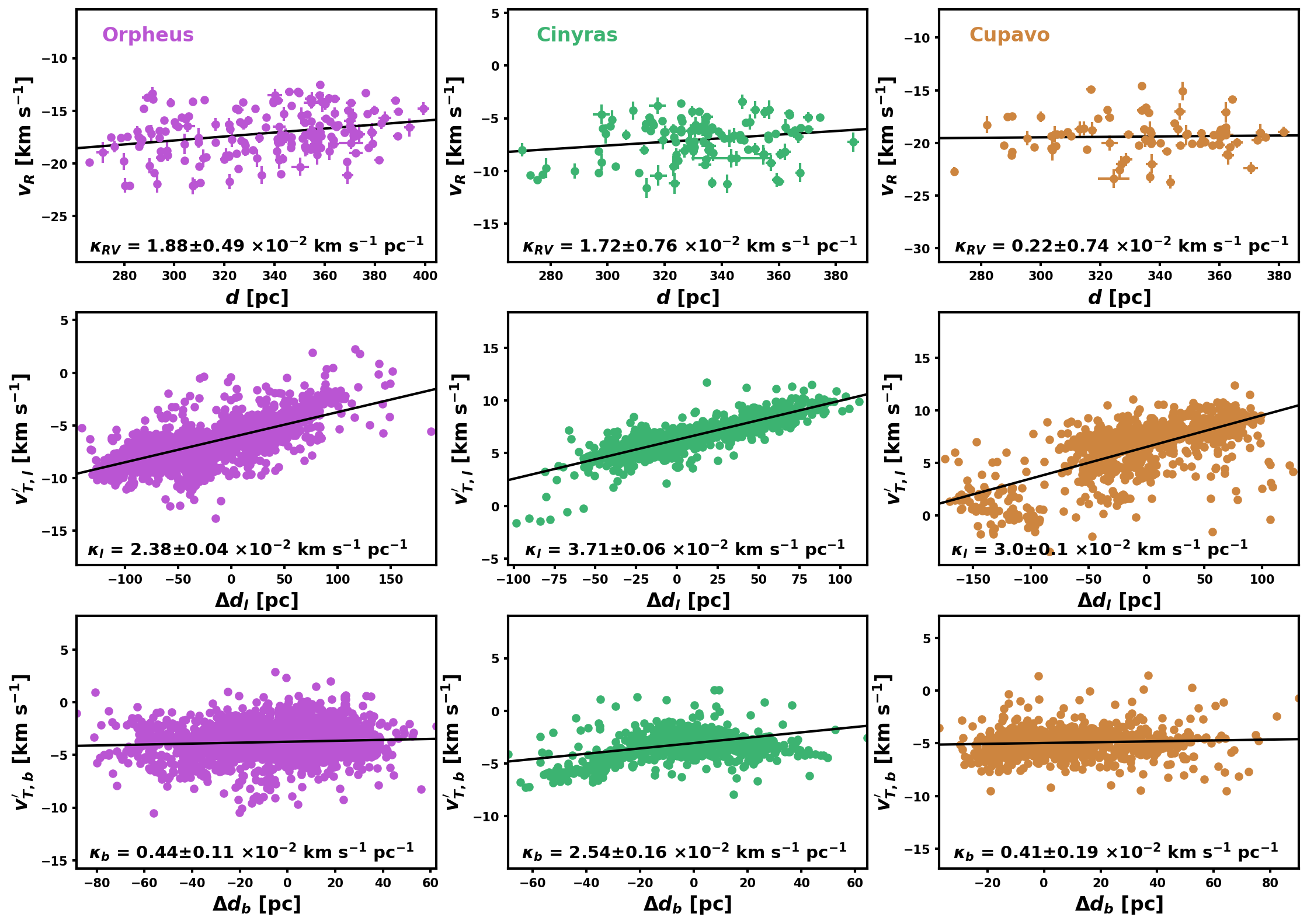}\hfill
\caption{Expansion trends in Orpheus, Cinyras, and Cupavo, shown as distance along an axis versus velocity along that axis. The top row shows distance vs RV, and the bottom and middle rows show the corrected transverse velocity in $l$ and $b$, respectively versus the linear distance in the $l$ and $b$ direction relative to the mean at the distance of each star. We include uncertainties for the distance axes, and error bars for the $l$ and $b$ axes are typically smaller than the data points. }
\label{fig:expansion}
\end{figure*}

By \citet{Wright18}, we can identify expansion through a positive trend in velocity along a given axis. Along the distance/RV axis, closer members of an expanding group have an RV component pointing away from the cluster center and towards the observer, while stars on the far edge have a velocity component away from the observer \citep{Wright18}. Along this axis and any other linear axis in 3D space, a line can be fit to the distribution of velocities along that axis, producing a rate of expansion, $\kappa$. In on-sky galactic $l$ and $b$ coordinates, this calculation is more complicated, as the sky-plane component of stellar motion is subject to a perspective effect in which stars moving towards the observer show virtual expansion and vice versa \citep{Kuhn19}. Since we have well-constrained RVs for all three associations in our sample, we can correct for this effect using the following formulae, following \citet{Brown97}:

\begin{align}
v^{\prime}_{T,l} &= v_{T,l} - \overline{v_{R}}*\cos b_c\sin(l_c-l)\\
v^{\prime}_{T,b} &= v_{T,b} - \overline{v_{R}}(\sin b_c \cos b - \sin b\cos b_c\cos(l_c -l))
\end{align}

\noindent Where $\overline{v_R}$ is the mean radial velocity, and $l_c$ and $b_c$ are the mean position of the association in galactic sky coordinates. This allows us to calculate the on-sky expansion rate in the $l$ and $b$ directions. 

Figure \ref{fig:expansion} shows the expansion trends along the distance axis and the on-sky galactic $l$/$b$ axes. For the distance axis, we plot distance against RV, while for the $l$ and $b$ axes we plot corrected transverse velocity against $\Delta d_l$ and $\Delta d_b$, the on-sky distance in pc between the mean value of $l$ and $b$ for the cluster and a given star at its distance. To each trend, we fit a line, which provides the expansion rate along that axis. For our fits to all axes, we remove binaries, as well as stars with $P_{\rm fin} < 0.8$ for Cinyras and Orpheus and $P_{\rm fin} < 0.5$ for Cupavo. These restrictions limit the presence of interlopers and apparent motions from stellar companions, with the looser cut in Cupavo justified by the underestimation of $P_{\rm fin}$ in this older population. For the distance/RV axes, we additionally require that $\sigma_{RV}<1$ km s$^{-1}$ to ensure that the stars we include have adequate RV data. All fits assume equal uncertainties, as most of the scatter seen is likely to be physical, rather than due to measurement uncertainties. 

The results show that 8 of the 9 subplots (3 axes across 3 associations) show statistically significant expansion at more than a 2-$\sigma$ level. The rest show a general trend of expansion, although often with very different rates along the three different axes. These varied expansion rates by axis are a common feature of studies like this, and may reflect spatially varied stellar feedback after the first stellar generation or other forms of substructure \citep[e.g.,][]{Wright18, CantatGaudin19, Pang21}. Galactic orbital effects are another potential cause, especially in the $b$ axis. As stars orbit the galaxy, the rise and fall through the disk. This results in stars reaching zero vertical velocity at the peaks of their epicycles before falling back towards the galactic plane. The result is that for an association with a wide range in galactic $Z$ and consistent epicyclic amplitudes for its members, stars will converge in $Z$ and $b$ space before diverging as the last stars begin heading back to the galactic midplane. This is the likely cause for the relatively weak divergence trend in $b$ that we display in Orpheus, which, with its high position in galactic $b$ and low $v_{T, b}$, is near the peak of its epicycle. A similar effect is visible in Cinyras, where we see an overall positive divergence trend in $b$ that turns negative for stars with the highest values of $b$.

\begin{figure}
\centering
\includegraphics[width=8cm]{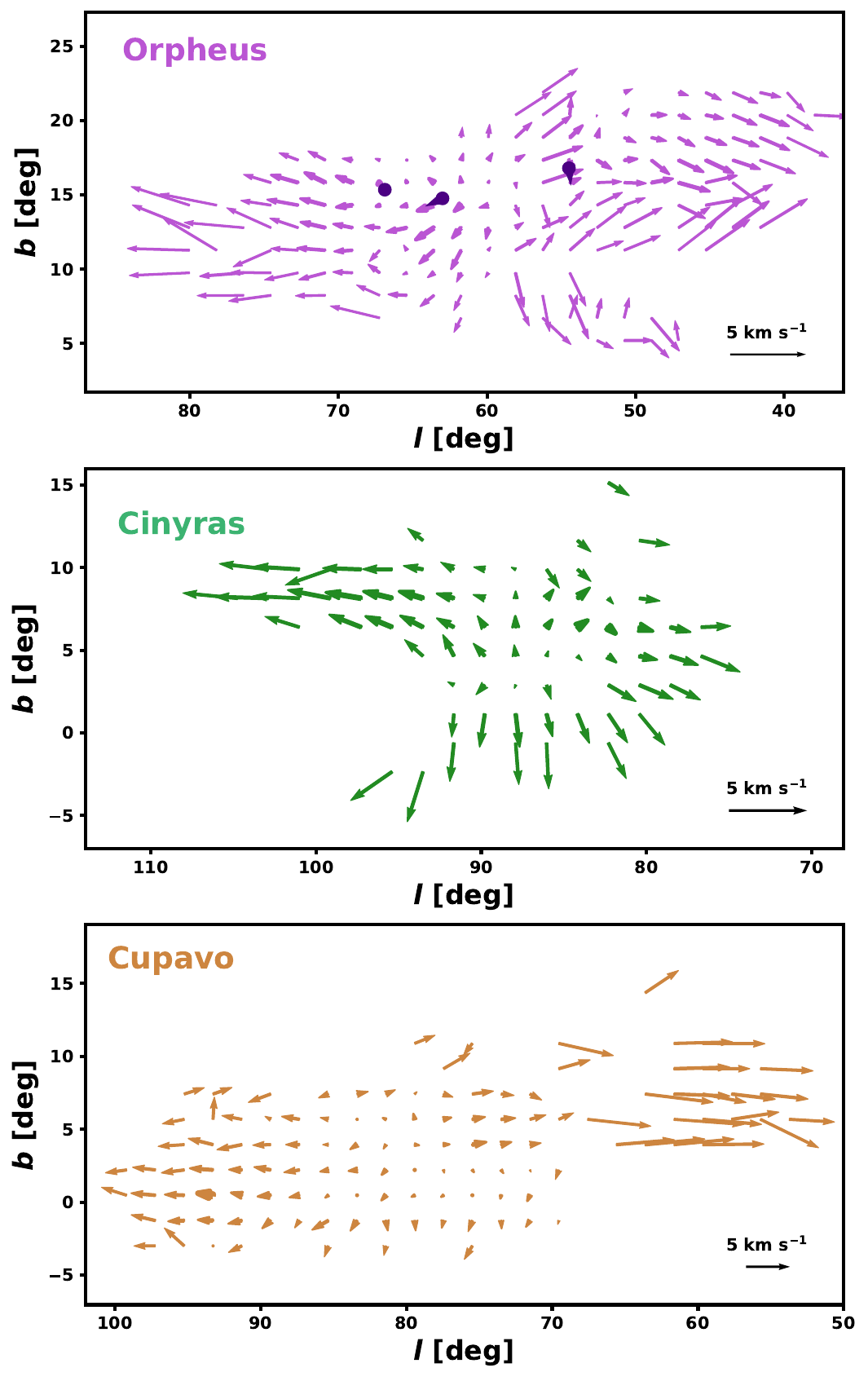}\hfill
\caption{Vector plots for the corrected transverse velocities of stars in Orpheus, Cinyras, and Cupavo, spatially averaged. A 5 km s$^{-1}$ velocity vector is shown in the bottom-right of each figure for reference. All three broadly show expansion from a common origin along at least one axis, with some evidence for further dynamical substructure. The arrow widths are proportional to the square root of the number of stars included in the velocity averages for each grid cell to indicate where substantial stellar overdensities lie. The dark dots and arrows in the Orpheus panel show the locations and velocity vectors of the three fully-bound subgroups (ORPH-15, 16, and 28).}
\label{fig:vectorplots}
\end{figure}

\subsection{Dynamical Coherence}

We have confirmed that all three associations in Cep-Her are expanding, however deviations within these overall velocity trends may indicate the presence of additional distinct structures at formation. We map these expansion trends in Figure \ref{fig:vectorplots}, which shows the spatial distribution of corrected on-sky transverse velocity vectors in Orpheus, Cinyras, and Cupavo, averaged by bin in $l$ and $b$. The conditions for inclusion in these average velocities are the same as those we use for our fits to $\Delta d$ and $v^{\prime}_{T}$ (Figure \ref{fig:expansion}). Velocities are shown relative to the group average. The result provides a visual representation of expansion that reveals regional patterns that are averaged over during bulk measurements of expansion. All three populations show visual indications of expansion from a common origin in at least some form, however they also show some anisotropies that reflect the anomalies visible in Figure \ref{fig:expansion}.

Cinyras shows especially consistent expansion in all directions on-sky, with most vectors pointing radial outward from an apparent origin. Stars with high $b$ are an exception to this trend, however, as the $+b$ component of the corrected transverse velocity decreases as it approaches $b = 10$ before turning over entirely. This is the same pattern shown in Figure \ref{fig:expansion}, and is likely attributable to motion near the peak of an epicycle. 

Cupavo also has relatively consistent expansion across most of its extent, however Figure \ref{fig:vectorplots} shows substantially higher rates of divergence in the galactic west ($-l$) of the population near Orpheus. These high velocity vectors correspond to CUPV-1, 2, and 4, which are all shown to have velocities differing by a mean of nearly $\sim8$ km s$^{-1}$ in Figure \ref{fig:cupavo}. While the subgroups CUPV-1, 2, and 4 are the most dynamically distinct, CUPV-7 has a similarly old age, and the stellar velocities found near it have a similar direction of motion to those found in those dynamically distinct groups to the galactic west. This may indicate the presence of another dynamically distinct population mixed in with Cupavo that has a fully independent star formation history, however traceback will be necessary to reveal whether the velocities found in these structures are consistent with other Cupavo substructures at formation. 

Orpheus has a more complex pattern of expansion. While the population as a whole is largely expanding, we also see two ridges of convergence within that appear to break the population into as many as three distinct expanding substructures. However, the open clusters $\delta$ Lyr (ORPH-28) and UBC 9 (ORPH-15), which are situated on opposite sides of one of these convergence boundaries and can be viewed as central within their local section of Orpheus, have transverse velocity anomalies separated by only $\sim 1$ km $^{-1}$ (see Fig. \ref{fig:orpheus}). The velocities  of these probable fully bound subgroups are similarly small in $v^{\prime}_{T}$-space, which we show using dark markers and arrows in Figure \ref{fig:vectorplots}. This pattern of consistent core velocities across multiple expansion origins suggests that what we see in Cep-Her is produced by multiple spatially-separated but dynamically similar star formation events, each of which giving rise to a series of subgroups that have dispersed independently since formation. 

The complexities we see in the expansion trends across all three Cep-Her associations demonstrate the potential risks in using linear expansion rates at a population-wide level for computing dynamical ages, as these may average over substructures within and ignore convergence effects that are not physical to the cluster. As discussed in SPYGLASS-II, subgroup-specific dynamical traceback is often required to produce reliable dynamical ages. However, due in part to the number of previously unknown structures identified in this paper, many regions in Cep-Her still do not have enough high-quality RVs ($\sigma_{RV} < 1$ km s$^{-1}$) to enable accurate traceback on the subgroup scale. A complete traceback and dynamical age analysis in Cep-Her is therefore beyond the scope of this paper. 

\subsection{Forward Simulations} \label{sec:simulations}

\begin{figure}
\centering
\includegraphics[width=8cm]{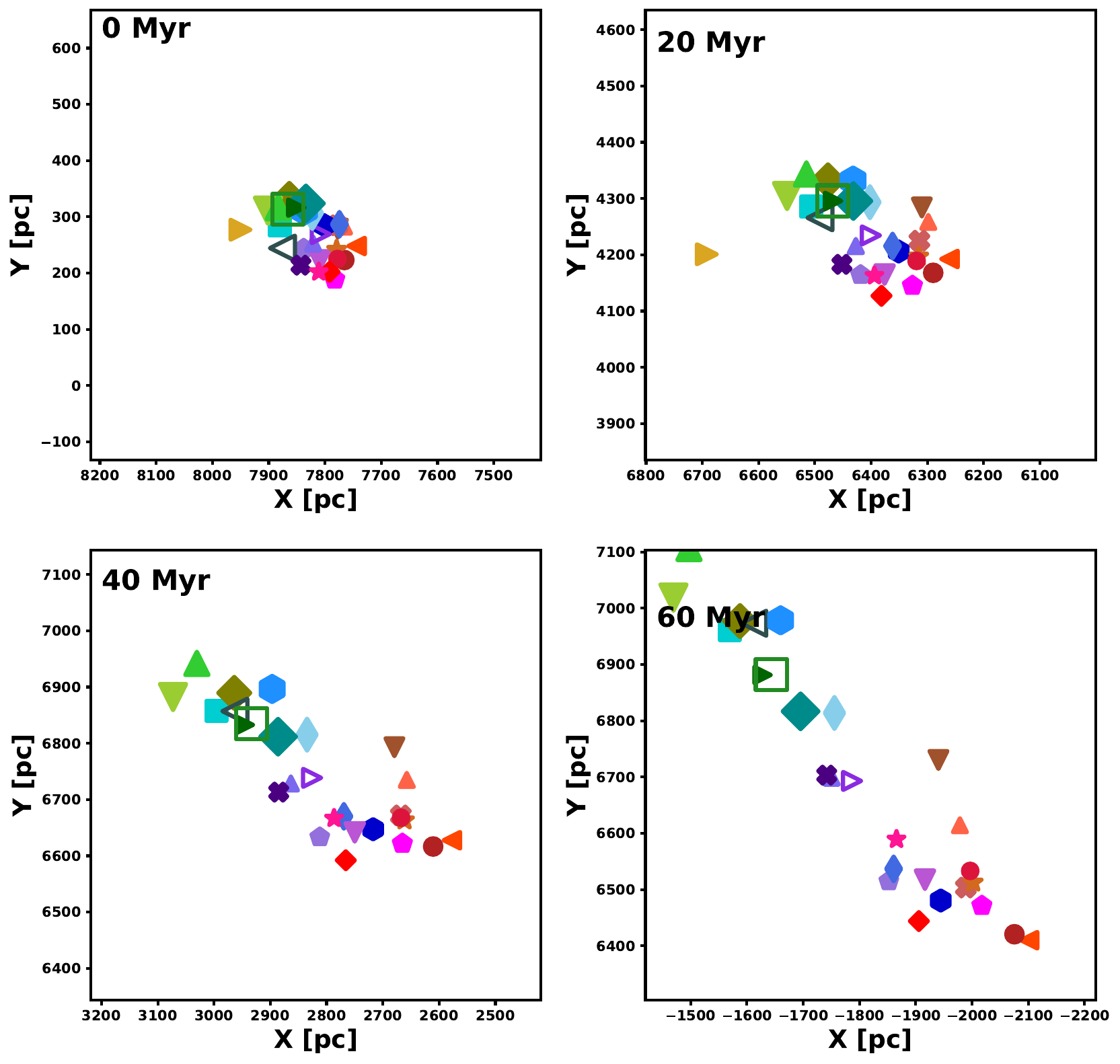}\hfill
\caption{Simulation of the evolution of the Orpheus Association over the next 60 Myr, presented in XY galactic coordinates. The markers match those used in Figure \ref{fig:orpheus}, where bound clusters or groups with a bound component are marked with an open icon. Each marker shows the mean location of a subgroup, with the marker size scaled with the group mass. The online-only version provides an interactive view of these results in three-dimensions, and also provides more time steps over a longer span of time. }
\label{fig:orph-sim}
\end{figure}

Dense concentrations of massive populations are found in the center of Orpheus and around RSG-5 in Cinyras, both of which are accompanied by low relative velocities. It is therefore worth considering whether the future evolution of these populations really is slow dispersal, or whether they will eventually fall into one another and merge. The question of whether star clusters form in singular events or through a gradual assembly of small populations into larger structures, which is known as hierarchical assembly, has long been a topic of discussion \citep[e.g.,][]{Lada84, VazquezSemadeni17}. In recent years, new observations have made it possible to perform accurate N-body simulations of real clusters. It is now possible to predict the future assembly of real young structures into a larger cluster, like in the case of $h$ and $\chi$ Per \citep{Dalessandro21, DellaCroce23}. With our new observations, we can perform a similar analysis in Cep-Her's component associations. Regardless of whether Orpheus and Cinyras in particular will experience hierarchical assembly, the evolving shape and structure of young associations colors our interpretations of their histories. Having a general method for evaluating the forward-evolution of populations with a self-consistent gravitational potential allows different associations to be compared in morphology at various stages of their evolution, which will ultimately enable concrete comparisons between populations that are currently in very different evolutionary states. We therefore forward-simulate the evolution of all three substructured associations in our sample to investigate their future morphology and the potential for hierarchical assembly. 

\begin{figure}
\centering
\includegraphics[width=8cm]{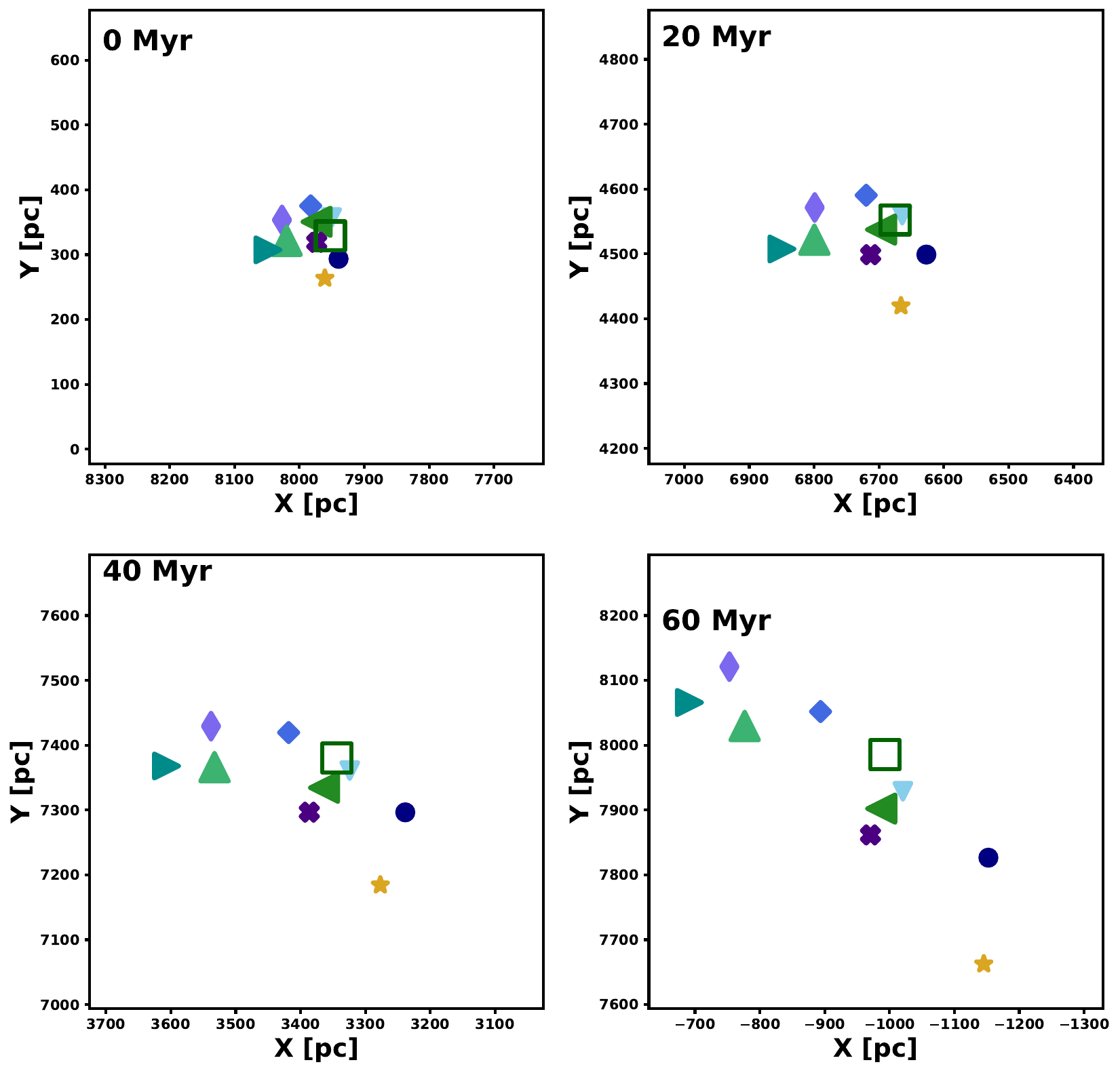}\hfill
\caption{The same as Figure \ref{fig:orph-sim}, but for the Cinyras Association.}
\label{fig:cinr-sim}
\end{figure}

\begin{figure}
\centering
\includegraphics[width=8cm]{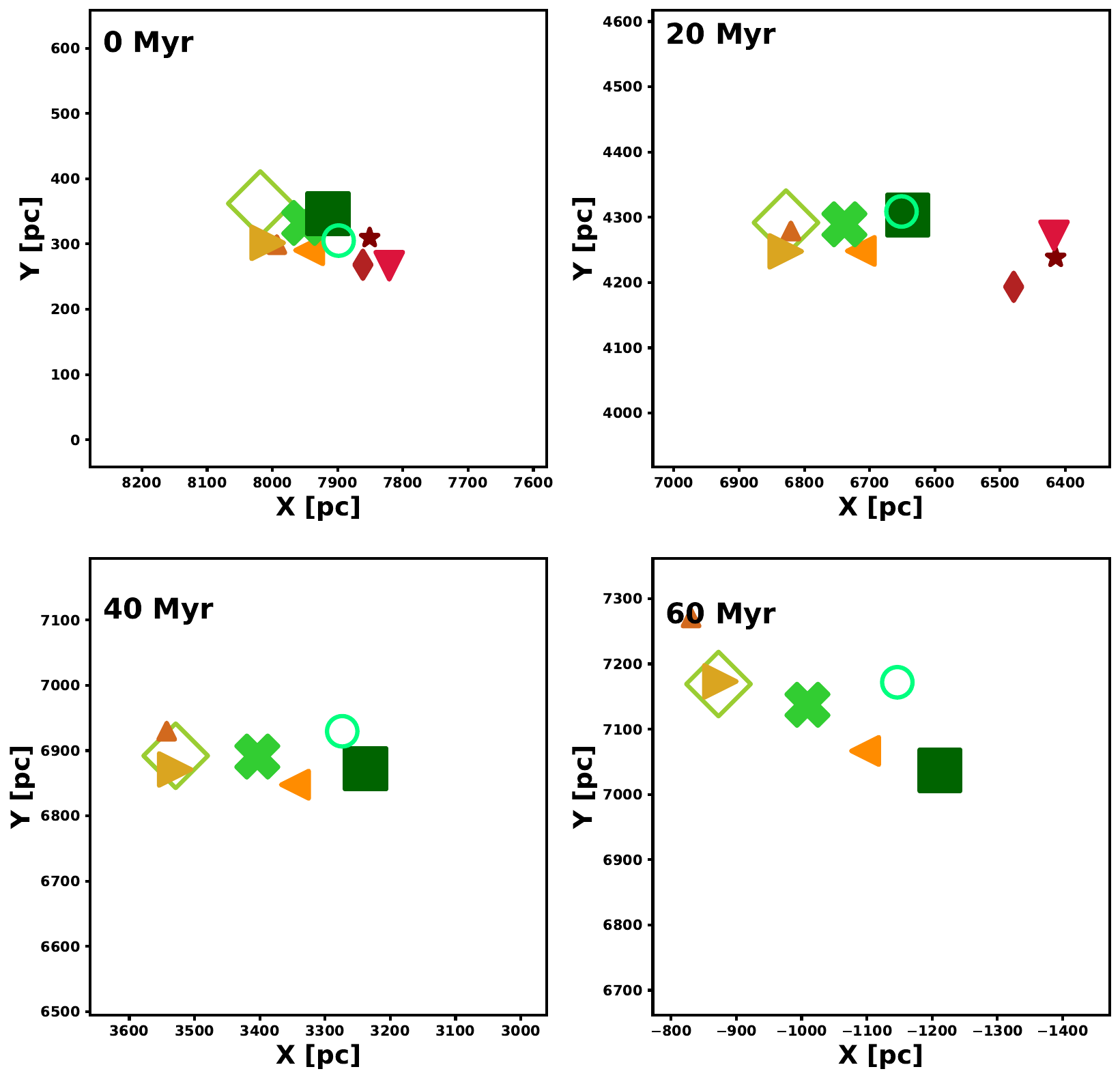}\hfill
\caption{The same as Figure \ref{fig:orph-sim}, but for the Cupavo Association.}
\label{fig:cupv-sim}
\end{figure}

We simulate the forward-evolution of our associations at the level of subgroups, modeling the motion of the subgroup as a whole rather than of individual stars within to more accurately track the location of subgroup centers. We evolve these subgroups using the Astrophysical Multipurpose Software Environment \citep[AMUSE;][]{PortegiesZwart18}, which allows us to combine the necessary gravitational codes for this task. We model the gravitational interactions between groups using the 4th order Hermite integrator \texttt{ph4}, and include the external influence of the Milky way using galpy's \texttt{MWPotential2014} gravitational potential \citep{Bovy15}. We assign each particle in the simulation the mass, mean position, and velocity of a subgroup. For the particle masses, we use the total mass of all stars that are closest to that group, following the method for defining closeness that we defined in Section \ref{sec:clustering}. This more generous inclusion of mass is important for producing an accurate potential in places like the center of Orpheus, where many populations are found near one another, and the exact affiliation of any one star is often ambiguous. This method therefore recognizes the contributions of these stars to the overall gravitational potential. The mean velocities for each subgroup are computed using a sigma-clipped mean of all members that are RV members and not binaries. We further restrict the selection to exclude stars with $\sigma_{RV}>1$, or RVs from sources with known systematic uncertainties at the 1 km s$^{-1}$ level such as Gaia and RAVE, provided that at least 3 stars remain after the cut. The use of subgroups rather than stars for these simulations is well-suited to investigate the shape of the association and the potential merger events that interest us. However, in future publications the same combination of AMUSE and galpy can be used to simulate stellar populations for each subgroup, and model the evolution of individual stars in the association over time. For this paper, we consider groups for merger if their cores pass within 20 pc, which is near the maximum half mass ratio for our populations.  

We therefore simulate the evolution of these structures over the next 250 Myr, and show their evolution over the next 60 Myr in Figures \ref{fig:orph-sim}, \ref{fig:cinr-sim}, and \ref{fig:cupv-sim}. We also vary their velocities within the uncertainties in their mean 100 times and re-run the positions for each to estimate uncertainty. The results show that all three populations are undergoing gradual expansion, as our results in Section \ref{sec:dyncohere} suggest. This expansion is especially pronounced along a diagonal axis consistent with orbital shear, although all three show expansion in all directions. In almost all cases, this divergence is more than sufficient to overcome the internal gravitational potentials in these associations. 

While this overall expansion dominates, we do see cases where dynamical evolution beyond simple expansion may be occurring. In Orpheus, while most populations readily diverge from each other, we see hints of convergence around $\delta$ Lyr (ORPH-28). The adjacent population of ORPH-27 has the strongest indication of convergence, with 75\% of uncertainty-varied simulations showing a mutual distance with $\delta$ Lyr of $<$20 pc beyond the first 20 Myr. $\delta$ Lyr also has a close encounter with the probable bound cluster ORPH-16 in 24\% of simulations, despite the current separation of 75 pc. The times of these potential encounters are all distributed throughout our 250 Myr simulation time. Cinyras sees a similar potential for future convergence, with CINR-4 and CINR-7 coming within 20 pc of the open cluster RSG-5 in 52\% and 16\% of uncertainty-varied simulations, respectively. Finally, Cupavo shows less potential for future evolution, with no groups showing close encounters beyond the first 20 Myr in more than 20\% of simulations. It therefore appears that if a phase of hierarchical assembly has occured in Cupavo, it has most likely already concluded, which is not surprising given its older age.

While these results suggest that other groups in Cinyras and Orpheus will have opportunities to merge with RSG-5 and $\delta$ Lyr, the velocities at closest approach tend to be quite high relative to their $\sigma_{virial}$ values of 0.23 and 0.41 km s$^{-1}$, respectively. Typical velocities at closest approach to one of our clusters are between 1 and 3 km s$^{-1}$, indicating that most of the stars in those groups are likely to survive their encounters without agglomeration into the cluster. Furthermore, over the long timescales between now and some of these potential interactions, cluster erosion by gravitational interactions with the field will substantially reduce the gravitational potential of these clusters, making them even less able to capture new stars. These two factors combined suggest that while parts of the central region of Cinyras and Orpheus shows signs of convergence, the groups there are unlikely to merge completely.





\section{Discussion}\label{sec:discussion}

\subsection{Relationship of Cep-Her Associations to Each Other} \label{sec:relbtwass}

In this paper, we have found that Cep-Her is not a singular association, but rather a complex that consists of at least four distinct young structures, including three dynamically coherent and highly substructured young associations. While these populations blend substantially in the 5D phase space in which Cep-Her was defined, their differences in age and velocity cast doubt on whether they had any effect on one another at formation. It is therefore important to consider whether the different components of Cep-Her have plausible connections to one another, or whether they can be treated as fully independent from one another in future studies of their star formation histories. 

We therefore compute bulk mean positions and velocities for each association in Cep-Her, and assess the possibility that they were close enough to interact during the process of formation. We use a stringent set of cuts that includes only velocity members with $P_{\rm fin}>0.8$, no stellar companions, and $\sigma_{RV}<1$ km s$^{-1}$. We use those stars compute a median UVW velocity vector of (U,V,W) = (-2.58, -17.47, -7.82) for Orpheus,  (-6.42, -6.48, -3.93) for Cinyras, (-9.76, -18.71, -6.14) for Cupavo, and (-9.83, -5.97, -5.95) for Roslund 6. The largest gap between the velocities of these groups is in the V component, where $\sim12$ km s$^{-1}$ separates the groups comprising Cep-Her North (Roslund 6 and Cinyras) from those comprising Cep-Her Main (Cupavo and Orpheus). Using these velocities, we can compute an approximate linear distance travelled since formation, which provides a reasonable approximation for their relative motion on timescales substantially shorter than the $\sim200$ Myr galactic rotation period near the sun. Since 1 km s$^{-1} = 1.023$ pc Myr$^{-1}$, and given a minimum time since formation of $\sim 25$ Myr , this $\sim12$ km s$^{-1}$ velocity difference propagates to a linear positional difference of at least 300 pc at formation. Since the centers of these populations are separated by less than 150 pc in the present day, this fact alone effectively excludes the possibility that Cupavo and Orpheus directly interacted with Cinyras and Roslund 6 at formation. 

However, the possibility remains for connections between Cinyras and Roslund 6, and between Cupavo and Orpheus. The velocity vectors for Cinyras and Roslund 6 are separated by only 4 km s$^{-1}$, however this still corresponds to a substantial relative motion of 114 pc since the last generation of star formation in Cinyras ended $\sim$28 Myr ago. Roslund 6 has a velocity offset from that of Cinyras primarily in the negative U direction, which, combined with its position on the +X side of Cinyras, implies that Roslund 6 is currently entering Cinyras, not diverging from it. Cinyras and Roslund 6 are therefore likely the closest they have ever been to one another, making connections at formation unlikely.  

Finally, Orpheus and Cupavo have velocity vectors separated by 7.47 km s$^{-1}$, with Cupavo's velocity offset relative to Orpheus being almost entirely in the -U direction. While this propagates to a position change of 191 pc since the formation of Orpheus' youngest generation, Orpheus is also offset from Cupavo by a similar distance in the +X direction, implying that Cupavo may have been passing near Orpheus's parent cloud around the time it was forming stars. This suggests that Cupavo may have had a role in the formation of Orpheus, potentially through triggering from an increase in stellar feedback as Cupavo approached Orpheus' parent cloud. This motivates a future study covering both Orpheus and Cupavo that combines revised ages with traceback through the galactic potential, which can assess whether the precise positions at the moment of formation are consistent with some sort of triggering relationship between the two associations.

\subsection{Age Structure}

Age trends in spatial coordinates have long been used to indicate potential star formation mechanisms \citep[e.g., ][]{Krause18, Posch23}. Patterns of increasing or decreasing age along an axis have garnered particular interest recently, as they have been taken as evidence for sequential star formation: a process in which stellar feedback compresses gas ahead of it, producing a new stellar generation that can trigger the next star formation event with its feedback \citep{Elmegreen77, Kerr21}. While the gold standard for understanding star formation histories is to reconstruct their history using ages and 3D motions, the ages must be very accurate on an absolute scale to allow the correct location at formation to be identified. Especially in the age range of Cep-Her, isochronal ages can disagree with one another by a factor of 2.5 or more \citep{Herczeg15}, so a dedicated study to better constrain the absolute age-scaling in Cep-Her will be necessary to reconstruct its star formation history. However, isochronal ages are much more accurate on a relative scale, allowing for the detection of patterns like those produced by sequential star formation.

\begin{figure}
\centering
\includegraphics[width=8cm]{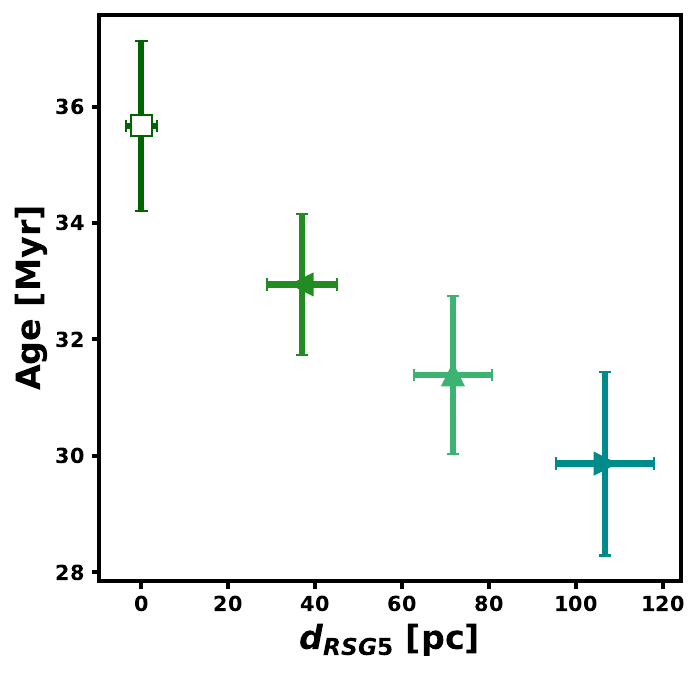}\hfill
\caption{The age sequence across the nearly linear structure in the galactic north of Cinyras, shown as the distance from RSG-5 vs Age. The uncertainty in distance we display is the half-mass radius from Section \ref{sec:boundedness}. The markers are the same as in Figures \ref{fig:cinyras} and \ref{fig:cinr-sim}}
\label{fig:cinr-aseq}
\end{figure}

Of our three substructured associations, Cinyras has the most evidence for meaningful age patterns. The galactic north ($+b$) of the association contains a structure that connects the subgroups CINR-6, 5, and 7 with a nearly straight line, before curving slightly and ending in RSG-5 (CINR-10) (see Figs. \ref{fig:CHNF_Ages}, \ref{fig:cinyras}). We show the age sequence as a function of distance from RSG-5 across this set of subgroups in Figure \ref{fig:cinr-aseq}. Ages along this sequence range from 36 Myr in RSG-5 in the galactic west ($-l$) to 30 Myr in CINR-6 in the galactic east. This section of Cinyras is therefore consistent with sequential star formation spanning 6 Myr starting at the open cluster RSG-5. Cinyras therefore appears to have very similar morphology to Vela OB2, where there is evidence for sequential star formation spanning $\sim4$ Myr originating from the $\gamma$ Vel cluster \citep{Pang21}. Outside of the star formation sequence in the galactic north of Cinyras, age sequences are less evident. However, most of the more scattered populations can likely be explained through either smaller bursts of star formation that precede the events that produced the main star formation sequence, or formation in feedback-driven gas fragments that did not get ejected along the main direction of sequential star formation propagation.

The star formation patterns in Orpheus are much less clear, as its ages are distributed in a much less orderly way. Star formation in this association formation takes place entirely within the 15 Myr range from 25 to 40 Myr, with very little indication of pauses in the star formation sequence over that period. Any spatial age trends are tenuous at best, and there is no indication of substantial age differences between the potentially independent regions of diverging stellar motion indicated in Figure \ref{fig:vectorplots}. Two of the open clusters, $\delta$ Lyrae and ORPH-16 both have ages above the median for the association, suggesting that they may have formed first, allowing their stellar feedback to drive the evolution of later generations. However, the other probable fully bound structure in Orpheus, ORPH-15, is one of the youngest groups in the entire association. Due to its unclear age trends in spatial coordinates, it appears that a complete dynamical reconstruction of Orpheus' star formation history like that provided in SPYGLASS-II will be necessary to fully unravel the patterns that guided its evolution. 

Age trends are particularly difficult to detect in the older Cupavo association, as the slower evolution of the pre-main sequence in its age range reduces the precision of the isochronal ages. However, we can split the population into two broad age categories: young (54-58 Myr; CUPV-3, 8, 9, and 10), and old ($>$63 Myr; CUPV-1, 2, 4, 5, 6, and 7). The young component consists of most of the more distant subgroups in Cupavo and also contains most of the association's mass, including the probable open clusters of NGC 7058 (CUPV-3) and CUPV-10. The old component is largely closer to the sun, and occupies a wider span of ages. While the mean ages of the old and young components are separated by 14 Myr, the youngest group in old component is only 4 Myr older than the oldest subgroup member of the young component. It is therefore possible that the young and old components emerged from the same star formation event, given the age spreads of 10-25 Myr seen in groups like CFN and Sco-Cen \citep[e.g.,][]{Pecaut16, Kerr21, Kerr22a}. The potential for a common origin between the young and old components is enforced by the very similar velocities seen between subgroups CUPV-5, 6, and 7 and those in the young component, as shown in Figure \ref{fig:cupavo}. However, velocities in CUPV-1, 2, and 4 differ from the young component much more dramatically, which may suggest that at least part of the old component may have a distinct origin. The combination of a large age scatter in the old component with the imprecise pre-main sequence ages available here leaves open the possibility of multiple generations of distinct star formation existing in Cupavo, motivating follow-up studies to test whether the old component truly is contiguous with the young component, or whether it is an older interloping dissolving structure. 

\begin{deluxetable*}{cccccccccccccc} 
\tablecolumns{14}
\tablewidth{0pt}
\tabletypesize{\scriptsize}
\tablecaption{Kepler planet candidate hosts in our catalog of Cep-Her members, including the planet ID, basic information on the planet host and its membership within Cep-Her, and some basic planet properties.}
\label{tab:planethosts}
\tablehead{
\colhead{Name} &
\colhead{Gaia DR3 ID} &
\colhead{AS} &
\colhead{SG} &
\colhead{RA} &
\colhead{Dec} &
\colhead{Disp.\tablenotemark{a}} &
\colhead{Score\tablenotemark{b}} &
\colhead{MES\tablenotemark{c}}&
\colhead{$P_{fin}$}&
\colhead{Im. Par.}&
\colhead{Per.}&
\colhead{Tr. Depth}&
\colhead{$R_{p}$}\\
\colhead{} &
\colhead{} &
\colhead{} &
\colhead{} &
\colhead{(deg)} &
\colhead{(deg)} &
\colhead{} &
\colhead{} &
\colhead{} &
\colhead{} &
\colhead{} &
\colhead{(days)} &
\colhead{(ppm)} &
\colhead{($R_{\oplus}$)}
}
\startdata
 K01121.01 & 2101855569017361280 & ORPH &      -1 & 290.3544 & 41.7691 &              FP &      0.000 &           6249.4 & 0.0585 &      1.1000 &        14.2 &    61625.0 &     53.25 \\
 K03399.01 & 2101855569017361280 & ORPH &      -1 & 290.3544 & 41.7691 &              FP &      0.000 &            563.9 & 0.0585 &      1.1990 &        14.2 &    21653.0 &     29.61 \\
 K02031.01 & 2103714637018491136 & CINR &      -1 & 283.8994 & 41.2216 &             CON &      1.000 &             22.2 & 0.2545 &      0.4880 &         9.3 &      637.3 &      1.67 \\
 K02752.01 & 2076615077014515968 & CINR &      -1 & 295.3240 & 40.9022 &              FP &      0.000 &             13.3 &     &      0.8190 &         1.3 &       36.5 &      0.87 \\
 K00061.01 & 2078912467894820096 & CINR &      -1 & 296.9360 & 44.1611 &              FP &      0.000 &             66.7 & 0.7042 &      1.1900 &         1.6 &      359.0 &     24.87 \\
 K03090.01 & 2077668752749671424 & CUPV &      -1 & 292.7377 & 42.0148 &             CON &      0.614 &             14.1 & 0.0335 &      0.7250 &         3.8 &      362.7 &      1.05 \\
 K03090.02 & 2077668752749671424 & CUPV &      -1 & 292.7377 & 42.0148 &              FP &      0.022 &              8.3 & 0.0335 &      1.1150 &        15.0 &      503.5 &      8.02 \\
 K07638.01 & 2099187088655077760 & ORPH &      -1 & 286.7015 & 38.0422 &             CAN &      0.760 &              8.4 & 0.0353 &      0.7560 &        25.0 &       64.1 &      1.47 \\
 K07655.01 & 2052871363926844928 & ORPH &      -1 & 290.1573 & 38.5244 &              FP &      0.000 &             11.6 & 0.8383 &      0.0640 &         3.0 &      259.5 &      1.02 \\
 K04336.01 & 2101037257488816640 & CUPV &      -1 & 289.7563 & 39.5764 &              FP &         &               & 0.0554 &      0.5122 &        41.5 &       29.9 &      1.09 \\
 K04600.01 & 2099191211824945792 & ORPH &      -1 & 286.3734 & 37.9190 &              FP &      0.000 &              8.3 &     &      0.9920 &         0.9 &       33.0 &      1.29 \\
 K02742.01 & 2076790552200353536 & ORPH &      -1 & 297.0738 & 41.2245 &             CON &      1.000 &             19.0 & 0.3205 &      0.3760 &         0.8 &      215.1 &      0.91 \\
 K04004.01 & 2103314346067173504 & CUPV &      -1 & 283.9951 & 39.9170 &             CON &      1.000 &             18.5 & 0.6020 &      0.9280 &         4.9 &      126.2 &      1.41 \\
 K05218.01 & 2076787562903136128 & ORPH &       3 & 297.1043 & 41.2199 &              FP &      0.000 &             10.6 &     &      1.1840 &         1.6 &      292.2 &     31.60 \\
 K05264.01 & 2104068095647479936 & ORPH &      25 & 285.1417 & 41.7116 &             CAN &         &               & 0.9898 &          &        52.7 &         &        \\
 K05024.01 & 2099646856314081024 & CINR &      -1 & 287.4060 & 39.0209 &              FP &         &               & 0.9738 &      0.5987 &       282.2 &      196.3 &      3.09 \\
 K05632.01 & 2080175085205622656 & CUPV &       7 & 296.3381 & 45.5451 &             CAN &      0.928 &             29.0 & 0.0834 &      0.1290 &       264.5 &      437.3 &      2.61 \\
 K05988.01 & 2100154452728605312 & ORPH &      -1 & 285.0551 & 38.9850 &              FP &      0.000 &            143.2 & 0.9760 &      1.2220 &        30.5 &    18441.0 &     23.87 \\
 K05997.01 & 2101298868241975424 & CUPV &      -1 & 288.4722 & 40.0875 &              FP &      0.000 &           6106.5 &     &      0.9360 &         4.4 &    25342.0 &     26.08 \\
 K07871.01 & 2075875792887385472 & CUPV &       9 & 301.0396 & 44.0361 &              FP &      0.000 &             10.0 & 0.6751 &      1.2230 &         1.2 &      966.7 &     18.39 \\
 K03774.01 & 2101650166495312512 & ORPH &      -1 & 291.9855 & 41.5564 &              FP &         &               &     &      1.2672 &         3.7 &     1114.6 &     22.78 \\
 K06277.01 & 2099133865420292992 & ORPH &      -1 & 286.8454 & 37.8159 &              FP &      0.000 &           3760.1 & 0.0352 &      1.1720 &         0.6 &    34770.0 &     72.32 \\
 K06812.01 & 2125788428159994368 & ORPH &      -1 & 291.5863 & 42.5099 &              FP &         &              7.7 & 0.6061 &      0.0063 &         0.6 &        7.5 &      0.27 \\
 K06437.01 & 2103497792709244544 & ORPH &      26 & 283.7055 & 39.8705 &              FP &      0.000 &             11.2 & 0.9782 &          &         1.2 &         &        \\
 K07375.01 & 2086715156462908672 & CINR &      -1 & 297.3044 & 48.1952 &             CAN &      0.996 &             12.8 & 0.9298 &      0.1910 &         4.8 &      554.7 &      1.51 \\
 K07572.01 & 2077884428827603456 & ORPH &      -1 & 295.5134 & 42.3188 &             CAN &         &              7.4 & 0.9915 &      0.5939 &        91.1 &       12.7 &      1.59 \\
 K08007.01 & 2129185923389399808 & ORPH &       3 & 291.2268 & 47.5500 &             CAN &      0.804 &              9.1 & 0.9768 &      1.2360 &        67.2 &     1689.9 &      9.66 \\
\enddata
\tablenotetext{a}{Kepler disposition: False positive (FP), Candidate (CAN), or Confirmed (CON)}
\tablenotetext{b}{Kepler disposition score}
\tablenotemark{c}{Maximum multiple event statistic, which has been used as a proxy for signal-to-noise for vetting transiting planet candidates \citep{Jenkins02}}
\end{deluxetable*}

\subsection{Membership of Planets}

The Cep-Her Complex overlaps substantially with the Kepler field, making it a compelling environment for discovering young transiting planets. \citet{Bouma22} used preliminary clustering based on the SPYGLASS-IV Cep-Her membership list to identify four new young planets. The three smallest objects, Kepler-1643 b ($R = 2.32 \pm 0.13 R_{\oplus}$), KOI-
7368 b ($R = 2.22 \pm 0.12  R_{\oplus}$), and KOI-7913 Ab ($R = 2.34 \pm 0.18  R_{\oplus}$) are all probable mini-Neptunes, and all three were the smallest planets ever discovered in their age bracket at the time of publication \citep{Bouma22}. The final planet is approximately the size of Neptune (Kepler-1627 Ab ; $R = 3.85 \pm 0.11 R_{\oplus}$). 

Assigning young planets to a parent stellar population is critical for contextualizing their discoveries, as the host association provides both ages, which place the planet within the sequence of rapid evolution that occurs within 100 Myr of formation, and the environment in which they formed \citep[e.g.,][]{Owen13, Ginzburg18}. \citet{Bouma22} originally placed Kepler-1643 in RSG-5, Kepler-1627 in $\delta$ Lyr, and both KOI-7368 and KOI-7913 in a low density preliminary Cep-Her subgroup that was referred to as ``CH-2''. Our new membership catalogs confirm RSG-5 (CINR-10) as the parent group for Kepler-1643, while assigning Kepler-1627 to ORPH-25, a smaller overdensity directly adjacent to $\delta$ Lyr. Finally, we do not identify a single population in the location of CH-2 in \citet{Bouma22}, but rather two populations that overlap in transverse velocity but separate cleanly in radial velocity: CINR-1 and ORPH-3. The result is that the planet host stars are assigned to not just different subgroups, but entirely different associations, with KOI-7368 in CINR-1 and KOI-7913 in ORPH-3. 

Our reassignment of the planets in Cep-Her to new parent populations therefore reveals that rather than having origins in one contiguous population, these planets are split between the Orpheus and Cinyras associations, with two planets in each. All of the planet-hosting subgroups have similar ages, so these subgroup reassignments do not substantially change the position of these planets within the planetary evolution sequence. However, it does emphasize the importance of radial velocity follow-up in assigning planets to parent populations, as in our complex galaxy, stellar populations that are seemingly contiguous in 5D phase space may not be in 6 dimensions. Furthermore, the assignment of these groups to different associations may inform key properties of the formation environment such as gas density and stellar feedback \citep[e.g.,][]{Adams06, Davies14, Winter18}, a question that can be further explored through future reconstructions of these associations' star formation histories. These association surveys may therefore have an important role in establishing the environmental dependence of star formation once larger planet samples are established. 

To locate any unexplored planet candidates in Cep-Her, we cross-match our stellar catalog with the KOI cumulative tables available on the NASA Exoplanet Archive \citep{NEA_Kepler}\footnote{Accessed on 2024-05-29 at 02:01, returning 9564 rows} and with the TOI catalog. We compile a list of KOI objects that overlap with our Cep-Her member catalog in Table \ref{tab:planethosts}, which includes 10 KOI objects that are assigned Kepler dispositions of ``Candidate'' and 17 that are marked as false positives. Four of these planets are flagged as confirmed, while an additional three candidate planets have Kepler disposition scores of 0.8 or higher, and 4 have $P_{\rm fin}>0.8$. Two candidate planets, KOI-7375.01 and KOI-8007.01, have both disposition scores above 0.8 and $P_{\rm fin}>0.8$. KOI-8007 is a particularly compelling planet host, with membership in ORPH-3 with $P_{\rm fin} = 0.98$, and it has a 9.66 $R_{\oplus}$ candidate planet around it identified based on a single eclipse with a disposition score of disposition score of 0.804. We also identify 5 TOI objects in our Cep-Her membership catalog: TOI-1286, TOI-3516, TOI-1162, TOI-5782, and TOI-3560, although all but TOI-1162 have $P_{\rm fin} \la 0.05$. TOI-1162 has $P_{\rm fin} = 0.96$ and membership in CUPV-8, which gives it an isochronal age of $55.7\pm4.2$ Myr. This TESS planet candidate, along with Kepler targets like KOI-8007, may therefore make compelling targets for future follow-up. The upcoming PLAnetary Transits and Oscillations (PLATO) mission's proposed northern Long-Duration Observing Phase (LOP) field covers more than half of the Cep-Her Complex, so more planet discoveries are likely once PLATO observations begin \citep{Nascimbeni22}. 

We separately investigated any connections with the planets identified in \citet{Barber22}, which are grouped into a $\sim$105 Myr association named MELANGE-3. None of these planets overlap with our core sample for Cep-Her, and they are therefore not assigned to any subgroups. However, one of them is in our maximally broad sample of $1.5\times10^5$ stars, suggesting that connections to MELANGE-3 may emerge with further exploration of the older populations in Cep-Her. 

\section{Conclusion} \label{sec:conclusion}

We have performed the first dedicated survey of the recently discovered and large ($N_{stars} > 20{,}000$) Cep-Her complex. By combining \textit{Gaia} photometry and astrometry with RVs from \textit{Gaia}, the literature, and our own new observations, we have generated a 6-dimensional view of the positions and velocities of stars in Cep-Her, which combine with PARSEC isochronal ages to reveal probable structures of common formation and evolution in Cep-Her. The result is a detailed structural and demographic survey of the Cep-Her complex. The key findings of this survey are as follows: 

\begin{enumerate}
    \item The Cep-Her complex consists of not one young population, but four. Three of these populations are highly-substructured young associations which we call Orpheus, Cinyras, and Cupavo, while the fourth is Roslund 6, which is dominated by a known and older ($> 100$ Myr) open cluster. 
    \item Two of the three substructured populations have similar ages, spanning 28-43 Myr in Cinyras and 25-40 Myr in Orpheus. Cupavo is somewhat older, with ages spanning from 54 Myr to our 80 Myr age limit. 
    \item The velocities of Cupavo and Orpheus indicate that Cupavo may have been close enough to influence the evolution of Orpheus' parent cloud. All other pairs of associations appear to have been more than 100 pc apart at formation. 
    \item All three substructured associations are among the largest known within 500 pc, with the largest, Orpheus, containing $9552 \pm 960$ members and $3881 \pm 390$ M$_{\odot}$, comparable in scale to Sco-Cen. Cinyras and Cupavo are smaller, with $3872 \pm 455$ and $8794 \pm 1827$ members, respectively, although they remain comparable in size to many large and well-known young associations, such as Vela OB2. 
    \item All three associations have dynamics consistent with gradual expansion. However, Orpheus appears to have more than one origin of expansion, suggesting the presence of multiple sites of star formation within the larger association, and some of the older populations in Cupavo may be dynamically distinct.  
    \item We see clear evidence for sequential star formation in Cinyras stretching from the 36 Myr old RSG-5 cluster to the  30 Myr old CINR-6 subgroup. 
    \item Forward simulations and velocities show slow or even convergent motions across massive central clusters and populations in both Orpheus and Cinyras, despite clear expansion among the more outlying and mainly unbound populations. This suggests that the outlying unbound populations were subject to accelerations from stellar feedback or gravitational collapse that did not affect the motions of the central populations and open clusters. 
    \item We update the membership of the four planet hosts in Cep-Her discussed in \citet{Bouma22}, confirming Kepler-1643's membership in RSG-5 (CINR-10) and assigning Kepler-1627 to ORPH-25, Kepler-1974 (KOI-7368) to CINR-1, and Kepler-1975 (KOI-7913) to ORPH-3. These new group assignments do not amend the currently proposed ages of these planets. We also find that Cep-Her may contain additional young planets, with 10 candidate or confirmed KOI objects, and one compelling TOI candidate. 
\end{enumerate}

Our results cement Cep-Her as one of the most massive and complex stellar populations in the solar neighborhood. The varied morphologies of the stellar associations within provide numerous laboratories for the further study of star formation processes, as well as planetary evolution. Cep-Her is therefore a strong candidate for follow-up age studies and the further refinement of its kinematics, which combine to enable a complete reconstruction of the associations' star formation histories. Such studies can shed further light on processes like sequential star formation, while revealing new star formation patterns that can help to refine our models of star formation. \\

RMPK was supported in part by funding from the Heising-Simons Foundation and by the NASA TESS grants 80NSSC21K0708 and 80NSSC22K0302 (PI Tofflemire). RMPK acknowledges the use of computational  resources  at  the  Texas  Advanced Computing Center (TACC) at the University of Texas at Austin, which was used for the reduction and analysis of our spectra. RMPK also acknowledges the support and mentorship of Aaron Rizzuto, which was essential to the development of the SPYGLASS program. 

%

\vspace{5mm}
\facilities{Gaia,  McDonald Observatory: Tull spectrograph/Harlan J. Smith Telescope, W. M. Keck Observatory: HIRES Spectrograph/Keck I Telescope}


\software{astropy \citep{Astropy13,Astropy18},  
          numpy \citep{numpy},
          saphires \citep{Tofflemire19},
          HDBSCAN \citep{McInnes2017}
          AMUSE \citep{PortegiesZwart09,Pelupessy13,PortegiesZwart13,PortegiesZwart18},
          galpy \citep{Bovy15},
          matplotlib \citep{Hunter07}
          }


\appendix
\section{Binaries} \label{app:bin}

\begin{deluxetable*}{ccccccc} 
\tablecolumns{7}
\tablewidth{0pt}
\tabletypesize{\scriptsize}
\tablecaption{Catalog of binaries in Cep-Her. Stars in the same system are assigned the same system ID. The results below show the first 12 entries, the full catalog is available in the online-only version of this paper. We also include some basic stellar properties, including on-sky position, distance, and Gaia apparent magnitude. }
\label{tab:binaries}
\tablehead{
\colhead{Gaia DR3 ID} &
\colhead{Sys ID} &
\colhead{$R$\tablenotemark{a}} &
\colhead{RA} &
\colhead{Dec} &
\colhead{d}&
\colhead{m$_G$} \\
\colhead{} &
\colhead{} &
\colhead{AU} &
\colhead{(deg)} &
\colhead{(deg)} &
\colhead{(pc)} &
\colhead{}
}
\startdata
2034781717635002112 &          0 &  2618 & 297.3814 & 33.4200 & 346.22180 & 18.16 \\
2034781713288926208 &          0 &     0 & 297.3811 & 33.4220 & 412.44040 & 17.76 \\
2034787318222313600 &          1 &     0 & 295.9200 & 32.2261 & 299.24880 & 20.41 \\
2034787318228665472 &          1 & 10833 & 295.9110 & 32.2215 & 338.49377 & 20.74 \\
2034833016686867328 &          2 &  8217 & 296.4336 & 32.8453 & 265.39767 & 20.72 \\
2034832947967381632 &          2 &     0 & 296.4438 & 32.8445 & 236.43692 & 20.64 \\
2034879161850632192 &          3 &     0 & 295.4506 & 32.9278 & 366.45180 & 11.63 \\
2034879157515942528 &          3 &   519 & 295.4506 & 32.9282 & 375.09125 & 12.96 \\
2034902831374133120 &          4 &  9455 & 296.0376 & 33.1534 & 321.04858 & 17.88 \\
2034902835693326720 &          4 &     0 & 296.0439 & 33.1597 & 384.79266 & 17.77 \\
2035025843553925376 &          5 &     0 & 296.2579 & 33.6914 & 370.74838 & 15.30 \\
2035025843574065408 &          5 &  2050 & 296.2579 & 33.6898 & 355.86722 & 18.43 \\
2035128231251935360 &          6 &     0 & 298.5529 & 33.6560 & 326.71090 & 18.65 \\
2035128166838684416 &          6 &  5000 & 298.5562 & 33.6521 & 291.28098 & 19.75 \\
\enddata
\tablenotetext{a}{Linear separation from the primary, assuming the distance to the primary holds for all companions. Primaries have a separation of zero.}
\vspace*{0.1in}
\end{deluxetable*}

Here we provide a complete catalog of potential resolved binaries in Cep-Her. As mentioned in Section \ref{sec:idbin}, we identify companions by searching a 10000 AU radius around each star, identifying companions as objects in that radius that have $\Delta v_T < 3$ km s$^{-1}$ and $\frac{\Delta\pi}{\pi} < 0.2$. This ensures that the objects are credibly co-spatial and co-moving at a level consistent with what is typically seen in binaries. We search for binary companions to all 152936 candidate members, which, for completeness, includes stars that fail our $P_{\rm spatial}>0.2$ cut. The complete catalog of potential binary components includes 10183 stars across 5032 systems. 



\bibliography{cepher}{}
\bibliographystyle{aasjournal}



\end{document}